\documentclass[12pt,draftclsnofoot,onecolumn]{IEEEtran}
\usepackage{caption}
\usepackage{subfigure}
\usepackage{moreverb}
\usepackage{epsfig}
\usepackage{amsmath,amssymb,amsthm,mathrsfs,amsfonts,dsfont}
\usepackage{adjustbox,lipsum}
\usepackage{algorithm,algorithmic}
\usepackage{amsfonts}
\usepackage{epsfig}
\usepackage{amssymb}
\usepackage{amsmath}
\usepackage{amsthm}
\usepackage{multirow}
\usepackage{rotating}
\usepackage{graphicx}
\usepackage{tabularx}
\usepackage{array}
\usepackage{anyfontsize}
\usepackage{color,soul}
\usepackage{graphicx,dblfloatfix}
\usepackage{epstopdf}
\usepackage{blindtext}
\usepackage{amsmath}
\usepackage{amsthm,amssymb,amsmath,bm}
\usepackage{amsfonts}
\usepackage{epsfig}
\usepackage{amssymb}
\usepackage{amsmath}
\usepackage{cite}
\usepackage{graphicx}
\usepackage{fancyhdr}
\usepackage[font={small}]{caption}
\usepackage{tabularx}
\usepackage{cite}
\usepackage{color}

\usepackage{colortbl}

\begin{document}
	
\title{Fairness and Transmission-Aware Caching and Delivery Policies in OFDMA-Based HetNets}
	\author{Sepehr Rezvani,
		Nader Mokari, \IEEEmembership{Member, IEEE},
        Mohammad R. Javan, \IEEEmembership{Member, IEEE},
		and Eduard A. Jorswieck, \IEEEmembership{Senior Member, IEEE}
		%\thanks{Manuscript received December 13, 2014.}
		\thanks{Sepehr~Rezvani and Nader~Mokari are with the Department of Electrical and Computer Engineering, Tarbiat Modares University, Tehran, Iran.} \thanks{Mohammad~R.~Javan is with the Department of Electrical and Robotics Engineering, Shahrood University of Technology, Shahrood, Iran.}
\thanks{Eduard A. Jorswieck is with the Dresden University of Technology, Communications Laboratory, Chair of Communication Theory, Dresden, Germany.}}
\maketitle
\begin{abstract}
Recently, wireless edge caching has been emerged as a promising technology for future wireless networks to cope with exponentially increasing demands for high data rate and low latency multimedia services by proactively storing contents at the network edge. Here, we aim to design efficient cache placement and delivery strategies for an orthogonal frequency division multiple access (OFDMA)-based cache-enabled heterogeneous cellular network (C-HetNet) which operates in two separated phases: caching phase (CP) and delivery phase (DP). Since guaranteeing fairness among mobile users (MUs) is not well investigated in cache-assisted wireless networks, we first propose two delay-based fairness schemes called proportional fairness (PF) and min-max fairness (MMF). The PF scheme deals with minimizing the total weighted latency of MUs while MMF aims at minimizing the maximum latency among them. In the CP, we propose a novel proactive fairness and transmission-aware cache placement strategy (CPS) corresponding to each target fairness scheme by exploiting the flexible wireless access and backhaul transmission opportunities. Specifically, we jointly perform the allocation of physical resources as storage and radio, and user association to improve the flexibility of the CPSs. Moreover, In the DP of each fairness scheme, an efficient delivery policy is proposed based on the arrival requests of MUs, CSI, and caching status. Numerical assessments demonstrate that our proposed CPSs outperform the total latency of MUs up to $27\%$ compared to the conventional baseline popular CPSs.
\newline
\emph{\textbf{Index Terms--}} Transmission-aware caching,\, delivery policy,\, fairness,\, latency,\, OFDMA,\, HetNet,\, 5G networks.
\end{abstract}

\section{Introduction}
\subsection{Introduction and Related Works}
\IEEEPARstart{S}{ervice} providers should handle the predicted data traffic growth in the future fifth-generation (5G) communication networks with low latency multimedia services \cite{wong_schober_ng_wang_2017,6871674}.
Research results on traffic explosion issue show that most of the global mobile data traffic is due to frequently downloading a modest number of contents from data centers (DCs)  \cite{7155502,7530876}. Besides, duplicated transferring contents through the backhaul links wastes more backhaul resources and also increases the latency of mobile users (MUs) \cite{6195469,7530876,7322204,7155502,7828114,CloudPrinciples}.

Recently, cache-enabled heterogeneous cellular networks (C-HetNets) have been proposed as a potential solution to cope with the backhaul capacity and availability bottleneck by equipping base stations (BSs) with storage capacity \cite{6871674,6195469,7155502}. By means of content caching at the network edge, popular contents can be prefetched from local caches, e.g., BSs, instead of duplicated prefetching from DCs through the scarce and expensive backhaul links \cite{6871674,6195469}.
In this way, the content delivery performance is significantly improved, especially in terms of latency \cite{6195469,7155502,8008769}, and in terms of backhaul data traffic which decreases the system delivery cost \cite{7322204,7828114}.

Generally, cache-enabled wireless networks operate in two phases called caching phase (CP) and delivery phase (DP) \cite{8008769,8030120,7155502,7558153}. The CP which runs at the off-peak times \cite{8030120,7828114,7155502} determines which content should be cached in which storage and the resulting cache placement strategy (CPS) is valid  during a long time where the popularity distribution information (PDI) of contents remains constant \cite{8030120,7558153}.
On the other hand, the DP operates during all the network serving time where the requested contents are delivered to MUs using the available transmission resources\footnote{Note that the PDI changes slowly and the CP results would be valid for a longer time, during which, the next DP is performed \cite{8030120,7558153}.} \cite{7828114,7805409,7558153}.

Fairness is another important issue which should be considered in all resource allocation strategies \cite{5580131,1424595,Aquantitativemeasure}. The concept of fairness has been discussed in many contexts such as
economics, computer, and telecommunication systems \cite{1424595,Aquantitativemeasure}. In the area of networking, software defined networking (SDN) enables the network programming capability via a logically central software defined (SD) controller and separates the control plane from the data plane \cite{7406764,7514219,7744833}. By utilizing SDN, some important areas as caching, transmission, and fairness can be jointly considered while the fast and efficient resource allocation in C-HetNets is guaranteed \cite{7406764,7925732,7744833,7514219}.

%The limitations of storage capacities in C-HetNets is on of the major challenges which should be carefully handled by the network operators.
Recently, to avoid duplicated content caching at nearby local caches, distributed CPSs have been developed where the opportunity of MUs to access to multiple storages is considered \cite{6195469,7227022,7155502,8030120,7510807,6600983}. In this regard, studying the benefits of transmission-aware CPSs by coupling the CPS and physical-layer transmission have attracted more attentions \cite{6195469,7150324,7227022,8008769,7155502,7536886,7530876,7406764,VASILAKOS2016306}. The accessibility of MUs to the edge devices seriously depends on the wireless channel capacities which can be improved by designing an efficient transmission policy. In this way, the performance of CPSs can be further improved by jointly optimizing the storage and radio resources, and MU association policy.
From the resource allocation perspective, fairness is critical in cases where a set of resources is shared among several individuals/users. It should be noted that considering fairness affects both the design of the CPSs \cite{Fairnessawarecooperative} and the delivery policy \cite{5580131,5463218}.

The recent works studying cache-assisted wireless networks fall into two main categories: 1) devising efficient delivery policies for a given CPS; 2) designing CPSs to have an efficient content delivery.
In the first category, some research works investigate the benefits of heuristic baseline popular CPSs and delivery policies in cache-aided wireless networks \cite{7558153,7828114,7805409,7150324,7925732}.
By adopting different heuristic CPSs in a fog-radio access network (RAN), the authors in \cite{7558153} design efficient delivery policies to maximize the delivery rate under fronthaul capacity and per-enhanced remote radio head (RRH) transmit power constraints. In \cite{7805409}, the authors devise a joint RRH selection and subcarrier and transmit power allocation algorithm for the DP of an orthogonal frequency division multiple access (OFDMA)-based cloud-RAN. The trade-off between system throughput and outage probability is also investigated in \cite{7297864,7150324}.
Besides, the authors in \cite{7828114} propose several power allocation algorithms for different objective functions in a cache-aided backhaul-limited small-cell network. In addition, the authors in \cite{7925732} devise a joint MU association and bandwidth allocation for a certain caching status.
Since the proposed solution is based on a centralized manner, a SD scheduler is applied to have a fast and efficient centralized resource management.
In the second category, i.e., developing a transmission-aware CPS, several research works have been published \cite{6195469,7227022,8008769,7322204,7155502,8030120,7536886,7530876,7406764,7417343,7158137,VASILAKOS2016306}.
The idea of femtocaching in C-HetNets was first proposed in \cite{6195469} where all femto BSs (FBSs) are treated as caching helper nodes with low-cost storages while connecting to a macro BS (MBS) with low-rate backhaul links. Moreover, the cache placement operation is assumed to be performed at low demand times. By using the alternating direction method of multipliers approach, the authors in \cite{7227022} propose a distributed CPS in mobile cellular networks which is based on the backhaul bandwidth consumption and storage capacities.
The works in \cite{7158137,VASILAKOS2016306} focus on devising mobility-aware CPSs based on the available PDI and mobility patterns of MUs.
In this way, the authors at first propose some methodology approaches to predict the mobility treatment of MUs. Although these works consider the mobility of MUs in different moments of DP, they do not exploit the benefits of jointly allocating storage and radio resources to improve the flexible access node selection opportunities in the system. In \cite{7322204}, with the availability of MU's requests, a transmission-aware CPS is proposed for a single-cell LTE system with device to device communications to reduce both access and backhaul traffic subject to storage capacity and delivery deadline constraints. In \cite{7155502}, the authors propose a CPS to minimize the average delay of MUs in C-HetNets by assuming an unlimited storage capacity for MBS and fixed channel state information (CSI) of access links which is not practical assumptions.
A cooperative CPS for fog-RANs is devised in \cite{8008769} with the aim of minimizing the average delay of MUs under storage capacity constraint.
However, the backhaul delay analysis and the effect of backhaul radio resource allocation (RRA) on the performance of CPSs is neglected. Moreover, they assume a fixed signal-to-interference-plus-noise ratio (SINR) and estimated interferences in RAN which reduces the flexibility and practicality of the proposed CPS. Fixed CSI and SINR are also considered in \cite{7530876} and \cite{8030120} for the design of CPSs in C-HetNets to minimize the total latency at MUs and maximize MUs data rates, respectively.

Different schemes have been developed for wireless networks to guarantee the fairness among MUs \cite{5580131,5463218,1424595,Aquantitativemeasure,Fairnessawarecooperative}. To consider fairness, some RRA algorithms are proposed for OFDMA-based cellular networks \cite{5580131,5463218}. More specifically, in \cite{5580131}, the authors propose RRA algorithm for the downlink of a cellular OFDMA-based system consisting of multiple BSs to maximize the weighted sum of the minimal MUs rates subject to per-BS transmit power constraint. In addition, the authors in \cite{Fairnessawarecooperative} design a heuristic fairness-aware cooperative caching approach in mobile social networks to improve the data access fairness. Although many existing transmission-aware CPSs in cache-enabled wireless networks efficiently improve the latency of MUs, they do not guarantee the fairness among MUs which could significantly affect the MUs latency.
	
\subsection{Contributions}
\begin{itemize}
  \item In this work, we first design a resource allocation framework for a two-tier multiuser OFDMA-based software defined C-HetNet with limited wireless backhaul links where the network operational time is divided into two separated phases as CP and DP.
  \item To guarantee the fairness in terms of delay among MUs, we first propose two delay-based fairness schemes, called proportional fairness (PF) and min-max fairness (MMF). To the best of our knowledge, we are the first to investigate the effect of delay-based fairness schemes in the wireless edge caching context.
  \item In the CP, we design novel fairness and transmission-aware CPSs (FTACPSs) based on the available
content popularity and channel stochastic information at the SD central scheduler. This novel strategy is proposed based on jointly allocating all available physical resources as storage, and both access and backhaul radio resources, and performing MU association to exploit the flexible physical layer transmission opportunities and the OFDMA technique in the CPS design. In the DP, we propose a delivery policy based on the arrival requests of MUs, CSI, and caching status.
  \item To solve the optimization problems corresponding to each fairness scheme, we propose low-complexity alternative optimization (AO) algorithms with a novel transformation method based on the well-known epigraph technique in order to tackle the non-convexity of access and backhaul delay functions. We also obtain the computational complexity of the proposed algorithms and analytically prove the convergence of each AO approach.
  \item In Simulation results, we show that our proposed FTACPS improves the system total latency up to $27\%$ compared to the conventional baseline popularity approaches. Moreover, we investigate the average delay of MUs with/without considering fairness schemes in order to authenticate our FTACPSs.
\end{itemize}

It is noteworthy that we are the first to investigate the benefits of jointly allocating physical resources as storage and radio to design efficient CPSs proactively in C-HetNets. We should emphasize that our proposed resource allocation framework and FTACPSs are not a panacea but a new attempt in the context of designing efficient CPSs in C-HetNets. There would be several interesting issues such as MUs mobility in the DP, variations of wireless channel fading through each time period of the DP, availability of the perfect PDI and CDI in the CP, and real system implementation. Addressing these concerns is considered as future works, and we hope this work will provide some guidelines to design efficient CPSs in cache-assisted wireless networks.
To preserve the readability of the paper, the abbreviations used are shown in Table \ref{Table ABBREVIATION}.
\begin{table*}[tp]
\centering
\caption{Abbreviations}
\begin{center} \label{Table ABBREVIATION}
\scalebox{0.85}{\begin{tabular}{|c|c||c|c|}
 \hline \rowcolor[gray]{0.850}
 \textbf{Abbreviation} & \textbf{Definition} & \textbf{Abbreviation} & \textbf{Definition} \\
\hline
\rowcolor[gray]{0.905}
 AO & Alternative Optimization & MBS & Macro Base Station \\
 \rowcolor[gray]{0.910}
 BS & Base Station & MIDCP & Mixed-Integer Disciplined Convex Programmin \\
 \rowcolor[gray]{0.915}
 CDI & Channel Distribution Information & MINLP & Mixed-Integer Nonlinear Programmin \\
 \rowcolor[gray]{0.920}
 C-HetNet & Cache-enabled Heterogeneous Cellular Network & MMF & Min-Max Fairness \\
 \rowcolor[gray]{0.925}
 CMP & Cache Most Popular & MU & Mobile User \\
 \rowcolor[gray]{0.930}
 CP & Caching Phase & NC & No Caching \\
 \rowcolor[gray]{0.935}
 CPS & Cache Placement Strategy & PDI & Popularity Distribution Information \\
 \rowcolor[gray]{0.940}
 CSI & Channel State Information & PF & Proportional Fairness \\
 \rowcolor[gray]{0.945}
 D.C. & Difference-of-Two-Concave-Functions & PRC & Popular Random Caching \\
 \rowcolor[gray]{0.950}
 DC & Data Center & PSD & Power Spectral Density \\
 \rowcolor[gray]{0.955}
 DCP & Disciplined Convex Programming & RAN & Radio Access Network \\
 \rowcolor[gray]{0.960}
 DP & Delivery Phase & RRA & Radio Resource Allocation \\
 \rowcolor[gray]{0.965}
 FTACPS & Fairness and Transmission-Aware CPS & RRH & Remote Radio Head \\
 \rowcolor[gray]{0.970}
 i.i.d. & Independent and Identically Distributed & SCA & Successive Convex Approximation \\
 \rowcolor[gray]{0.975}
 INLP & Integer Nonlinear Programming & SD & Software Defined \\
 \rowcolor[gray]{0.983}
 IPM & Interior-Point Method & SDN & Software Defined Networking \\
 \rowcolor[gray]{0.990}
 IS & Interfering Source & URC & Uniform Random Caching \\
 \hline
\end{tabular}}
\end{center}
\end{table*}

\subsection{Paper Organization}
The remainder of the paper is organized as follows. Section \ref{Section system model and Problem Formulations} describes the systems setup, the proposed fairness schemes, and the optimization problem statements in each scheme. Section \ref{Section Solution} proposes solutions for the main problems. In addition, Section \ref{section computational complexity} investigates the computational complexity of the proposed solution algorithms. In Section \ref{section simulation results}, we present numerical results to evaluate the performance of CPSs. We also present the conclusion of this paper in Section \ref{section Conclusion}.

\section{System Model and Problem Formulations}\label{Section system model and Problem Formulations}
\begin{figure}
\centering
\includegraphics[scale=0.34]{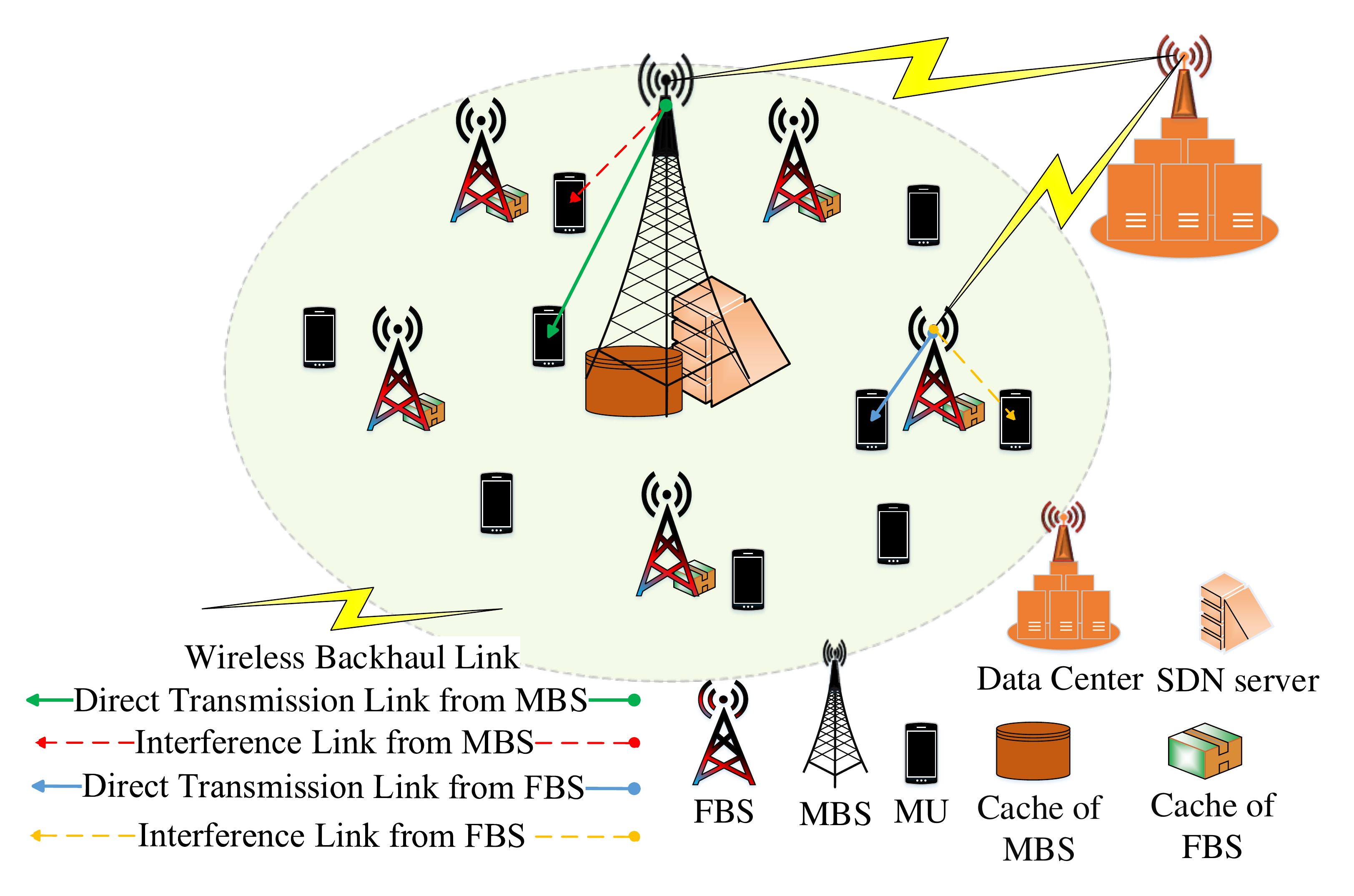}
\caption{An illustration of a software defined C-HetNet, where the MBS and all FBSs are under the control of a central SD scheduler, and all BSs are connected to a DC via limited wireless backhaul links.}
\label{Fig01SysModel}
\end{figure}

Consider the downlink of a wireless software defined C-HetNet consisting of a single MBS and $B$ FBSs connected to a DC via wireless backhaul links. Moreover, there exist $U$ MUs in the network such that each MU can be associated to only one BS. Fig. \ref{Fig01SysModel} illustrates the considered network model. In the MBS, a SD controller with a global view of the network is deployed \cite{7406764,7925732,7744833,7514219}. Furthermore, the SD controller enables a fast and efficient control over networks devices by using the OpenFlow interface \cite{7514219,7744833,7925732}.
Denoted by $\mathcal{B}=\{0,1,2,\dots,B\}$, the set of BSs in the network where 0 represents the MBS and $\{1,2,\dots,B\}$ is the set of FBSs. Let $\mathcal{U}=\{1,2,\dots,U\}$ be the set of MUs in the network. We assume that there are $C$ contents in the network whose set is represented by $\mathcal{C}=\{1,2,\dots,C\}$. The size of content $c$ is indicated by $s_c$ which is modeled by Log-normal distribution with the mean value $\mu_s$ and variance $\sigma^2_s$ \cite{7322204,Sobkowicz2013}.
The PDI of contents are the same among all MUs in the network and also modeled as Zipf distributed \cite{7530876,7155502}. Hence, the probability of requesting the $c^\text{th}$ popular content, i.e., rank $c$, is $\Delta_c = \frac{1/c^{\zeta_1}}{\sum_{c=1}^{C} 1/c^{\zeta_1}}$
where $\zeta_1$ is the Zipf parameter and tunes the skewness of the distribution.
We assume that BSs are able to cache contents where $M^\text{max}_b$ denotes the maximum storage capacity of BS $b$. Furthermore, we assume that the DC has all contents in its library. Let $\rho_{b,c} \in \{0,1\}$ be the binary content placement indicator where if content $c$ is cached by BS $b$, $\rho_{b,c}=1$ and otherwise, $\rho_{b,c}=0$.

The considered system operates in two separated phases as CP and DP \cite{7805409,7155502,7558153} which are described in the following:
\begin{enumerate}
  \item Once the content placement process in the CP is completed, all BSs and DC are ready to serve the requests of MUs, immediately. It is assumed that this system is fast enough to switch between the CP and DP \cite{8030120,7155502,7558153}.
Assume that the DP is divided into several time periods, each of which has the time length of $T$ seconds which is much smaller than the whole DP \cite{7558153}. We suppose that each MU requests one content at the beginning of each time period \cite{7805409,7828114,7322204}.
In this scheme, with assuming that MU is associated to BS $b$, if the requested content $c$ is cached by BS $b$, BS $b$ sends content $c$ to the MU, immediately. Otherwise, the request of content $c$ is forwarded to the DC, and then, the DC disseminates it to BS $b$. After BS $b$ received content $c$, it sends the content to the MU.
Same as most of the prior works, we assume a block fading model for access and backhaul channel gains in each time period\footnote{In this work, we assume that MUs do not move during the DP. Devising a mobility-aware CPS based on jointly optimizing the storage and radio resources is considered as a future work.} \cite{7322204,7406764,7536886}.
We assume that channel fadings and IGRs remain constant within one block of length $T$ and change from block to block independently \cite{7322204,7406764,7536886}. Moreover, all IGRs of a time period are served within the current time period \cite{7322204,7406764}.
This delivery model is widely used in many related research works in the context of wireless edge caching \cite{7322204,7406764,7536886,7828114,7805409,7558153,8030120}.
  \item In the CP, BSs cache contents via wireless backhaul links.
  Practically, operators are able to collect/monitor the number of requests for each content and CSI of MUs through each long-term delivery phase to obtain PDI and channel distribution information (CDI) in the next CP, respectively. Following related works, we assume that PDI and CDI are known at a central SD scheduler \cite{7406764,7925732,7744833,7514219}, while the CSI and IGRs of MUs are unknown\footnote{Actually, in contrast to most of the prior works which assume that the CSI is fixed and also available in the CP \cite{7155502,7322204,7536886,7510807,7406764}, we consider a more practical scenario where CSI is not available while the CDI which is averaged over many CSI samples are available at the scheduler.}.
Moreover, the received PDI and CDI in the CP remain constant during the next DP\footnote{In contrast to the prior works where CSI is assumed to be fixed in all networks operational time, we only suppose that CDI remains fixed and CSI may vary between different time periods.} \cite{7155502,7322204,7536886,7510807,7406764,7805409,7828114}.
\end{enumerate}

\begin{figure}
	\centering
	\includegraphics[scale=0.34]{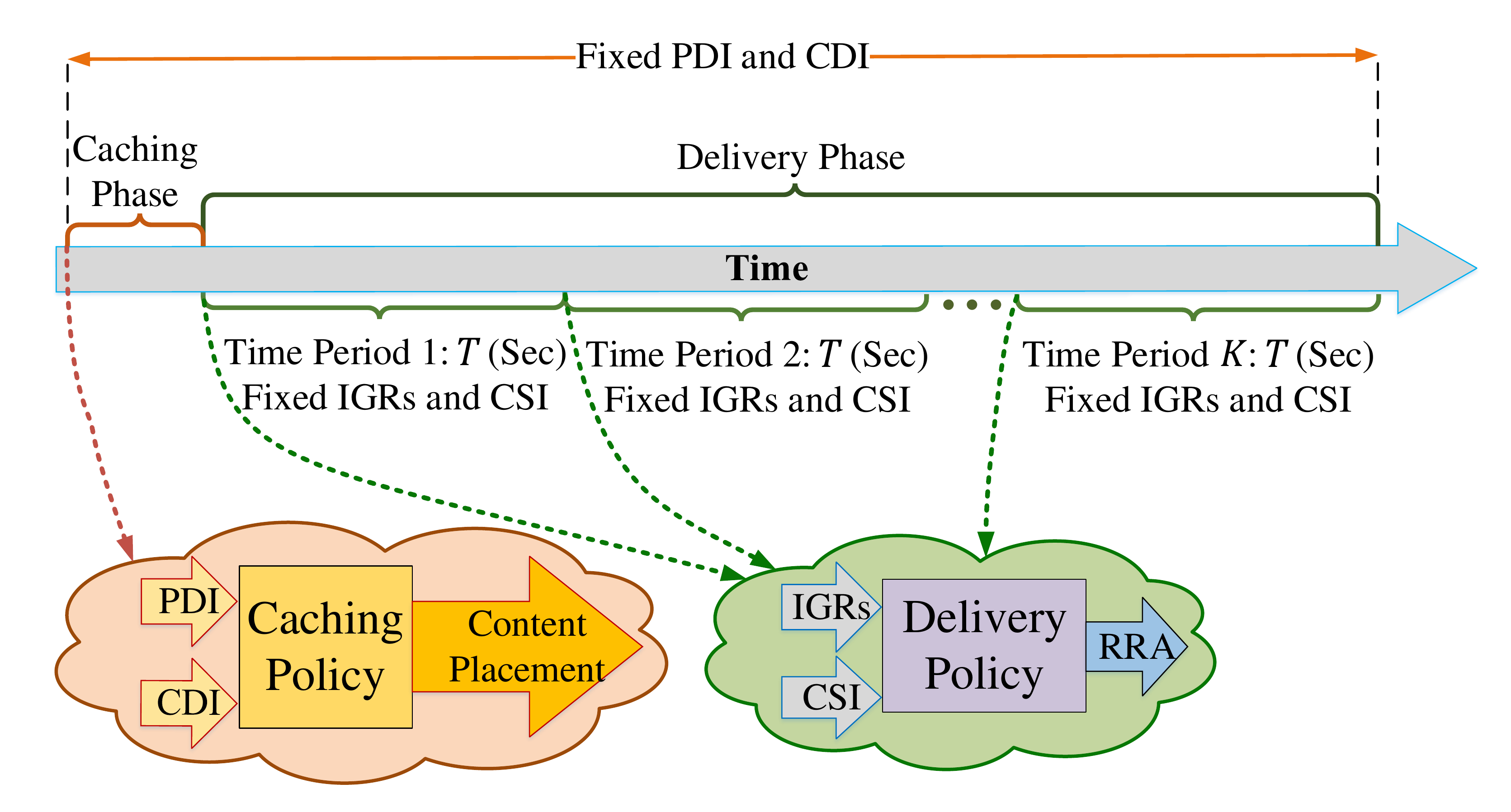}
	\caption{The main structure of the proposed resource allocation framework. The finite time length of the DP is divided into $K$ equal time periods with the length of $T$.}
	\label{Fig00structure}
\end{figure}

For this system, we first propose a two-phase resource allocation framework based on the available stochastic and deterministic information in the CP and DP, respectively. In the CP, we aim to design an efficient FTACPSs at the off-peak time based on the PDI and CDI by coupling access, backhaul and storage resources. The proposed FTACPSs also cover each time period of the DP, since the RRA is only based on the fixed CDI of joint access and backhaul links, and is referred to joint access and backhaul ergodic RRA. The validation of our proposed FTACPSs are guaranteed until the PDI and/or the CDI change.
In the DP, we apply a RRA for instantaneous performance optimization at the beginning of each time period based on the IGRs of MUs, caching status, and CSI which are known at the central scheduler. Since all IGRs of MUs are served within the time period, the proposed delivery policies for different time periods are completely independent.
The main structure of our caching and delivery policies is illustrated in Fig. \ref{Fig00structure}.

Denote by $\theta_{b,u}$ the MU association indicator where $\theta_{b,u}=1$ if MU $u$ is associated to BS $b$ and otherwise, $\theta_{b,u}=0$.
Let $W_\text{Ac}$ and $W_\text{BH}$ be the access frequency bandwidth and out-of-band backhauling, respectively. By utilizing the OFDMA technique for both the access and backhaul transmissions, the set of subcarriers of access and backhaul links are expressed by $\mathcal{N}_\text{Ac}=\{ 1,2,\dots,N_\text{Ac} \}$ and $\mathcal{N}_\text{BH}=\{ 1,2,\dots,N_\text{BH} \}$, respectively, each of whose bandwidth is $W_\text{S}$.
The binary subcarrier assignment indicator for the access link is defined as $\gamma_{b,u}^{n_\text{Ac}}$: if subcarrier $n_\text{Ac}$ is assigned to the channel from BS $b$ to MU $u$, $\gamma_{b,u}^{n_\text{Ac}}=1$ and otherwise, $\gamma_{b,u}^{n_\text{Ac}}=0$.
Denoted by $h_{b,u}^{n_\text{Ac}}$, the channel power gain from BS $b$ to MU $u$ on subcarrier $n_\text{Ac} \in \mathcal{N}_\text{Ac}$. The transmit power of BS $b$ to MU $u$ on subcarrier $n_\text{Ac}$ is also indicated by $p_{b,u}^{n_\text{Ac}}$. Hence, the instantaneous received data rate at MU $u$ from BS $b$ on subcarrier $n_\text{Ac}$ is given by \cite{6678362}
\begin{align}\label{rate Ac subn}
r_{b,u}^{n_\text{Ac}}  =  W_\text{S} \log_2 \left( 1+\frac { p_{b,u}^{n_\text{Ac}} h_{b,u}^{n_\text{Ac}} }   { \sum\limits_{i \in \mathcal{B}/ \{b\}} \sum\limits_{j \in \mathcal{U}/\{u\}} \gamma_{i,j}^{n_\text{Ac}} p_{i,j}^{n_\text{Ac}} h_{i,u}^{n_\text{Ac}} + \sigma_{u}^{n_\text{Ac}} } \right),
\end{align}
where $\sigma_{u}^{n_\text{Ac}}$ is the additive white Gaussian noise (AWGN) power at MU $u$ on subcarrier $n_\text{Ac}$, and $\sum\limits_{i \in \mathcal{B}/ \{b\}} \sum\limits_{j \in \mathcal{U}/ \{u\}} \gamma_{i,j}^{n_\text{Ac}} p_{i,j}^{n_\text{Ac}} h_{i,u}^{n_\text{Ac}}$ is the worst-case inter-cell interference signal at MU $u$ on subcarrier\footnote{In this case, we assume that all access links are active and all BSs are disseminating the requested contents.} $n_\text{Ac}$. Therefore, the data rate at MU $u$ from BS $b$ is given by
$r_{b,u}^{\text{Ac}}  = \sum_{n_\text{Ac}=1}^{N_\text{Ac}} \gamma_{b,u}^{n_\text{Ac}} r_{b,u}^{n_\text{Ac}}$.
The instantaneous received data rate from DC at BS $b$ on subcarrier $n_\text{BH}$ is given by
\begin{align}\label{rate BH sub}
r_{b}^{n_\text{BH}}  = W_\text{S} \log_2 \left( 1 + \frac { p_{b}^{n_\text{BH}} h_{b}^{n_\text{BH}} } { \sigma_{b}^{n_\text{BH}} } \right),
\end{align}
where $h_{b}^{n_\text{BH}}$ is the channel power gain from DC to BS $b$ on subcarrier $n_\text{BH}$, $p_{b}^{n_\text{BH}}$ is the transmit power of DC to BS $b$ on subcarrier $n_\text{BH}$, and $\sigma_{b}^{n_\text{BH}}$ is the combined AWGN noise power and received interference from other DCs at BS $b$ on subcarrier $n_\text{BH}$\footnote{We assume that an external node acts as an interfering source (IS) and operates on the backhaul frequency bandwidth. Hence, the received interference of other DCs at BSs can be modeled by the IS.}.
Moreover, indicated by $\gamma_{b}^{n_\text{BH}}$ the binary subcarrier assignment variable, where if subcarrier $n_\text{BH}$ is assigned to BS $b$, $\gamma_{b}^{n_\text{BH}}=1$ and otherwise, $\gamma_{b}^{n_\text{BH}}=0$.
Therefore, the achievable data rate at BS $b$ can be obtained by
$r_{b}^{\text{BH}}  =  \sum_{n_\text{BH}=1}^{N_\text{BH}} \gamma_{b}^{n_\text{BH}} r_{b}^{n_\text{BH}}.$

\subsection{Proportional Fairness Scheme}\label{subsection No Fair Latency}
In this subsection, we aim at minimizing the total weighted latency of MUs in the DP. Hence, we devise a FTACPS and delivery policy based on the proposed resource allocation framework presented in Fig. \ref{Fig00structure}.

\subsubsection{Caching Phase}\label{subsection Caching Policy total delay}
In this phase, we address the problem of minimizing the total weighted average latency of MUs and find an efficient FTACPS by jointly allocating storage and radio resources. Based on the available CDI, we denote $R_{b,u}^{\text{Ac}}=\mathbb{E}_{\boldsymbol{h}} \big{ \{ } r_{b,u}^{\text{Ac}} \big{ \} }$ and $R_{b}^{\text{BH}}=\mathbb{E}_{\boldsymbol{h}} \big{ \{ } r_{b}^{\text{BH}} \big{ \} }$ as the received ergodic data rates at MU $u$ from BS $b$, and BS $b$ from the DC, respectively, where $\mathbb{E}_{\boldsymbol{h}} \{ \cdot  \} $ is the expectation operator. Note that even though we have a slowly block fading channel model, from the point of view of the central SD scheduler, the relevant quantity is the average (or ergodic) rate, because it has only CDI. The average access latency of each MU $u$ to receive a content from BS $b$ can thus be obtained by
\begin{align}\label{latency each user probability}
D_{b,u}^{\text{Pr}}=\sum\limits_{c \in \mathcal{C}} \Delta_c \theta_{b,u} \left(
\frac{s_c}{ R_{b,u}^{\text{Ac}} } +
\frac{\left(1-\rho_{b,c}\right) s_c}{ \beta_{b,c} R_{b}^{\text{BH}} } \right),
\end{align}
where the term $\frac{s_c}{ R_{b,u}^{\text{Ac}} }$ represents the average access latency of MU $u$ to receive content $c$ from BS $b$ and the term $\frac{\left(1-\rho_{b,c}\right) s_c}{ \beta_{b,c} R_{b}^{\text{BH}} }$ represents the average backhaul latency of BS $b$ to receive the un-cached content $c$ from the DC. The parameter $0 \leq \beta_{b,c} \leq 1$ also represents the portion of the achievable backhaul data rate $R_{b}^{\text{BH}}$ dedicated for downloading content $c$ at BS $b$. The definition of $\beta_{b,c}$ comes from the fact that based on the IGRs and caching status, each BS may need more than one content from the DC during a period of time in the next DP. Hence, the available backhaul rates should be efficiently distributed over the scheduled contents. The total weighted average latency of MUs can thus be formulated as follows:
$D^{\text{Pr},\text{tot}} = \sum\limits_{u \in \mathcal{U}} \omega_{u} \sum\limits_{b \in \mathcal{B}} D_{b,u}^{\text{Pr}},$
where $\omega_{u}$ is the determined weight for MU $u \in \mathcal{U}$.

In this two-hop transmission system, a delivery deadline constraint is applied to ensure that the content delivery is done within the current time period. Due to the non-homogeneity of content sizes, we consider a minimum required data rate constraint at each MU to guarantee the QoS of MUs.
For notational convenience, we denote $\boldsymbol{p}=[\boldsymbol{p}^\text{Ac},\boldsymbol{p}^\text{BH}]$, $\boldsymbol{p}^\text{Ac}=[p_{b,u}^{n_\text{Ac}}]$, $\boldsymbol{p}^\text{BH}=[p_{b}^{n_\text{BH}}]$, $\boldsymbol{\gamma}=[\boldsymbol{\gamma}^\text{Ac},\boldsymbol{\gamma}^\text{BH}]$, $\boldsymbol{\gamma}^\text{Ac}=[\gamma_{b,u}^{n_\text{Ac}}]$, $\boldsymbol{\gamma}^\text{BH}=[\gamma_{b}^{n_\text{BH}}]$, $\boldsymbol{\theta}=[\theta_{b,u}]$, $\boldsymbol{\beta}=[\beta_{b,c}]$, and $\boldsymbol{\rho}=[\rho_{b,c}]$.
Therefore, we formulate the CP-PF optimization problem as follows:
\allowdisplaybreaks
\begin{subequations}\label{total latency Problem caching}
\begin{align}\label{obf total latency Problem caching}
\text{CP-PF:}~~&\min_{ \boldsymbol{p} , \boldsymbol{\gamma} , \boldsymbol{\theta} , \boldsymbol{\rho} , \boldsymbol{\beta} }\hspace{.0 cm}	
~~ D^{\text{Pr},\text{tot}}
\\
\label{Constraint cache size}
\textrm{s.t.}\hspace{.0cm}~~
& \sum_{c=1}^{C} \rho_{b,c} s_c \leq M^\text{max}_b, \forall b \in \mathcal{B},
\\
\label{Constraint latency user}
&
\sum_{b \in \mathcal{B}} D_{b,u}^{\text{Pr}} \leq T , \forall u \in \mathcal{U},
\\
\label{Constraint minimum rate}
&
\sum_{b \in \mathcal{B}} R_{b,u}^{\text{Ac}} \geq R^\text{min}_{u} , \forall u \in \mathcal{U},
\\
\label{Constraint power BS allowance}
&
\mathbb{E}_{\boldsymbol{h}} \left\{ \sum_{u \in \mathcal{U}} \sum_{n_\text{Ac} = 1}^{N_\text{Ac}}  \gamma_{b,u}^{n_\text{Ac}}  p_{b,u}^{n_\text{Ac}} \right\} \leq P^\text{max}_b, \forall b \in \mathcal{B},
\\
\label{Constraint power data center allowance}
&
\mathbb{E}_{\boldsymbol{h}} \left\{ \sum\limits_{b \in \mathcal{B}} \sum_{n_\text{BH} = 1}^{N_\text{BH}}  \gamma_{b}^{n_\text{BH}}   p_{b}^{n_\text{BH}} \right\} \leq P^\text{max}_\text{DC},
\\
\label{Constraint user association}
&
\sum_{b \in \mathcal{B}} \theta_{b,u} \leq 1, \forall u \in \mathcal{U},
\\
\label{Constraint subcarrier user association}
&
\sum_{n_\text{Ac} = 1}^{N_\text{Ac}} \gamma_{b,u}^{n_\text{Ac}} \leq N_\text{Ac} \theta_{b,u}, \forall b \in \mathcal{B}, u \in \mathcal{U},
\\
\label{Constraint interference Access}
&
\sum_{u \in \mathcal{U}} \gamma_{b,u}^{n_\text{Ac}} \leq 1, \forall b \in \mathcal{B}, \forall n_\text{Ac} \in \mathcal{N}_\text{Ac},
\\
\label{Constraint interference BH}
&
\sum\limits_{b \in \mathcal{B}}  \gamma_{b}^{n_\text{BH}} \leq 1, \forall n_\text{BH} \in \mathcal{N}_\text{BH},
\\
\label{Constraint beta}
&
\sum_{c=1}^{C} \beta_{b,c} \leq 1, \forall b \in \mathcal{B},~~~~~ 0 \leq \beta_{b,c}\leq 1,
\\
\label{Constraint indicator gamma}
&
\gamma_{b}^{n_\text{BH}},\gamma_{b,u}^{n_\text{Ac}}, \theta_{b,u} \in \{0,1\},
\\
\label{Constraint positive powers}
&
p_{b,u}^{n_\text{Ac}},p_{b}^{n_\text{BH}} \geq 0,
\\
\label{Constraint indicator rho}
&
\rho_{b,c} \in \{0,1\},
\end{align}
\end{subequations}
where \eqref{Constraint cache size} represents the maximum storage capacity constraint of each BS. \eqref{Constraint latency user} is the delivery deadline constraint at each MU $u$ based on the allocated ergodic data rates.
\eqref{Constraint minimum rate} is the minimum required data rate constraint of each MU and $R^\text{min}_{u}$ is the minimum data rate of MU $u$. \eqref{Constraint power BS allowance} and \eqref{Constraint power data center allowance} are the maximum allowable transmit power constraints for each BS and DC, respectively. $P^\text{max}_b$ and $P^\text{max}_\text{DC}$ denote the maximum transmit power of each BS $b$ and DC, respectively.
\eqref{Constraint user association} ensures us each MU can be associated to at most one BS. Moreover, \eqref{Constraint subcarrier user association} represents that each MU can only take access subcarriers from the BS that is associated with it.
In addition, \eqref{Constraint interference Access} and \eqref{Constraint interference BH} are the exclusive subcarrier allocation constraints for access and backhaul channels, respectively, due to the OFDMA assumption. Constraint \eqref{Constraint beta} represents that the summation of all portions of the backhaul data rate $R_{b}^{\text{BH}}$ should not exceed $1$.

\subsubsection{Delivery Phase}\label{subsection Delivery Policy total delay}
In this step, based on the available IGRs, CSI, and caching status (see Fig. \ref{Fig00structure}) we design a delivery policy in each time period with the aim of minimizing the total weighted latency of MUs subject to the mentioned constraints in Subsection \ref{subsection Caching Policy total delay}.
We introduce the binary request indicator $\delta^{c}_{u} \in \{0,1\}$ where if MU $u$ requests content $c$, $\delta^{c}_{u}=1$ and otherwise, $\delta^{c}_{u}=0$. The IGRs of MUs at a time period are inherently independent from each other, and only follow the PDI of contents \cite{7322204,7558153}.
The instantaneous delivery latency of MU $u$ for receiving content $c$ from BS $b$ in the network can thus be obtained by
\begin{align}\label{latency each user delivery}
D_{b,u}^{\text{Del},c} = \delta^{c}_{u} \theta_{b,u}  \left(
\frac{s_c}{ r_{b,u}^{\text{Ac}} } +  \frac{(1-\rho_{b,c}) s_c}{ \beta_{b,c} r_{b}^{\text{BH}} }
\right).
\end{align}
Accordingly, the total weighted delivery latency of MUs in the time period is given by
$
D^{\text{Del},\text{tot}} = \sum\limits_{u \in \mathcal{U}} \omega_{u} \sum_{c=1}^{C} \sum\limits_{b \in \mathcal{B}} D_{b,u}^{\text{Del},c}.
$
Since each MU has only one request in a time period, i.e., $\sum_{c \in \mathcal{C}} \delta^{c}_{u}=1,\forall b, u \in \mathcal{U}$, and each MU can be associated to at most one BS, only one term in $\sum_{c=1}^{C} \sum\limits_{b \in \mathcal{B}} D_{b,u}^{\text{Del},c}$ has a non-zero value and the rest of them are zero. Hence, in each time period, we formulate an optimization problem to minimize the total weighted delivery latency of MUs as follows:
\begin{subequations}\label{total latency Problem delivery}
\begin{align}\label{obf total latency Problem delivery}
\text{DP-PF:}~~&\min_{ \boldsymbol{p} , \boldsymbol{\gamma} , \boldsymbol{\theta} , \boldsymbol{\beta}  }\hspace{.0 cm}
~~ D^{\text{Del},\text{tot}}
\\
\textrm{s.t.}\hspace{.0cm}~~
& \text{\eqref{Constraint user association}-\eqref{Constraint positive powers},} \nonumber
\\
\label{Constraint latency user delivery}
&
\sum\limits_{b \in \mathcal{B}} \sum\limits_{c \in \mathcal{C}} D_{b,u}^{\text{Del},c} \leq T , \forall u \in \mathcal{U},
\\
\label{Constraint minimum rate delivery}
&
\sum\limits_{b \in \mathcal{B}} r_{b,u}^{\text{Ac}} \geq R^\text{min}_{u} , \forall u \in \mathcal{U},
\\
\label{Constraint power BS allowance delivery}
&
\sum_{u \in \mathcal{U}} \sum_{n_\text{Ac} = 1}^{N_\text{Ac}}  \gamma_{b,u}^{n_\text{Ac}}  p_{b,u}^{n_\text{Ac}} \leq P^\text{max}_b, \forall b \in \mathcal{B},
\\
\label{Constraint power data center allowance delivery}
&
\sum\limits_{b \in \mathcal{B}} \sum_{n_\text{BH} = 1}^{N_\text{BH}}  \gamma_{b}^{n_\text{BH}}   p_{b}^{n_\text{BH}} \leq P^\text{max}_\text{DC},
\end{align}
\end{subequations}
where \eqref{Constraint latency user delivery} and \eqref{Constraint minimum rate delivery} are the per-MU delivery deadline and minimum required instantaneous data rate constraints, respectively. In addition, \eqref{Constraint power BS allowance delivery} and \eqref{Constraint power data center allowance delivery} are the maximum allowable transmit power of each BS and DC, respectively.

\subsection{Min-Max Fairness Scheme}\label{subsection Fairness Aware Latency}
In order to guarantee the fairness among MUs in the design of both the CPSs and delivery policies, we address the problem of minimizing the maximal latency of MUs for each phase. Consequently, in the CP, we formulate the following min-max optimization problem:
\begin{subequations}\label{total latency Problem fair caching}
\begin{align}\label{obf total latency Problem fair caching}
\text{CP-MMF:}~~\min_{ \boldsymbol{p} , \boldsymbol{\gamma} , \boldsymbol{\theta} , \boldsymbol{\rho} , \boldsymbol{\beta} }\hspace{.0 cm}	
&
~~ \max\limits_{u \in \mathcal{U}} \sum\limits_{b \in \mathcal{B}} D_{b,u}^{\text{Pr}}
\\
\textrm{s.t.}\hspace{.0cm}~~
& \text{\eqref{Constraint cache size}-\eqref{Constraint indicator rho}.} \nonumber
\end{align}
\end{subequations}
In the DP, we formulate the following optimization problem for a fixed $\boldsymbol{\rho}$ as
\begin{subequations}\label{total latency Problem delivery fair}
\begin{align}\label{obf total latency Problem delivery fair}
\text{DP-MMF:}~~\min_{ \boldsymbol{p} , \boldsymbol{\gamma} , \boldsymbol{\theta} , \boldsymbol{\beta}  }\hspace{.0 cm}
& 	
~~ \max\limits_{u \in \mathcal{U}} \sum\limits_{c \in \mathcal{C}} \sum\limits_{b \in \mathcal{B}} D_{b,u}^{\text{Del},c}
\\
\textrm{s.t.}\hspace{.0cm}~~
& \text{\eqref{Constraint user association}-\eqref{Constraint positive powers}, \eqref{Constraint latency user delivery}-\eqref{Constraint power data center allowance delivery}.} \nonumber
\end{align}
\end{subequations}
For the ease of reference, we put all the notations used in the paper in Table \ref{table main notations}.
\begin{table*}[tp]
\centering
\caption{Main notations}
\begin{center} \label{table main notations}
\scalebox{0.85}{\begin{tabular}{|c c c c c c|}
    \hline \rowcolor[gray]{0.85}
    \hline \rowcolor[gray]{0.85}
    \hline \rowcolor[gray]{0.85}
    \hline \rowcolor[gray]{0.85}
    \textbf{Description} & \textbf{Notation} & \textbf{Phase} & \textbf{Description} & \textbf{Notation} & \textbf{Phase} \\
    \hline \rowcolor[gray]{0.940}
    \hline \rowcolor[gray]{0.940}
    \hline \rowcolor[gray]{0.940}
    \hline  \rowcolor[gray]{0.940}
    Number of FBSs & $B$ & (CP/DP) & MU's weight & $\omega_{u}$ & (CP/DP) \\
     \rowcolor[gray]{0.945}
    Number of MUs & $U$ & (CP/DP) & Minimum access data rate of MU $u$ & $R^\text{min}_{u}$ & (CP/DP) \\
     \rowcolor[gray]{0.950}
    Number of contents & $C$ & (CP/DP) & Maximum power of BS $b$ & $P^\text{max}_b$ & (CP/DP) \\
     \rowcolor[gray]{0.955}
    Size of content $c$ & $s_c$ & (CP/DP) & Maximum power of DC & $P^\text{max}_\text{DC}$ & (CP/DP)  \\
     \rowcolor[gray]{0.960}
    Content placement indicator & $\rho_{b,c}$ & (CP/DP) & Popularity of content $c$ & $\Delta_c$ & CP \\
     \rowcolor[gray]{0.963}
    Time duration & $T$ & (CP/DP) & Zipf parameter & $\zeta_1$ & CP \\
     \rowcolor[gray]{0.966}
    MU association indicator & $\theta_{b,u}$ & (CP/DP) & Cache size of BS $b$ & $M^\text{max}_b$ & CP \\
     \rowcolor[gray]{0.969}
    Access frequency bandwidth & $W_\text{Ac}$ & (CP/DP) & Average access data rate & $R_{b,u}^{\text{Ac}}$ & CP \\
     \rowcolor[gray]{0.972}
    Out of band backhauling & $W_\text{BH}$ & (CP/DP) & Average backhaul data rate & $R_{b}^{\text{BH}}$ & CP \\
     \rowcolor[gray]{0.975}
    Number of access subcarriers & $N_\text{Ac}$ & (CP/DP) & Average latency of MU & $D_{b,u}^{\text{Pr}}$ & CP \\
     \rowcolor[gray]{0.978}
    Number of backhaul subcarriers & $N_\text{BH}$ & (CP/DP) &  Total weighted average latencies & $D^{\text{Pr},\text{tot}}$ & CP \\
     \rowcolor[gray]{0.981}
    Subcarrier frequency bandwidth & $W_\text{S}$ & (CP/DP) & Instantaneous access channel gain & $h_{b,u}^{n_\text{Ac}}$ & DP \\
     \rowcolor[gray]{0.984}
    Access subcarrier assignment indicator & $\gamma_{b,u}^{n_\text{Ac}}$ & (CP/DP) & Instantaneous access data rate over subcarrier & $r_{b,u}^{n_\text{Ac}}$ & DP \\
     \rowcolor[gray]{0.987}
    Access transmit power & $p_{b,u}^{n_\text{Ac}}$ & (CP/DP) & Instantaneous access data rate of MU & $r_{b,u}^{\text{Ac}}$ & DP \\
     \rowcolor[gray]{0.990}
    AWGN noise power at MU & $\sigma_{u}^{n_\text{Ac}}$ & (CP/DP) & Instantaneous backhaul data rate over subcarrier & $r_{b}^{n_\text{BH}}$ & DP \\
     \rowcolor[gray]{0.993}
    AWGN noise power at BS & $\sigma_{b}^{n_\text{BH}}$ & (CP/DP) & Instantaneous backhaul data rate of BS & $r_{b}^{\text{BH}}$ & DP \\
    Access transmit power & $\gamma_{b}^{n_\text{BH}}$ & (CP/DP) & Request of MU & $\delta^{c}_{u}$ & DP \\
     \rowcolor[gray]{0.996}
    AWGN noise power at MU & $p_{b}^{n_\text{BH}}$ & (CP/DP) & Instantaneous latency of MU & $D_{b,u}^{\text{Del},c}$ & DP \\
     \rowcolor[gray]{0.999}
    AWGN noise power at BS & $\beta_{b,c}$ & (CP/DP) & Total weighted instantaneous latencies & $D^{\text{Del},\text{tot}}$ & DP \\
    \hline
  \end{tabular}}
\end{center}
\end{table*}

\section{Characterization of Solution and Algorithm}\label{Section Solution}
In this section, we solve problems \eqref{total latency Problem caching}, \eqref{total latency Problem delivery}, \eqref{total latency Problem fair caching} and \eqref{total latency Problem delivery fair}. The solution algorithms are presented in the following subsections.

\subsection{Solving the CP-PF problem}
Here, we aim to solve the non-convex problem \eqref{total latency Problem caching}. Problem \eqref{total latency Problem caching} is a mixed-integer nonlinear programming (MINLP) which is NP-hard, and hence, it is difficult to find an optimal solution for it \cite{6678362,5580131,8008769,7417343,8030120,7925732}. To make \eqref{total latency Problem caching} tractable, we propose a three-step AO approach in which $(\boldsymbol{\rho} , \boldsymbol{\beta} , \boldsymbol{\theta})$, $(\boldsymbol{p})$, and $(\boldsymbol{\gamma} )$ are iteratively obtained \cite{6678362,5580131}.
These iterations are applied until accuracy is obtained.
The pseudo code of the proposed AO algorithm is summarized in Alg. \ref{Alg iterative caching total}.
\begin{algorithm}[tp]
 \caption{The proposed AO algorithm for solving \eqref{total latency Problem caching}.}\label{Alg iterative caching total}
 \begin{algorithmic}[1]
  \STATE Initialize $\boldsymbol{\rho}_0$, $\boldsymbol{\beta}_0$, $\boldsymbol{p}_0$, $\boldsymbol{\theta}_0$ and $\boldsymbol{\gamma}_0$ to feasible values.
  \\ \textbf{repeat}
  \STATE Find $\boldsymbol{\rho}_{t_1}$, $\boldsymbol{\beta}_{t_1}$ and $\boldsymbol{\theta}_{t_1}$ by solving \eqref{total latency Problem caching} for a fixed $(\boldsymbol{p}_{t_1-1},\boldsymbol{\gamma}_{t_1-1})$.
  \STATE Find $\boldsymbol{p}_{t_1}$ by solving \eqref{total latency Problem caching} for a fixed $(\boldsymbol{\rho}_{t_1},\boldsymbol{\beta}_{t_1},\boldsymbol{\theta}_{t_1},\boldsymbol{\gamma}_{t_1-1})$.
  \STATE Find $\boldsymbol{\gamma}_{t_1}$ by solving \eqref{total latency Problem caching} for a fixed $(\boldsymbol{\rho}_{t_1},\boldsymbol{\beta}_{t_1},\boldsymbol{\theta}_{t_1},\boldsymbol{p}_{t_1})$.
  \STATE Set $t_1=t_1+1$.
  \\ \textbf{Until} Convergence of $\boldsymbol{\rho}$, $\boldsymbol{\beta}$, $\boldsymbol{\theta}$, $\boldsymbol{p}$ and $\boldsymbol{\gamma}$.
  \STATE $\boldsymbol{\rho}$, $\boldsymbol{\beta}$, $\boldsymbol{\theta}$, $\boldsymbol{p}$ and $\boldsymbol{\gamma}$ are the outputs of the algorithm.
 \end{algorithmic}
\end{algorithm}

\emph{Proposition 1}: In the proposed Alg. \ref{Alg iterative caching total}, the objective function is either improved (lowered) or remains constant after each iteration $t_1$. It is lower bounded by zero and hence, Alg. \ref{Alg iterative caching total} converges to a local optimum solution.
\begin{proof}
Please see Appendix \ref{appendix proposition1}.
\end{proof}

\subsubsection{Initialization} \label{subsubsection Initialization caching phase total latency}
In this subsection, we initialize the optimization variables in \eqref{total latency Problem caching}. Before initializing, we note the difficulty of satisfying QoS constraints \eqref{Constraint latency user} and \eqref{Constraint minimum rate}.
To initialize $\boldsymbol{\theta}$, we assume that each MU is associated to the nearest FBS within a determined distance threshold $d^\text{max}$. If there is no FBS within $d^\text{max}$, the MU chooses MBS.
For initializing $\boldsymbol{\rho}$, we first assume that all most popular contents are stored in the cache of BSs one by one until the storages are filled. This proactive strategy, which is known as the cache most popular (CMP) method, is used in many previous works \cite{7558153,7805409}. Then, we assume that the transmit power of each BS is equally distributed over each subcarrier \cite{7406764,7456319}. Hence, we have $p_{b,u}^{n_\text{Ac}}=\frac{P^\text{max}_b} {N_\text{Ac} U}, \forall b \in \mathcal{B}, u \in \mathcal{U}, n_\text{Ac} \in \mathcal{N}_\text{Ac}$. In addition, we assume the same equal power allocation approach for the DC which means $p_{b}^{n_\text{BH}} = \frac{P^\text{max}_\text{DC}} {N_\text{BH}}, \forall b \in \mathcal{B}, n_\text{BH} \in \mathcal{N}_\text{BH}$.
For initializing $\boldsymbol{\beta}$, at first we note that in the CP, all un-cached contents in BS $b$ are sent from the DC to BS $b$ with a pre-defined probability. According to \eqref{latency each user probability}, the delivery latency of each content is a linear function of its popularity. Therefore, we set $\beta_{b,c} = \frac{(1-\rho_{b,c}) \Delta_c} {\sum_{c=1}^{C} (1-\rho_{b,c}) \Delta_c} , \forall b \in \mathcal{B}, c \in \mathcal{C}$ to allocate the backhaul portion data rate to each content based on its popularity and placement. When $\sum_{c=1}^{C} (1-\rho_{b,c})=0$, it means all contents are cached at BS $b$. In this situation, BS $b$ does not request any content from the DC and we set $\beta_{b,c}=0,\forall c \in \mathcal{C}$.
After initializing $\boldsymbol{\rho}$, $\boldsymbol{\beta}$, $\boldsymbol{\theta}$ and $\boldsymbol{p}$, we find $\boldsymbol{\gamma}$ by solving \eqref{total latency Problem caching} for the given $(\boldsymbol{\rho},\boldsymbol{\beta},\boldsymbol{\theta},\boldsymbol{p})$ using the relaxation method which is presented in Subsection \ref{Subsubsection find gamma p caching PF}. This approach ensures to meet all the constraints in \eqref{total latency Problem caching}.

\subsubsection{Finding $\boldsymbol{\rho}$, $\boldsymbol{\beta}$ and $\boldsymbol{\theta}$ for Fixed $(\boldsymbol{p},\boldsymbol{\gamma})$}
In order to find $\boldsymbol{\rho}$, $\boldsymbol{\beta}$ and $\boldsymbol{\theta}$, we solve the following problem:
\begin{subequations}\label{problem rho total latency}
\begin{align}\label{obf problem rho total latency}
\min_{ \boldsymbol{\rho} , \boldsymbol{\beta}, \boldsymbol{\theta} }\hspace{.0 cm}
& 	
~~ D^{\text{Pr},\text{tot}}
\\
\textrm{s.t.}\hspace{.0cm}~~
& \eqref{Constraint cache size}, \eqref{Constraint latency user}, \eqref{Constraint user association}, \eqref{Constraint subcarrier user association}, \eqref{Constraint beta}, \eqref{Constraint indicator gamma}, \eqref{Constraint indicator rho}. \nonumber
\end{align}
\end{subequations}
Each term in \eqref{obf problem rho total latency} and \eqref{Constraint latency user} are ratios of two functions which result in non-convex functions with respect to optimization variables $\boldsymbol{\rho}$, $\boldsymbol{\beta}$ and $\boldsymbol{\theta}$. Hence, the combinatorial optimization problem \eqref{problem rho total latency} can be classified as a nonlinear fractional programming \cite{Jorswieckfractional} with binary and nonlinear constraints. It seems that \eqref{problem rho total latency} can not be approximated directly. Hence, we apply some steps to make \eqref{problem rho total latency} tractable in the following. We first utilize the epigraph technique \cite{Boydconvex} by introducing new variable $y^{\text{BH},c}_{b,u}$, where
$
\theta_{b,u} (1-\rho_{b,c}) \frac{s_c} {  \beta_{b,c} R_{b}^{\text{BH}}  } \leq y^{\text{BH},c}_{b,u}, \forall y^{\text{BH},c}_{b,u} \geq 0,
$
which represents that for each MU $u$ such that $\theta_{b,u}=1$, the average latency of BS $b$ for receiving the un-cached content $c$ from DC is upper-bounded by $y^{\text{BH},c}_{b,u}$. Hence, \eqref{problem rho total latency} is simplified to
\begin{subequations}\label{problem rho total latency equivalent form}
\begin{align}\label{obf problem rho total latency equivalent form}
\min_{ \boldsymbol{\rho},\boldsymbol{\beta},\boldsymbol{\theta},\boldsymbol{y} }\hspace{.0 cm}
& 	
~~ \sum\limits_{u \in \mathcal{U}} \omega_{u} \sum_{c=1}^{C} \Delta_c \sum\limits_{b \in \mathcal{B}}
\left( \theta_{b,u} \frac{s_c}{ R_{b,u}^{\text{Ac}} } +  y^{\text{BH},c}_{b,u} \right)
\\
\textrm{s.t.}\hspace{.0cm}~~
& \eqref{Constraint cache size}, \eqref{Constraint user association}, \eqref{Constraint subcarrier user association}, \eqref{Constraint beta}, \eqref{Constraint indicator gamma}, \eqref{Constraint indicator rho}, \nonumber
\\
\label{Constraint latency user caching epi2}
& y^{\text{BH},c}_{b} \beta_{b,c} R_{b}^{\text{BH}} \geq \theta_{b,u} (1-\rho_{b,c}) s_c, \forall b \in \mathcal{B}, u \in \mathcal{U}, c \in \mathcal{C},
\\
\label{Constraint latency user caching epi3}
& \sum\limits_{b \in \mathcal{B}} \sum\limits_{c \in \mathcal{C}} \Delta_c \left( \theta_{b,u} \frac{s_c}{ R_{b,u}^{\text{Ac}} } + y^{\text{BH},c}_{b} \right) \leq T, \forall u \in \mathcal{U},
\end{align}
\end{subequations}
where $\boldsymbol{y}=[y_{b,u}^{\text{BH},c}]$. Problem \eqref{problem rho total latency equivalent form} is a MINLP with mixed-integer nonlinear constraint \eqref{Constraint latency user caching epi2} which is due to the bilinear and binary bilinear products $y^{\text{BH},c}_{b,u} \beta_{b,c}$ and $\theta_{b,u} (1-\rho_{b,c})$, respectively.
Since $\ln(x)$ is an increasing function of $x>0$, we transform \eqref{Constraint latency user caching epi2} to
\begin{align}\label{Constraint latency user caching epi2 ln}
\ln \left( y^{\text{BH},c}_{b,u} \right)+ \ln \left( \beta_{b,c} \right) + \ln \left( R_{b}^{\text{BH}} \right) \geq \ln \left( s_c \right) + \left( 1-\theta_{b,u} + \rho_{b,c} \right) \ln\left(\epsilon\right), \forall b, u, c,
\end{align}
which is due to the fact that for a binary variable $z$ and proper small positive value $\epsilon$, we have $\ln\left(1-z\right)=z \ln\left(\epsilon\right)$ and $\ln\left(z\right)=(1-z) \ln\left(\epsilon\right)$\footnote{Under the assumption $\ln\left(\epsilon\right) < \ln \left( s_c \right)$, \eqref{Constraint latency user caching epi2 ln} is equivalent to \eqref{Constraint latency user caching epi2}.}. By substituting \eqref{Constraint latency user caching epi2 ln} in  \eqref{problem rho total latency equivalent form}, we have
\begin{subequations}\label{problem rho total latency equivalent form 2}
\begin{align}\label{obf problem rho total latency equivalent form 2}
\min_{ \boldsymbol{\rho},\boldsymbol{\beta},\boldsymbol{\theta},\boldsymbol{y} }\hspace{.0 cm}
& 	
~~ \sum\limits_{u \in \mathcal{U}} \omega_{u} \sum_{c=1}^{C} \Delta_c \sum\limits_{b \in \mathcal{B}}
\left( \theta_{b,u} \frac{s_c}{ R_{b,u}^{\text{Ac}} } +  y^{\text{BH},c}_{b,u} \right)
\\
\textrm{s.t.}\hspace{.0cm}~~
& \eqref{Constraint cache size}, \eqref{Constraint user association}, \eqref{Constraint subcarrier user association}, \eqref{Constraint beta}, \eqref{Constraint indicator gamma}, \eqref{Constraint indicator rho}, \eqref{Constraint latency user caching epi3}, \eqref{Constraint latency user caching epi2 ln}, \nonumber
\end{align}
\end{subequations}
which is a mixed-integer disciplined convex programming (MIDCP) and can be solved by utilizing available standard optimization softwares such as CVX with the internal solver MOSEK \cite{7100916,CVXmatlab}. MOSEK is able to solve mixed-integer linear, conic and quadratic optimization problems using the Branch\&bound\&cut algorithm \cite{MOSEKsolver}.

\subsubsection{Finding $\boldsymbol{p}$ for Fixed $(\boldsymbol{\rho},\boldsymbol{\beta},\boldsymbol{\theta},\boldsymbol{\gamma})$}\label{Subsubsection find gamma p caching PF}
In this step, we solve the following optimization problem:
\begin{subequations}\label{problem power total latency}
\begin{align}\label{obf problem power total latency}
\min_{ \boldsymbol{p} }\hspace{.0 cm}
& 	
~~ D^{\text{Pr},\text{tot}}
\\
\textrm{s.t.}\hspace{.0cm}~~
& \text{\eqref{Constraint latency user}-\eqref{Constraint power data center allowance}, \eqref{Constraint positive powers}.} \nonumber
\end{align}
\end{subequations}
Each term in \eqref{obf problem power total latency} and \eqref{Constraint latency user} are ratios of two functions which are non-convex functions. Hence, \eqref{problem power total latency} can be classified as a nonlinear fractional programming \cite{Jorswieckfractional,6251827} with nonlinear constraints. We first utilize a transformation method based on the epigraph technique presented in Appendix \ref{appendix EquivalentTransformation} to deal with the non-convexity of delay functions in \eqref{obf problem power total latency} and \eqref{Constraint latency user}. In doing so, \eqref{problem power total latency} is transformed into the following equivalent form:
\begin{subequations}\label{problem power total latency equivalent epi 2}
\begin{align}\label{obf problem power total latency equivalent epi 2}
\min_{ \boldsymbol{p}, \boldsymbol{x}, \hat{\boldsymbol{x}} }\hspace{.0 cm}
& 	
~~ \sum\limits_{b \in \mathcal{B}} \sum\limits_{u \in \mathcal{U}} \sum_{c=1}^{C} \omega_{u} \Delta_c \left( \hat{x}_{b,u}^{\text{Ac},c}  + \hat{x}^{\text{BH},c}_{b,u}  \right)
\\
\textrm{s.t.}\hspace{.0cm}~~
& \text{\eqref{Constraint minimum rate}-\eqref{Constraint power data center allowance}, \eqref{Constraint positive powers},} \nonumber  \\
\label{Constraint latency user epi1 power}
&
R_{b,u}^{\text{Ac}} \geq \theta_{b,u} s_c x_{b,u}^{\text{Ac},c}, \forall b \in \mathcal{B}, u \in \mathcal{U}, c \in \mathcal{C},
\\
\label{Constraint latency user epi2  power}
&
R_{b}^{\text{BH}} \geq \frac { \theta_{b,u} \left(1-\rho_{b,c}\right) s_c x^{\text{BH},c}_{b,u}}  { \beta_{b,c} }, \forall b \in \mathcal{B}, u \in \mathcal{U}, c \in \mathcal{C},
\\
\label{Constraint latency user epi3  power}
&
\sum\limits_{b \in \mathcal{B}} \sum\limits_{c \in \mathcal{C}} \Delta_c \left( \hat{x}_{b,u}^{\text{Ac},c} + \hat{x}^{\text{BH},c}_{b,u}\right) \leq T, u \in \mathcal{U},
\\
\label{Constraint latency user epi4  power}
&
\ln \left( \hat{x}_{b,u}^{\text{Ac},c} \right) + \ln \left( x_{b,u}^{\text{Ac},c} \right) \geq 0, \forall b \in \mathcal{B}, u \in \mathcal{U}, c \in \mathcal{C},
\\
\label{Constraint latency user epi5  power}
&
\ln \left( \hat{x}^{\text{BH},c}_{b,u} \right) + \ln \left( x^{\text{BH},c}_{b,u} \right) \geq 0, \forall b \in \mathcal{B}, u \in \mathcal{U}, c \in \mathcal{C},
\end{align}
\end{subequations}
where $\boldsymbol{x}=[\boldsymbol{x}^\text{Ac},\boldsymbol{x}^\text{BH}]$, $\boldsymbol{x}^\text{Ac}=[x_{b,u}^{\text{Ac},c}]$, $\boldsymbol{x}^\text{BH}=[x_{b,u}^{\text{BH},c}]$, $\hat{\boldsymbol{x}}=[\hat{\boldsymbol{x}}^\text{Ac},\hat{\boldsymbol{x}}^\text{BH}]$, $\hat{\boldsymbol{x}}^\text{Ac}=[\hat{x}_{b,u}^{\text{Ac},c}]$ and $\hat{\boldsymbol{x}}^\text{BH}=[\hat{x}_{b,u}^{\text{BH},c}]$. $R_{b,u}^{\text{Ac}}$ in \eqref{Constraint minimum rate} and \eqref{Constraint latency user epi1 power} makes \eqref{problem power total latency equivalent epi 2} non-convex. In the following, we use the successive convex approximation (SCA) algorithm to approximate $R_{b,u}^{\text{Ac}}$ in a concave form based on the difference-of-two-concave-functions (D.C.) approximation method \cite{6678362,7100916}. In this line, we first update the approximation parameter based on $\boldsymbol{p}_{t_2-1}$ at iteration $(t_2-1)$. Then, we approximate $R_{b,u}^{\text{Ac}}$ to a concave form. Finally, we solve the approximated convex problem and obtain $\boldsymbol{p}_{t_2}$. To approximate $R_{b,u}^{\text{Ac}}$ to a concave form, we first formulate the non-concave data rate \eqref{rate Ac subn} in a D.C. form as
$
r_{b,u}^{n_\text{Ac}} = f_{b,u}^{n_\text{Ac}} - g_{b,u}^{n_\text{Ac}},
$
where the concave functions $f_{b,u}^{n_\text{Ac}}$ and $g_{b,u}^{n_\text{Ac}}$ are formulated by
$
W_\text{S} \log_2 \left(  \sum\limits_{i \in \mathcal{B}/ \{b\}} \sum\limits_{j \in \mathcal{U}/\{u\}} \gamma_{i,j}^{n_\text{Ac}} p_{i,j}^{n_\text{Ac}} h_{i,u}^{n_\text{Ac}} + \sigma_{u}^{n_\text{Ac}} + p_{b,u}^{n_\text{Ac}} h_{b,u}^{n_\text{Ac}} \right)
$ and
$
W_\text{S} \log_2 \left(  \sum\limits_{i \in \mathcal{B}/ \{b\}} \sum\limits_{j \in \mathcal{U}/\{u\}} \gamma_{i,j}^{n_\text{Ac}} p_{i,j}^{n_\text{Ac}} h_{i,u}^{n_\text{Ac}} + \sigma_{u}^{n_\text{Ac}} \right)
$, respectively.
Subsequently, we approximate $g_{b,u}^{n_\text{Ac}} \left(\boldsymbol{p}^\text{Ac}_{t_2}\right)$ at each iteration $t_2$ by using the following linear approximation approach as \cite{6678362}:
\begin{align}\label{g approximated}
g_{b,u}^{n_\text{Ac}} \left(\boldsymbol{p}^\text{Ac}_{t_2}\right) \approx  g_{b,u}^{n_\text{Ac}} \left(\boldsymbol{p}^\text{Ac}_{t_2-1}\right) +
\nabla g_{b,u}^{n_\text{Ac}} \left(\boldsymbol{p}^\text{Ac}_{t_2-1}\right)
\left(\boldsymbol{p}^\text{Ac}_{t_2} - \boldsymbol{p}^\text{Ac}_{t_2-1}\right),
\end{align}
for a given $\boldsymbol{p}^\text{Ac}_{t_2-1}$ from previous iteration $t_2-1 \geq 0$. In addition, $\nabla g_{b,u}^{n_\text{Ac}} \left(\boldsymbol{p}^\text{Ac}\right)$ is a vector of length $BU$ and its entry is obtained by
\begin{equation}\label{g nabla}
\nabla g_{b,u}^{n_\text{Ac}} (\boldsymbol{p}^\text{Ac})  =
\left\{
  \begin{array}{ll}
    0, & \hbox{$\forall i=b$}, \\
    \frac{W_\text{s} \gamma_{i,j}^{n_\text{Ac}} h_{i,u}^{n_\text{Ac}}  }{ (\ln2) \sum\limits_{v \in \mathcal{B}/ \{b\}} \sum\limits_{k \in \mathcal{U}/\{u\}} \gamma_{v,k}^{n_\text{Ac}} p_{v,k}^{n_\text{Ac}} h_{v,u}^{n_\text{Ac}} + \sigma_{u}^{n_\text{Ac}} }, & \hbox{$\forall i \neq b, j \in \mathcal{U}/\{u\}$}.
  \end{array}
\right.
\end{equation}
Hence, the approximated concave ergodic access data rate at MU $u$ over subcarrier $n_\text{Ac}$ at each iteration $t_2$ is given by
\begin{align}\label{rate approximate}
\hat{R}_{b,u}^{n_\text{Ac}} (\boldsymbol{p}^\text{Ac}_{t_2}) \approx \mathbb{E}_{\boldsymbol{h}} \left\{ f_{b,u}^{n_\text{Ac}} (\boldsymbol{p}^\text{Ac}_{t_2}) - g_{b,u}^{n_\text{Ac}} (\boldsymbol{p}^\text{Ac}_{t_2-1}) -
\nabla g_{b,u}^{n_\text{Ac}} (\boldsymbol{p}^\text{Ac}_{t_2-1})
\left(\boldsymbol{p}^\text{Ac}_{t_2} - \boldsymbol{p}^\text{Ac}_{t_2-1}\right) \right\}.
\end{align}
Therefore, by substituting the approximated concave data rate $\hat{R}_{b,u}^{n_\text{Ac}} (\boldsymbol{p}^\text{Ac}_{t_2})$ in \eqref{problem power total latency equivalent epi 2}, the convex approximated problem at each iteration $t_2$ is formulated as
\begin{subequations}\label{power Problem approximated}
\begin{align}\label{obf power Problem approximated}
\min_{ \boldsymbol{p}_{t_2}, \boldsymbol{x}_{t_2}, \hat{\boldsymbol{x}}_{t_2} }\hspace{.0 cm}
& 	
~~ \sum\limits_{b \in \mathcal{B}} \sum\limits_{u \in \mathcal{U}} \sum_{c=1}^{C} \omega_{u} \Delta_c \left( \hat{x}_{b,u}^{\text{Ac},c,(t_2)}  + \hat{x}^{\text{BH},c,(t_2)}_{b,u}  \right)
\\
\textrm{s.t.}\hspace{.0cm}~~
& \text{\eqref{Constraint power BS allowance}, \eqref{Constraint power data center allowance}, \eqref{Constraint positive powers}, \eqref{Constraint latency user epi2  power}-\eqref{Constraint latency user epi5  power},} \nonumber  \\
\label{constraint minrate approxDC}
& \sum\limits_{b \in \mathcal{B}} \hat{R}_{b,u}^{\text{Ac}} (\boldsymbol{p}^\text{Ac}_{t_2}) \geq R^\text{min}_{u} , \forall u \in \mathcal{U},
\\
\label{constraint minrate approxDC2}
& \hat{R}_{b,u}^{\text{Ac}} (\boldsymbol{p}^\text{Ac}_{t_2}) \geq  \theta_{b,u} s_c x_{b,u}^{\text{Ac},c,(t_2)}, \forall b \in \mathcal{B}, u \in \mathcal{U}, c \in \mathcal{C}.
\end{align}
\end{subequations}
The disciplined convex programming (DCP) problem \eqref{power Problem approximated} can be easily solved by using the
available optimization toolboxes, such as CVX \cite{CVXmatlab}.
The CVX package with SeDuMi and SDPT3 solvers can be used to solve the DCP problems \cite{CVXmatlab}. In this regard, CVX employs geometric programming (GP) with the interior point method (IPM) \cite{7100916,Boydconvex}. The default solver for solving the DCP problems is currently SDPT3. However, SeDuMi is faster for most DCP problems \cite{CVXmatlab}.
Moreover, the Lagrange dual method can be easily applied to solve the convex programming \eqref{power Problem approximated} \cite{6678362,7828114,Jorswieckfractional,7805409}.
The pseudo code of the proposed SCA algorithm with the D.C. approximation method is summarized in Alg. \ref{Alg SCA power caching PF}.
\begin{algorithm}[tp]
 \caption{The proposed SCA algorithm with the D.C. approximation method for solving \eqref{problem power total latency equivalent epi 2}} \label{Alg SCA power caching PF}
 \begin{algorithmic}[1]
 \STATE Initialize $\boldsymbol{p}_{0}$.
 \\ \textbf{repeat}
  \STATE Formulate $\hat{R}_{b,u}^{\text{Ac}} (\boldsymbol{p}^\text{Ac}_{t_2})$ using \eqref{rate approximate}.
  \STATE Obtain $\boldsymbol{p}_{t_2}$ by solving \eqref{power Problem approximated}.
  \STATE Set ${t_2}={t_2}+1$
  \\ \textbf{Until} Convergence of $\boldsymbol{p}$.
  \STATE The transmit power allocation $\boldsymbol{p}$ is the output of the algorithm.
 \end{algorithmic}
\end{algorithm}

\emph{Proposition 2}: The proposed SCA algorithm with the D.C. approximation method improves the objective function \eqref{obf problem power total latency} or remains constant at each iteration. Hence, the proposed algorithm converges to a locally optimal solution at each iteration.
\begin{proof}
Please see Appendix \ref{appendix proposition3}.
\end{proof}

\subsubsection{Finding $\boldsymbol{\gamma}$ for Fixed $(\boldsymbol{\rho},\boldsymbol{\beta},\boldsymbol{\theta},\boldsymbol{p})$}\label{Subsubsection find gamma caching PF}
To find $\boldsymbol{\gamma}$, we solve the following problem:
\begin{subequations}\label{problem joint gamma beta total latency}
\begin{align}\label{obf problem joint gamma beta total latency}
\min_{ \boldsymbol{\gamma} }\hspace{.0 cm}
& 	
~~ D^{\text{Pr},\text{tot}}
\\
\textrm{s.t.}\hspace{.0cm}~~
& \text{\eqref{Constraint latency user}-\eqref{Constraint power data center allowance}, \eqref{Constraint subcarrier user association}-\eqref{Constraint interference BH}, \eqref{Constraint indicator gamma}.} \nonumber
\end{align}
\end{subequations}
\eqref{problem joint gamma beta total latency} is an integer nonlinear programming (INLP) problem which is NP-hard. To solve \eqref{problem joint gamma beta total latency}, we first use the well-known relaxation method to relax the combinatorial constraint \eqref{Constraint indicator gamma} \cite{6251827,1658226}. To this end, $\gamma_{b,u}^{n_\text{Ac}}$ and $\gamma_{b}^{n_\text{BH}}$ are set to be real values between $0$ and $1$, and known as time sharing factors for all MUs associated to BS $b$ for transmitting requested contents through subcarrier $n_\text{Ac}$ and all BSs over subcarrier $n_\text{BH}$, respectively. Then, we again apply the transformation method presented in Appendix \ref{appendix EquivalentTransformation}. In this regard, we reformulate \eqref{problem joint gamma beta total latency} as follows:
\begin{subequations}\label{problem subcarrier total latency equivalent epi 2}
\begin{align}\label{obf problem subcarrier total latency equivalent epi 2}
\min_{ \boldsymbol{\gamma}, \boldsymbol{z}, \hat{\boldsymbol{z}} }\hspace{.0 cm}
& 	
~~ \sum\limits_{b \in \mathcal{B}} \sum\limits_{u \in \mathcal{U}} \sum_{c=1}^{C} \omega_{u} \Delta_c \left( \hat{z}_{b,u}^{\text{Ac},c}  + \hat{z}^{\text{BH},c}_{b,u}  \right)
\\
\textrm{s.t.}\hspace{.0cm}~~
& \text{\eqref{Constraint minimum rate}-\eqref{Constraint power data center allowance}, \eqref{Constraint subcarrier user association}-\eqref{Constraint interference BH}, \eqref{Constraint indicator gamma},} \nonumber  \\
\label{Constraint latency user epi1 2z}
&
R_{b,u}^{\text{Ac}} \geq \theta_{b,u} s_c z_{b,u}^{\text{Ac},c}, \forall b \in \mathcal{B}, u \in \mathcal{U}, c \in \mathcal{C},
\\
\label{Constraint latency user epi2 2z}
&
R_{b}^{\text{BH}} \geq \frac { \theta_{b,u} (1-\rho_{b,c}) s_c z^{\text{BH},c}_{b,u}} {\beta_{b,c}}, \forall b \in \mathcal{B}, u \in \mathcal{U}, c \in \mathcal{C},
\\
\label{Constraint latency user epi3 2z}
&
\sum\limits_{b \in \mathcal{B}} \sum\limits_{c \in \mathcal{C}} \Delta_c \left( \hat{z}_{b,u}^{\text{Ac},c} + \hat{z}^{\text{BH},c}_{b,u} \right) \leq T, \forall u \in \mathcal{U},
\\
\label{Constraint latency user epi4 2z}
&
\ln \left( \hat{z}_{b,u}^{\text{Ac},c} \right) + \ln \left( z_{b,u}^{\text{Ac},c} \right) \geq 0, \forall b \in \mathcal{B}, u \in \mathcal{U}, c \in \mathcal{C},
\\
\label{Constraint latency user epi5 2z}
&
\ln \left( \hat{z}^{\text{BH},c}_{b,u} \right) + \ln \left( z^{\text{BH},c}_{b,u} \right) \geq 0, \forall b \in \mathcal{B}, u \in \mathcal{U}, c \in \mathcal{C},
\end{align}
\end{subequations}
where $\boldsymbol{z}=[\boldsymbol{z}^\text{Ac},\boldsymbol{z}^\text{BH}]$, $\boldsymbol{z}^\text{Ac}=[z_{b,u}^{\text{Ac},c}]$, $\boldsymbol{z}^\text{BH}=[z_{b,u}^{\text{BH},c}]$, $\hat{\boldsymbol{z}}=[\hat{\boldsymbol{z}}^\text{Ac},\hat{\boldsymbol{z}}^\text{BH}]$, $\hat{\boldsymbol{z}}^\text{Ac}=[\hat{z}_{b,u}^{\text{Ac},c}]$ and $\hat{\boldsymbol{z}}^\text{BH}=[\hat{z}_{b,u}^{\text{BH},c}]$.
Note that the access data rate function $R_{b,u}^{\text{Ac}}$ is still non-concave in $\boldsymbol{\gamma}$. To tackle this issue, we again use the SCA approach with the D.C. method and approximate $R_{b,u}^{\text{Ac}} \left( \boldsymbol{\gamma}^\text{Ac} \right)$ to a concave form. This approach and its approximation formulations are very similar to Alg. \ref{Alg SCA power caching PF}. It is noteworthy that the resulting approximated DCP problem of finding $\gamma_{b,u}^{n_\text{Ac}}$ at each SCA iteration can be solved by using CVX or the Lagrange dual method.

\subsection{Solving the DP-PF problem}\label{subsection solving delivery nonfair}
Here, we solve \eqref{total latency Problem delivery} and obtain $\left(\boldsymbol{\theta},\boldsymbol{p},\boldsymbol{\gamma},\boldsymbol{\beta}\right)$ for a given $\boldsymbol{\rho}$ from the prior CP. Since both the CP-PF and DP-PF problems have the same structure, from the viewpoint of objective functions and constraints, we can utilize the same approach. The main difference in the solution algorithm for \eqref{total latency Problem delivery} is the first step of Alg. \ref{Alg iterative caching total}, i.e., finding $\boldsymbol{\beta}$ and $\boldsymbol{\theta}$ for a fixed $\boldsymbol{\rho}$. In this step, we also use the logarithmic transformation approach which is proposed for \eqref{Constraint latency user caching epi2}. Hence, we first formulate the optimization problem of finding $\boldsymbol{\beta}$ and $\boldsymbol{\theta}$ for fixed $\boldsymbol{\rho}$, $\boldsymbol{p}$ and $\boldsymbol{\gamma}$ as follows:
\begin{subequations}\label{problem beta delivery total latency equivalent form}
\begin{align}\label{obf problem beta delivery total latency equivalent form}
\min_{ \boldsymbol{\beta},\boldsymbol{\theta},\boldsymbol{y} }\hspace{.0 cm}
& 	
~~ \sum\limits_{u \in \mathcal{U}} \omega_{u} \sum_{c=1}^{C} \delta^{c}_{u} \sum\limits_{b \in \mathcal{B}}
\left( \theta_{b,u} \frac{s_c}{ r_{b,u}^{\text{Ac}} } +  y^{\text{BH},c}_{b,u} \right)
\\
\textrm{s.t.}\hspace{.0cm}~~
& \eqref{Constraint user association}, \eqref{Constraint subcarrier user association}, \eqref{Constraint beta}, \eqref{Constraint indicator gamma}, \nonumber
\\
&
\label{Constraint latency user caching epi2 trans deli2}
y^{\text{BH},c}_{b,u} \beta_{b,c} r_{b}^{\text{BH}} \geq \delta^{c}_{u} \theta_{b,u} (1-\rho_{b,c}) s_c, \forall b \in \mathcal{B}, u \in \mathcal{U},  c \in \mathcal{C},
\\
&
\label{Constraint latency user caching epi3 trans deli3}
\sum\limits_{b \in \mathcal{B}} \sum_{c \in \mathcal{C}} \left( \delta^{c}_{u} \theta_{b,u} \frac{s_c}{ r_{b,u}^{\text{Ac}} } + y^{\text{BH},c}_{b,u} \right)  \leq T, \forall u \in \mathcal{U}.
\end{align}
\end{subequations}
Constraint \eqref{Constraint latency user caching epi2 trans deli2} always holds when $\rho_{b,c}=1$ and/or $\delta^{c}_{u}=0$. This is because $y^{\text{BH},c}_{b,u}$, $\beta_{b,c}$, $\theta_{b,u}$ and $r_{b}^{\text{BH}}$ are lower-bounded by zero. For the case that $\rho_{b,c}=0$ and $\delta^{c}_{u}=1$, this constraint appears.
By using the logarithmic technique, \eqref{Constraint latency user caching epi2 trans deli2} is transformed into an equivalent form as
\begin{align}\label{Constraint latency user caching epi2 trans ln}
\ln \left( y^{\text{BH},c}_{b,u} \right) + \ln \left(\beta_{b,c}\right)+ \ln \left(r_{b}^{\text{BH}}\right) \geq \ln \left(s_c\right) + \theta_{b,u} \ln\left(\epsilon\right), \forall b, u, c , \delta^{c}_{u}(1-\rho_{b,c})=1.
\end{align}
The resulting MIDCP problem can be solved by using CXV with its internal solver MOSEK. To find $\boldsymbol{p}$ and $\boldsymbol{\gamma}$, we apply the same transformation strategy in Appendix \ref{appendix EquivalentTransformation}. Then, we again apply the SCA approach with the D.C. approximation method to tackle the non-concave access data rates. The approximated DCP problem of finding $\boldsymbol{p}$ at each SCA iteration $t_3$ is formulated as follows:
\begin{subequations}\label{problem power delivery total latency equivalent epi 2}
\begin{align}\label{obf problem power delivery total latency equivalent epi 2}
\min_{ \boldsymbol{p}_{t_3}, \boldsymbol{x}_{t_3}, \hat{\boldsymbol{x}}_{t_3} }\hspace{.0 cm}
& 	
~~ \sum\limits_{b \in \mathcal{B}} \sum\limits_{u \in \mathcal{U}} \sum_{c=1}^{C} \omega_{u} \left( \hat{x}_{b,u}^{\text{Ac},c,(t_3)}  + \hat{x}^{\text{BH},c,(t_3)}_{b,u}  \right)
\\
& \textrm{s.t.}\hspace{.0cm}~~ \text{\eqref{Constraint power BS allowance}, \eqref{Constraint power data center allowance}, \eqref{Constraint positive powers},} \nonumber \\
& \sum\limits_{b \in \mathcal{B}} \hat{r}_{b,u}^{\text{Ac}} (\boldsymbol{p}^\text{Ac}_{t_3}) \geq R^\text{min}_{u} , \forall u \in \mathcal{U},
\\
& \hat{r}_{b,u}^{\text{Ac}} (\boldsymbol{p}^\text{Ac}_{t_3}) \geq \delta^{c}_{u} \theta_{b,u} s_c x_{b,u}^{\text{Ac},c,(t_3)}, \forall b \in \mathcal{B}, u \in \mathcal{U}, c \in \mathcal{C},
\\
& r_{b}^{\text{BH}} \geq \frac {\delta^{c}_{u} \theta_{b,u} (1-\rho_{b,c}) s_c x^{\text{BH},c,(t_3)}_{b,u}} { \beta_{b,c} }, \forall b \in \mathcal{B}, u \in \mathcal{U}, c \in \mathcal{C},
\\
& \sum\limits_{b \in \mathcal{B}} \sum_{c \in \mathcal{C}} \left( \hat{x}_{b,u}^{\text{Ac},c,(t_3)} + \hat{x}^{\text{BH},c,(t_3)}_{b,u} \right) \leq T, \forall u \in \mathcal{U},
\\
& \ln \left( \hat{x}_{b,u}^{\text{Ac},c,(t_3)} \right) + \ln \left( x_{b,u}^{\text{Ac},c,(t_3)} \right) \geq 0, \forall b \in \mathcal{B}, u \in \mathcal{U}, c \in \mathcal{C},
\\
& \ln \left( \hat{x}^{\text{BH},c,(t_3)}_{b,u} \right) + \ln \left( x^{\text{BH},c,(t_3)}_{b,u} \right) \geq 0, \forall b \in \mathcal{B}, u \in \mathcal{U}, c \in \mathcal{C}.
\end{align}
\end{subequations}
\eqref{problem power delivery total latency equivalent epi 2} can be solved by using the CVX software or the Lagrange dual method. The proof of the convergence of the proposed algorithm to find $\boldsymbol{p}$ is very similar to \emph{Proposition 2}. Besides, similar to \eqref{problem subcarrier total latency equivalent epi 2}, the relaxed approximated problem of finding $\boldsymbol{\gamma}_{t_4}$ is formulated by
\begin{subequations}\label{problem subcarrier total latency equivalent epi 2 delivery}
\begin{align}\label{obf problem subcarrier total latency equivalent epi 2 delivery}
\min_{ \boldsymbol{\gamma}_{t_4}, \boldsymbol{z}_{t_4}, \hat{\boldsymbol{z}}_{t_4} }\hspace{.0 cm}
& 	
~~ \sum\limits_{b \in \mathcal{B}} \sum\limits_{u \in \mathcal{U}} \sum_{c=1}^{C} \omega_{u} \left( \hat{z}_{b,u}^{\text{Ac},c,(t_4)}  + \hat{z}^{\text{BH},c,(t_4)}_{b,u}  \right)
\\
\textrm{s.t.}\hspace{.0cm}~~
& \text{\eqref{Constraint subcarrier user association}-\eqref{Constraint interference BH}, \eqref{Constraint indicator gamma}, \eqref{Constraint power BS allowance delivery}, \eqref{Constraint power data center allowance delivery},} \nonumber  \\
& \sum\limits_{b \in \mathcal{B}} \hat{r}_{b,u}^{\text{Ac}} (\boldsymbol{\gamma}^\text{Ac}_{t_4}) \geq R^\text{min}_{u} , \forall u \in \mathcal{U},
\\
& \hat{r}_{b,u}^{\text{Ac}} (\boldsymbol{\gamma}^\text{Ac}_{t_4}) \geq \delta^{c}_{u} \theta_{b,u} s_c z_{b,u}^{\text{Ac},c,(t_4)}, \forall b \in \mathcal{B}, u \in \mathcal{U}, c \in \mathcal{C},
\\
& r_{b}^{\text{BH}} \geq \frac{\delta^{c}_{u} \theta_{b,u} (1-\rho_{b,c}) s_c z^{\text{BH},c,(t_4)}_{b,u}} { \beta_{b,c} }, \forall b \in \mathcal{B}, u \in \mathcal{U}, c \in \mathcal{C},
\\
& \sum\limits_{b \in \mathcal{B}} \sum_{c \in \mathcal{C}} \left( \hat{z}_{b,u}^{\text{Ac},c,(t_4)} + \hat{z}^{\text{BH},c,(t_4)}_{b,u} \right) \leq T, \forall u \in \mathcal{U},
\\
& \ln \left( \hat{z}_{b,u}^{\text{Ac},c,(t_4)} \right) + \ln \left( z_{b,u}^{\text{Ac},c,(t_4)} \right) \geq 0, \forall b \in \mathcal{B}, u \in \mathcal{U}, c \in \mathcal{C},
\\
& \ln \left( \hat{z}^{\text{BH},c,(t_4)}_{b,u} \right) + \ln \left( z^{\text{BH},c,(t_4)}_{b,u} \right) \geq 0, \forall b \in \mathcal{B}, u \in \mathcal{U}, c \in \mathcal{C},
\\
\label{relaxed constraints sub}
& 0 \leq \gamma_{b,u}^{n_\text{Ac}} \leq 1, ~~~~~0 \leq \gamma_{b}^{n_\text{BH}} \leq 1,
\end{align}
\end{subequations}
in which we first used the transformation technique presented in Appendix \ref{appendix EquivalentTransformation}, then we relaxed all binary variables in $\boldsymbol{\gamma}_{t_4}$ and after that, we applied the SCA approach with the D.C. method to approximate the non-concave data rate $r_{b,u}^{\text{Ac}} (\boldsymbol{\gamma}^\text{Ac}_{t_4})$ at each iteration $t_4$ to a concave form. The DCP problem \eqref{problem subcarrier total latency equivalent epi 2 delivery} can be efficiently solved using CVX or the Lagrange dual method.

To initialize $\boldsymbol{\theta}$, $\boldsymbol{p}$, $\boldsymbol{\beta}$, and $\boldsymbol{\gamma}$ in the DP, we use the following information-centric approach.
Each MU selects the nearest FBS within $d^\text{max}$ if the FBS has the requested content. If there is no FBS within $d^\text{max}$ which has the requested content, the MU is associated to MBS, if the MBS has the requested content. Otherwise, the MU is associated to the nearest FBS within $d^\text{max}$. If there is no FBS within $d^\text{max}$, the MU is associated to MBS.
Then, we apply the equal power allocation approach between MUs and active BSs, i.e., BSs which request at least one content from the DC. Based on binary IGRs of MUs at each time period in the DP, we note that all the un-cached requested contents of MUs in cell $b$ should be sent to BS $b$ from the DC. By equal allocation of non-zero values in $\boldsymbol{\beta}$, we have
$
	\beta_{b,c} = \frac{ \min\{ \sum_{u \in \mathcal{U}} \delta^{c}_{u} \theta_{b,u} (1-\rho_{b,c}) , 1 \}  }
	{ \sum_{u \in \mathcal{U}} \sum_{c=1}^{C} \delta^{c}_{u} \theta_{b,u} (1-\rho_{b,c}) },
$
which represents that the portions of backhaul data rate for the un-cached requested contents in each cell are equally distributed.
Note that in each time period, each un-cached requested content is sent only once to the BS. After finding $\boldsymbol{\theta}$, $\boldsymbol{p}$ and $\boldsymbol{\beta}$, we solve \eqref{total latency Problem delivery} and find $\boldsymbol{\gamma}$.

\subsection{Solving the CP-MMF problem}\label{subsection solution fair caching}
In this subsection, we solve \eqref{total latency Problem fair caching} and design a FTACPS based on the MMF scheme. In doing so, we first transform \eqref{total latency Problem fair caching} into the following equivalent problem as
\begin{subequations}\label{problem fair caching equ1}
\begin{align}\label{obf problem fair caching equ1}
\min_{ \boldsymbol{\rho} , \boldsymbol{\beta}, \boldsymbol{\theta} , \boldsymbol{p} , \boldsymbol{\gamma} , v }\hspace{.0 cm}
& 	
~~ v
\\
\textrm{s.t.}\hspace{.0cm}~~
& \text{\eqref{Constraint cache size}-\eqref{Constraint indicator rho},} \nonumber  \\
\label{constraint max}
& \sum\limits_{b \in \mathcal{B}} D_{b,u}^{\text{Pr}} \leq v, \forall u \in \mathcal{U}.
\end{align}
\end{subequations}
Since \eqref{Constraint latency user} and \eqref{constraint max} state that the average latency of each MU $u \in \mathcal{U}$ should not exceed the considered threshold time lengths, \eqref{problem fair caching equ1} can be rewritten as
\begin{subequations}\label{problem fair caching equ2}
\begin{align}\label{obf problem fair caching equ2}
\min_{ \boldsymbol{\rho} , \boldsymbol{\beta}, \boldsymbol{\theta} , \boldsymbol{p} , \boldsymbol{\gamma} , v }\hspace{.0 cm}
& 	
~~ v
\\
\textrm{s.t.}\hspace{.0cm}~~
& \text{\eqref{Constraint cache size}, \eqref{Constraint minimum rate}-\eqref{Constraint indicator rho}, \eqref{constraint max},} \nonumber  \\
\label{constraint v threshold}
& v \leq T,
\end{align}
\end{subequations}
where the convex constraint \eqref{constraint v threshold} represents that the maximal latency of MUs should not exceed $T$.
In order to solve the MINLP problem \eqref{problem fair caching equ2}, we propose an AO algorithm which is similar to Alg. \ref{Alg iterative caching total} for solving the CP-PF problem. Specifically, in each iteration, we first obtain $\boldsymbol{\rho}$, $\boldsymbol{\beta}$, $\boldsymbol{\theta}$ and $v$ for a pre-defined $(\boldsymbol{p} , \boldsymbol{\gamma})$. After that, we find $(\boldsymbol{p} ,v)$ and $(\boldsymbol{\gamma} ,v)$ in separated steps.
These iterations are repeatedly applied until the proposed AO algorithm converges. The proof of the convergence of the proposed algorithm is similar to \emph{Proposition 1}. To initialize $\boldsymbol{\rho}_0$, $\boldsymbol{p}_0$, $\boldsymbol{\beta}_0$ and $\boldsymbol{\theta}_0$, we use the same initialization method presented in Subsection \ref{subsubsection Initialization caching phase total latency} for the CP. Then, we solve \eqref{problem fair caching equ2} and find $\boldsymbol{\gamma}_0$. After initializing, we find $(\boldsymbol{\rho},\boldsymbol{\beta},\boldsymbol{\theta},v)$ by solving the following MINLP problem:
\begin{subequations}\label{problem fair caching rho v}
\begin{align}\label{obf problem fair caching rho v}
\min_{ \boldsymbol{\rho} , \boldsymbol{\beta}, \boldsymbol{\theta}, v  }\hspace{.0 cm}
& 	
~~ v
\\
\textrm{s.t.}\hspace{.0cm}~~
& \eqref{Constraint cache size}, \eqref{Constraint user association}, \eqref{Constraint subcarrier user association}, \eqref{Constraint beta}, \eqref{Constraint indicator gamma}, \eqref{Constraint indicator rho}, \eqref{constraint max}, \eqref{constraint v threshold}, \nonumber
\end{align}
\end{subequations}
which can be solved by using the proposed approach utilized to obtain $\left(\boldsymbol{\rho} , \boldsymbol{\beta}, \boldsymbol{\theta}\right)$ in the CP-PF problem.
The MIDCP form of \eqref{problem fair caching rho v} is given by
\begin{subequations}\label{problem fair caching rho v epi1}
\begin{align}\label{obf problem fair caching rho v epi1}
\min_{  \boldsymbol{\rho} , \boldsymbol{\beta}, \boldsymbol{\theta}, v, \boldsymbol{y} }\hspace{.0 cm}
& 	
~~ v
\\
\textrm{s.t.}\hspace{.0cm}~~
& \text{ \eqref{Constraint cache size}, \eqref{Constraint user association}, \eqref{Constraint subcarrier user association}, \eqref{Constraint beta}, \eqref{Constraint indicator gamma}, \eqref{Constraint indicator rho}, \eqref{Constraint latency user caching epi2 ln}, \eqref{constraint v threshold},} \nonumber
\\
\label{constraint max epi3 trans}
&
\sum\limits_{b \in \mathcal{B}} \sum\limits_{c \in \mathcal{C}} \Delta_c \left( \theta_{b,u} \frac{s_c}{ R_{b,u}^{\text{Ac}} } + y^{\text{BH},c}_{b} \right)  \leq v, \forall u \in \mathcal{U},
\end{align}
\end{subequations}
which can be solved by utilizing CVX with the internal solver MOSEK. After finding $\boldsymbol{\rho}$, $\boldsymbol{\beta}$ and $\boldsymbol{\theta}$, we solve the following optimization problem to obtain $(\boldsymbol{p},v)$ as
\begin{subequations}\label{problem fair caching joint powerV}
\begin{align}\label{obf problem fair caching joint powerV}
\min_{ \boldsymbol{p} , v }\hspace{.0 cm}
& 	
~~ v
\\
\textrm{s.t.}\hspace{.0cm}~~
& \text{\eqref{Constraint minimum rate}-\eqref{Constraint power data center allowance}, \eqref{Constraint positive powers}, \eqref{constraint max}, \eqref{constraint v threshold}.} \nonumber
\end{align}
\end{subequations}
Similar to \eqref{problem power total latency}, \eqref{problem fair caching joint powerV} is highly non-convex. Therefore, we again use the epigraph method which is presented in Appendix \ref{appendix EquivalentTransformation} and subsequently the SCA approach with the D.C. approximation method to approximate the access data rates to concave forms. At each SCA iteration, the result approximated DCP problem is
\begin{subequations}\label{problem fair caching joint powerV deadline concavity}
\begin{align}\label{obf problem fair caching joint powerV}
\min_{ \boldsymbol{p}, v, \boldsymbol{x}, \hat{\boldsymbol{x}} }\hspace{.0 cm}
& 	
~~ v
\\
\textrm{s.t.}\hspace{.0cm}~~
& \text{\eqref{Constraint power BS allowance}, \eqref{Constraint power data center allowance}, \eqref{Constraint positive powers}, \eqref{Constraint latency user epi2  power}, \eqref{Constraint latency user epi4  power}, \eqref{Constraint latency user epi5  power}, \eqref{constraint minrate approxDC}, \eqref{constraint minrate approxDC2}, \eqref{constraint v threshold}}, \nonumber \\
& \sum\limits_{b \in \mathcal{B}} \sum\limits_{c \in \mathcal{C}} \Delta_c \left( \hat{x}_{b,u}^{\text{Ac},c} + \hat{x}^{\text{BH},c}_{b,u} \right) \leq v, \forall u \in \mathcal{U},
\end{align}
\end{subequations}
which can be solved by using CVX or the Lagrange dual method. The convergence of the proposed solution for \eqref{problem fair caching joint powerV} is also similar to \emph{Proposition 2}.

To find $\boldsymbol{\gamma}$ and $v$, similar to \eqref{problem subcarrier total latency equivalent epi 2}, the relaxed approximated DCP problem of \eqref{problem fair caching joint powerV} is formulated by
\begin{subequations}\label{problem joint gamma beta fair latency equ}
\begin{align}\label{obf problem joint gamma beta fair latency equ}
\min_{ \boldsymbol{\gamma},v , \boldsymbol{z}, \hat{\boldsymbol{z}} }\hspace{.0 cm}
& 	
~~ v
\\
\textrm{s.t.}\hspace{.0cm}~~
& \text{\eqref{Constraint power BS allowance}, \eqref{Constraint power data center allowance}, \eqref{Constraint subcarrier user association}-\eqref{Constraint interference BH}, \eqref{Constraint latency user epi2 2z}, \eqref{Constraint latency user epi4 2z}, \eqref{Constraint latency user epi5 2z}, \eqref{relaxed constraints sub}, \eqref{constraint v threshold},} \nonumber    \\
\label{minrate gamma approximate}
&
\sum\limits_{b \in \mathcal{B}} \hat{R}_{b,u}^{\text{Ac}}\left(\boldsymbol{\gamma}^\text{Ac}\right) \geq R^\text{min}_{u} , \forall u \in \mathcal{U},
\\
\label{Constraint latency user epi1 2z approx}
&
\hat{R}_{b,u}^{\text{Ac}}\left(\boldsymbol{\gamma}^\text{Ac}\right) \geq \theta_{b,u} s_c z_{b,u}^{\text{Ac},c}, \forall b \in \mathcal{B}, u \in \mathcal{U}, c \in \mathcal{C},
\\
\label{Constraint latency user epi3 epi}
&
\sum\limits_{b \in \mathcal{B}} \sum\limits_{c \in \mathcal{C}} \Delta_c \left( \hat{z}_{b,u}^{\text{Ac},c} + \hat{z}^{\text{BH},c}_{b} \right) \leq v, \forall u \in \mathcal{U},
\end{align}
\end{subequations}
in which we first used the epigraph technique and the relaxation method to tackle the fractional and combinatorial constraints, respectively, and then, we applied the SCA approach with the D.C. method to approximate $R_{b,u}^{\text{Ac}}\left(\boldsymbol{\gamma}^\text{Ac}\right)$ to a concave form. \eqref{problem joint gamma beta fair latency equ} can also be solved by CVX or the Lagrange dual method.

\subsection{Solving the DP-MMF problem}
Similar to Subsection \ref{subsection solving delivery nonfair}, the CP-MMF and DP-MMF problems have the same structure (for a fixed $\boldsymbol{\rho}$). Therefore, to solve \eqref{total latency Problem delivery fair}, we use the proposed algorithm which is devised to solve CP-MMF problem.
To initialize $\boldsymbol{p}_0$ and $\boldsymbol{\beta}_0$, we use the same initialization approach as in Subsection \ref{subsection solving delivery nonfair}. Then, we find $\boldsymbol{\gamma}_0$ by solving \eqref{total latency Problem delivery fair} using the same approach proposed to find $\boldsymbol{\gamma}$ in the CP-MMF problem.

\section{Computational Complexity}\label{section computational complexity}
Here, we obtain the computational complexity of the proposed AO algorithms. The complexity of each algorithm is a linear function of total number of iterations performed, and the complexity of each subproblem. The total number of main iterations depends on the adopted stopping criterion or the accuracy of the algorithm.

\subsection{Computational complexity of solving the CP-PF problem}\label{subsection cc total latency caching}
The proposed Alg. \ref{Alg iterative caching total} for solving the CP-PF problem operates in three main steps. The first step is finding $\boldsymbol{\rho}$, $\boldsymbol{\beta}$, and $\boldsymbol{\theta}$. In this line, we first transformed \eqref{problem rho total latency} into the equivalent problem \eqref{problem rho total latency equivalent form 2} which is solved by using  CVX with the internal solver MOSEK. CVX is fundamentally based on IPM \cite{CVXmatlab,7100916,Boydconvex}. Hence, the computational complexity of solving \eqref{problem rho total latency equivalent form 2} is formulated as follows:
\begin{equation}\label{Complexrho PF}
  \Psi_\text{1}^\text{CP,PF} = \frac {  \log\left(  T_\text{1}^\text{CP,PF} /\left(t^0 \varrho\right) \right) }
        {  \log \xi },
\end{equation}
where $T_\text{1}^\text{CP,PF} = 2+2B+(B+3)U+(B+1)UC$ is the total number of constraints in \eqref{problem rho total latency equivalent form 2}, $t^0$ is the initial point for approximating the accuracy of the IPM, $0< \varrho \ll \infty$ is the stopping criterion for the IPM, and $\xi$ is used for updating the accuracy of the IPM \cite{Boydconvex,7100916}. It is noteworthy that the complexity estimation formulation \eqref{Complexrho PF} does not assume sparsity or any special structures in the data matrices which greatly affects the complexity of solving problems. In the first step, $\boldsymbol{\rho}$ and $\boldsymbol{\theta}$ are very sparse matrixes due to constraints \eqref{Constraint cache size} and \eqref{Constraint user association}. Moreover, $\boldsymbol{\beta}$ is related to $\boldsymbol{\rho}$ and is another sparse matrix.

In the second step of the Alg. \ref{Alg iterative caching total}, we find $\boldsymbol{p}$ by solving \eqref{problem power total latency} using the SCA approach with the D.C. approximation method, where the approximated DCP problem \eqref{power Problem approximated} at each iteration $t_2$ is solved by using CVX. The complexity of solving \eqref{power Problem approximated} at each iteration $t_2$ is given by
$\Psi_\text{2}^\text{CP,PF} = \frac {  \log(  T_\text{2}^\text{CP,PF} /t^0 \varrho ) } {  \log( \xi ) },$
where $T_\text{2}^\text{CP,PF} = 1+B+2U+4(B+1)UC$. Finally, we obtained $\boldsymbol{\gamma}$ in the same way as $\boldsymbol{p}$. The computational complexity of obtaining $\boldsymbol{\gamma}$ at each SCA iteration is thus on the order of
$
  \Psi_\text{3}^\text{CP,PF} = \frac {  \log(  T_\text{3}^\text{CP,PF} /t^0 \varrho ) }
        {  \log( \xi ) },
$
where $T_\text{3}^\text{CP,PF} = 1+N_\text{BH}+2U+(B+1)\left(1+U+N_\text{Ac}\right)+4(B+1)UC$. Therefore, the total computational complexity of solving \eqref{total latency Problem caching} using Alg. \ref{Alg iterative caching total} is on the order of $\Psi_\text{tot}^\text{CP,PF}=I_\text{tot}\left(
\Psi_\text{1}^\text{CP,PF} + I^\text{SCA}_\text{Pow} \Psi_\text{2}^\text{CP,PF} + I^\text{SCA}_\text{Sub} \Psi_\text{3}^\text{CP,PF} \right)$ where $I_\text{tot}$, $I^\text{SCA}_\text{Pow}$ and $I^\text{SCA}_\text{Sub}$ are the number of main iterations, SCA iterations for finding $\boldsymbol{p}$ and SCA iterations for finding $\boldsymbol{\gamma}$, respectively. Note that the sparsity of $\boldsymbol{p}$ and $\boldsymbol{\gamma}$ are not considered in the complexity formulations. Moreover, the special structures of finding new added variables such as $\boldsymbol{x}$ and $\hat{\boldsymbol{x}}$ for finding $\boldsymbol{p}$ are not considered in the formulations. Actually, new variables $\boldsymbol{x}$ and $\hat{\boldsymbol{x}}$ for finding $\boldsymbol{p}$ can be easily found, since they are only the upper-bound delay variables.

\subsection{Computational complexity of solving the DP-PF problem}\label{subsection cc total latency delivery}
Both the proposed algorithms for solving the CP-PF and DP-PF problems have the same fundamental structure for a fixed $\boldsymbol{\rho}$. The main difference between the proposed algorithms is removing the constraints \eqref{Constraint cache size} and \eqref{Constraint indicator rho} from \eqref{problem rho total latency equivalent form 2} in the DP. The complexity of finding $\boldsymbol{\beta}$ and $\boldsymbol{\theta}$ is
$
  \Psi_\text{1}^\text{DP,PF} = \frac {  \log\left(  T_\text{1}^\text{DP,PF} /\left(t^0 \varrho\right) \right) }
        {  \log \xi },
$
where $T_\text{1}^\text{DP,PF} = T_\text{1}^\text{CP,PF}-(B+1)$. Besides, the complexity of finding $\boldsymbol{p}$ at each SCA iteration can be obtained by
$
  \Psi_\text{2}^\text{DP,PF} = \frac {  \log(  T_\text{2}^\text{DP,PF} /t^0 \varrho ) }
        {  \log( \xi ) },
$
where $T_\text{2}^\text{DP,PF} = T_\text{2}^\text{CP,PF}$. Similarly, the complexity of obtaining $\boldsymbol{\gamma}$ at each SCA iteration is on the order of
$
  \Psi_\text{3}^\text{DP,PF} = \frac {  \log(  T_\text{3}^\text{DP,PF} /t^0 \varrho ) }
        {  \log( \xi ) },
$
where $T_\text{3}^\text{DP,PF} = T_\text{3}^\text{CP,PF}$.

\subsection{Computational complexity of solving the CP-MMF problem}\label{subsection cc fair caching}
The main structure of the proposed algorithm for solving the CP-MMF problem is similar to the proposed Alg. \ref{Alg iterative caching total} to solve the CP-PF problem. The main difference between them is finding $v$ at each step. The complexity of finding $\left(\boldsymbol{\rho},\boldsymbol{\beta},\boldsymbol{\theta}\right)$ is on the order of
$
  \Psi_\text{1}^\text{CP,MMF} = \frac {  \log\left(  T_\text{1}^\text{CP,MMF} /\left(t^0 \varrho\right) \right) }
        {  \log \xi },
$
where $T_\text{1}^\text{CP,MMF} = T_\text{1}^\text{CP,PF}+1$. In addition, the computational complexity of finding $\boldsymbol{p}$ at each SCA iteration is given by
$
  \Psi_\text{2}^\text{CP,MMF} = \frac {  \log(  T_\text{2}^\text{CP,MMF} /t^0 \varrho ) }
        {  \log( \xi ) },
$
where $T_\text{2}^\text{CP,MMF} = T_\text{2}^\text{CP,PF}+1$. Furthermore, the complexity of obtaining $\boldsymbol{\gamma}$ at each SCA iteration is formulated by
$
  \Psi_\text{3}^\text{CP,MMF} = \frac {  \log(  T_\text{3}^\text{CP,MMF} /t^0 \varrho ) }
        {  \log( \xi ) },
$
where $T_\text{3}^\text{CP,MMF} = T_\text{3}^\text{CP,PF}+1$.

\subsection{Computational complexity of solving the DP-MMF problem}\label{subsection cc delivery fair}
Similar to Subsection \ref{subsection cc total latency delivery}, both the proposed algorithms for solving the CP-MMF and DP-MMF problems have the same main structure. Accordingly, the computational complexity of finding $(\boldsymbol{\beta},\boldsymbol{\theta})$, $\boldsymbol{p}$ at each SCA iteration and $\boldsymbol{\gamma}$ at each SCA iteration are formulated, respectively by
$
  \Psi_\text{1}^\text{DP,MMF} = \frac {  \log\left(  T_\text{1}^\text{DP,MMF} /\left(t^0 \varrho\right) \right) }
        {  \log \xi },
$
where $T_\text{1}^\text{DP,MMF} = T_\text{1}^\text{CP,MMF}-(B+1)$,
$
  \Psi_\text{2}^\text{DP,MMF} = \frac {  \log(  T_\text{2}^\text{DP,MMF} /t^0 \varrho ) }
        {  \log( \xi ) },
$
where $T_\text{2}^\text{DP,MMF} = T_\text{2}^\text{CP,MMF}$, and
$
  \Psi_\text{3}^\text{DP,MMF} = \frac {  \log(  T_\text{3}^\text{DP,MMF} /t^0 \varrho ) }
        {  \log( \xi ) },
$
in which $T_\text{3}^\text{DP,MMF} = T_\text{3}^\text{CP,MMF}$.
We also summarize the computational complexity of the proposed AO algorithms in Table \ref{Table Complexity}. Since we assumed that all subproblems are solved using the CVX solver, the difference of the computational complexity of subproblems comes from the difference of the total number of constraints. To this end, we only present the total number of constraints in each subproblem in Table \ref{Table Complexity}.
\begin{table*}[tp]
\centering
\caption{Total number of constraints in each subproblem}
\begin{center} \label{Table Complexity}
\begin{tabular}{| c | >{\centering\arraybackslash}m{1in} | c | >{\centering\arraybackslash}m{1in} | >{\centering\arraybackslash}m{1.5in} |}
    \hline \rowcolor[gray]{0.85}
    \hline \rowcolor[gray]{0.85}
    \hline \rowcolor[gray]{0.85}
    \hline \rowcolor[gray]{0.85}
     \textbf{Main Problem} &  \textbf{Finding} $(\boldsymbol{\rho},\boldsymbol{\beta},\boldsymbol{\theta})$ &
     \textbf{Finding} $(\boldsymbol{\beta},\boldsymbol{\theta})$ & \textbf{Finding} $\boldsymbol{p}$ & \textbf{Finding} $\boldsymbol{\gamma}$ \\
    \hline \rowcolor[gray]{0.93}
    \hline \rowcolor[gray]{0.93}
    \hline \rowcolor[gray]{0.93}
    \hline  \rowcolor[gray]{0.93}
     CP-PF
     & \begingroup\makeatletter\def\f@size{6}\check@mathfonts\def\maketag@@@#1{\hbox{\m@th\large
     \normalfont#1}}
     $T_\text{1}^\text{CP,PF} = 2+2B+(B+3)U+(B+1)UC$ \endgroup
     & \begingroup\makeatletter\def\f@size{6}\check@mathfonts\def\maketag@@@#1{\hbox{\m@th\large
     \normalfont#1}}
     $-$ \endgroup
     & \begingroup\makeatletter\def\f@size{6}\check@mathfonts\def\maketag@@@#1{\hbox{\m@th\large
     \normalfont#1}}
     $T_\text{2}^\text{CP,PF} = 1+B+2U+4(B+1)UC$ \endgroup
     & \begingroup\makeatletter\def\f@size{6}\check@mathfonts\def\maketag@@@#1{\hbox{\m@th\large
     \normalfont#1}}
     $T_\text{3}^\text{CP,PF} = 1+N_\text{BH}+2U+(B+1)\left(1+U+N_\text{Ac}\right)+4(B+1)UC)$ \endgroup \\
     \hline \rowcolor[gray]{0.95}
     DP-PF
     & \begingroup\makeatletter\def\f@size{6}\check@mathfonts\def\maketag@@@#1{\hbox{\m@th\large
     \normalfont#1}}
     $-$ \endgroup
     & \begingroup\makeatletter\def\f@size{6}\check@mathfonts\def\maketag@@@#1{\hbox{\m@th\large
     \normalfont#1}}
     $T_\text{1}^\text{CP,PF}-(B+1)$ \endgroup
     & \begingroup\makeatletter\def\f@size{6}\check@mathfonts\def\maketag@@@#1{\hbox{\m@th\large
     \normalfont#1}}
     $T_\text{2}^\text{CP,PF}$ \endgroup
     & \begingroup\makeatletter\def\f@size{6}\check@mathfonts\def\maketag@@@#1{\hbox{\m@th\large
     \normalfont#1}}
     $T_\text{3}^\text{CP,PF}$ \endgroup \\
     \hline \rowcolor[gray]{0.97}
     CP-MMF
     & \begingroup\makeatletter\def\f@size{6}\check@mathfonts\def\maketag@@@#1{\hbox{\m@th\large
     \normalfont#1}}
     $T_\text{1}^\text{CP,PF}+1$ \endgroup
     & \begingroup\makeatletter\def\f@size{6}\check@mathfonts\def\maketag@@@#1{\hbox{\m@th\large
     \normalfont#1}}
     $-$ \endgroup
     & \begingroup\makeatletter\def\f@size{6}\check@mathfonts\def\maketag@@@#1{\hbox{\m@th\large
     \normalfont#1}}
     $T_\text{2}^\text{CP,PF}+1$ \endgroup
     & \begingroup\makeatletter\def\f@size{6}\check@mathfonts\def\maketag@@@#1{\hbox{\m@th\large
     \normalfont#1}}
     $T_\text{3}^\text{CP,PF}+1$ \endgroup \\
     \hline \rowcolor[gray]{0.99}
     DP-MMF
     & \begingroup\makeatletter\def\f@size{6}\check@mathfonts\def\maketag@@@#1{\hbox{\m@th\large
     \normalfont#1}}
     $-$ \endgroup
     & \begingroup\makeatletter\def\f@size{6}\check@mathfonts\def\maketag@@@#1{\hbox{\m@th\large
     \normalfont#1}}
     $T_\text{1}^\text{CP,PF}-B$ \endgroup
     & \begingroup\makeatletter\def\f@size{6}\check@mathfonts\def\maketag@@@#1{\hbox{\m@th\large
     \normalfont#1}}
     $T_\text{2}^\text{CP,PF}+1$ \endgroup
     & \begingroup\makeatletter\def\f@size{6}\check@mathfonts\def\maketag@@@#1{\hbox{\m@th\large
     \normalfont#1}}
     $T_\text{3}^\text{CP,PF}+1$ \endgroup \\
     \hline
  \end{tabular}
\end{center}
\end{table*}

\section{Simulation Results}\label{section simulation results}
\begin{figure}
\centering
\includegraphics[scale=0.50]{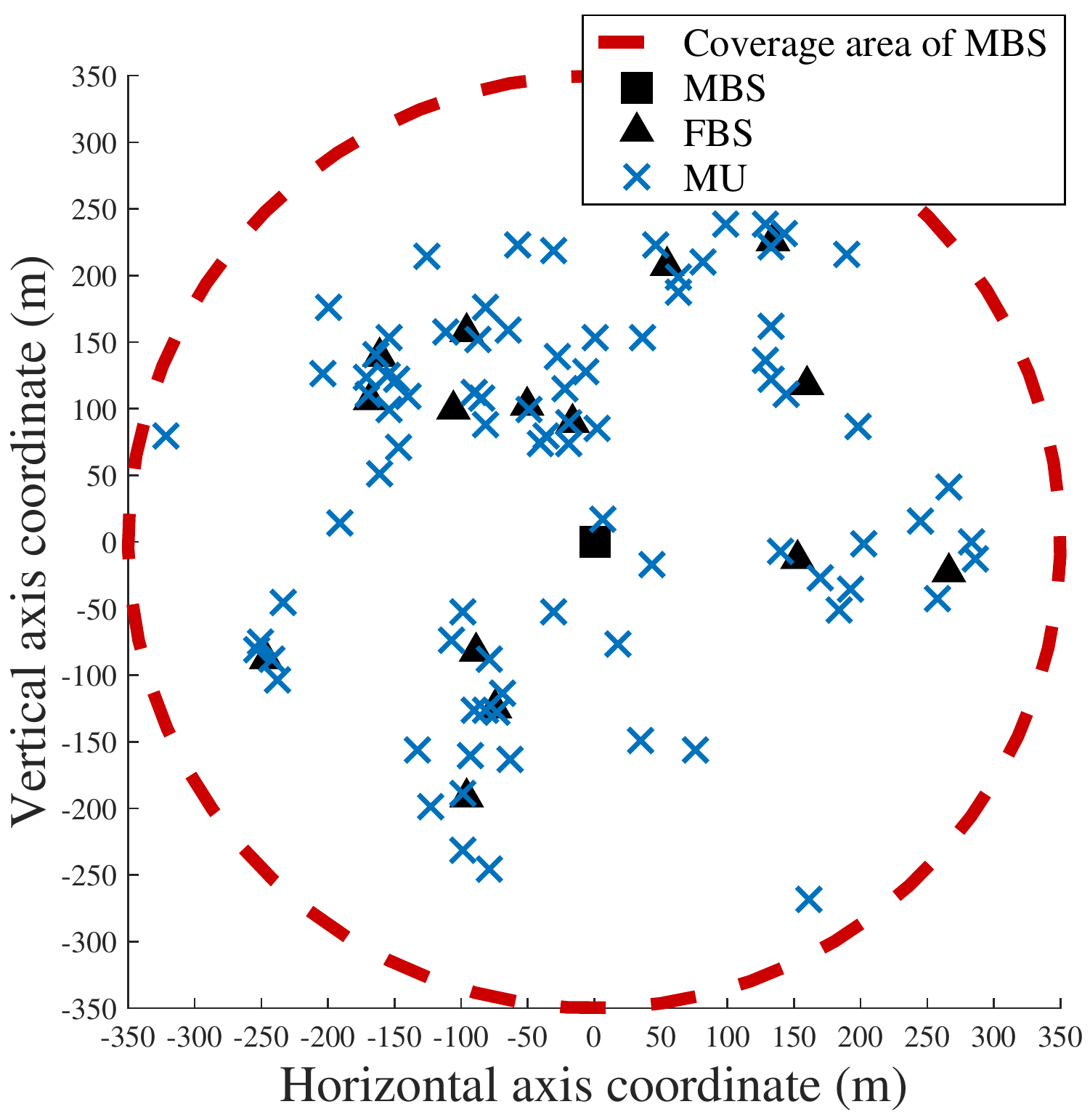}
\caption{Exemplary network topology and users placement in our numerical assessments.}
\label{Fig03Locations}
\end{figure}

Here, we provide the Monte Carlo-based simulation results to evaluate the performance of our proposed FTACPSs via a Matlab-based simulator system. All numerical examples are conducted with Matlab R2014b on an x64-based laptop equipped with an Intel(R) Core(TM) i7-5500U CPU of speed 2.40 GHz and a memory of 8 GB. In these simulations, a single MBS is positioned at the center of a circular area with radius $350$ m (typical of urban macrocell \cite{7530876,6600983,6883210}). $15$ FBSs each having radii of $70$ m are uniformly distributed in the coverage area of MBS \cite{7530876,6600983}. Moreover, there are $5$ and $10$ MUs uniformly distributed in the coverage area of each FBS and MBS, respectively \cite{6678362}.
Fig. \ref{Fig03Locations} illustrates an exemplary network topology in our numerical results.
The distance between MBS and DC is also set to $2$ km.

The wireless access and backhaul frequency bands are set to $W_\text{Ac}=20$ \cite{7530876,6600983,7536886,6883210,7322204} and $W_\text{BH}=20$ MHz with the total number of subcarriers $N_\text{Ac}=N_\text{BH}=64$, and a carrier frequency of 2 GHz, following the Third Generation Partnership Project (3GPP) Long Term Evolution-Advanced (LTE-A) standard \cite{7805409}. Hence, the bandwidth of each subcarrier is $W_\text{S}=312.5$ KHz.
The wireless access and backhaul channels consist of both the large-scale and small-scale fadings.
The large-scale fading consists of path loss and shadowing, and is modeled as $128.1 + 37.6 \log_{10} d_{i,j} + \chi_{i,j}$ in dB, where $d_{i,j} > 0$ in (Km) and $\chi_{i,j}$ in (dB) are the distance and shadowing between transmitter $i$ and receiver $j$, respectively. Moreover, $\chi_{i,j}$ is modeled as a log-normal random variable with the standard deviation of 8 dB \cite{7530876}. The small-scale fading is modeled as independent and identically distributed (i.i.d.) Rayleigh fading with variance $1$ \cite{7530876,7805409}. The power spectral density (PSD) of AWGN noise is set to $-174$ dBm/Hz. Without loss of generality, we assume that the IS acts as an AWGN noise with the PSD of $-120$ dBm/Hz.

Assume that there are $C=1000$ contents in the network \cite{7530876,6600983}. The popularity of contents is also modeled as Zipf distribution with the Zipf parameter $\zeta_1=0.56$ \cite{7530876,6600983}.
Similar to \cite{7322204}, the size of contents is modeled as Log-normal distribution in terms of MByte where $\mu_s=0.5$ and $\sigma^2_s=1.5$ \cite{7322204}.
We assume that our cache sizes are percentages of the total content sizes \cite{7558153}. The relative cache size percentage of each FBS and MBS are set to $3\%$ and $10\%$, respectively \cite{7530876,6600983,6883210}. Following \cite{7530876,6600983}, we assume the transmission power of DC, MBS and each FBS are set to $50$ Watts (W), $40$ W, and $2$ W, respectively.
The minimum required data rate of each MU is $R^\text{min}_{u}=3~\text{Mbps},~\forall u \in \mathcal{U}$.
The time length of each time period, i.e., delivery deadline, is $T=300$ seconds (Sec) \cite{7322204}.
%The DP is also divided into $1000$ time periods with equal time length $T=300$ seconds (Sec) \cite{7322204}.
Without loss of generality, we assume a symmetric fairness in the PF scheme \cite{5580131}.
The system parameters are summarized in Table \ref{Table parameters}.
\begin{table*}
\centering
\caption{System parameters}
\begin{center} \label{Table parameters}
\scalebox{0.80}{\begin{tabular}{|c c c|c c c|}
    \hline \rowcolor[gray]{0.87}
    \hline \rowcolor[gray]{0.87}
    \hline \rowcolor[gray]{0.87}
    \hline \rowcolor[gray]{0.87}
    \textbf{Description} & \textbf{Notation} & \textbf{Value} & \textbf{Description} & \textbf{Notation} & \textbf{Value} \\
    \hline \rowcolor[gray]{0.950}
    \hline \rowcolor[gray]{0.950}
    \hline \rowcolor[gray]{0.950}
    \hline \rowcolor[gray]{0.950}
    Radius of MBS  & - & $350$ m & Large-scale fading model & - & $128.1 + 37.6 \log_{10} d_{i,j} + \chi_{i,j}$ (dB), $d_{i,j}$ (Km) \\
    \hline \rowcolor[gray]{0.955}
    Number of FBSs & $B$ & $15$ & Shadowing standard deviation & $\chi_{i,j}$ & $8$ dB \\
    \hline \rowcolor[gray]{0.960}
    Number of MUs & $U$ & $85$ & Small-scale fading model & - & Rayleigh fading with variance $1$ \\
    \hline \rowcolor[gray]{0.965}
    Distance between MBS and DC & - & 2 Km & AWGN noise density & - & $-174$ dBm/Hz \\
    \hline \rowcolor[gray]{0.970}
    Wireless access bandwidth & $W_\text{Ac}$ & 20 MHz  & AWGN IS level & - & $-120$ dBm/Hz  \\
    \hline \rowcolor[gray]{0.975}
    Wireless backhaul bandwidth & $W_\text{BH}$ & 20 MHz  & Number of contents & $C$ & 1000 \\
    \hline \rowcolor[gray]{0.980}
    Number of access subcarriers & $N_\text{Ac}$ & 64 & Size of contents & $\left(\mu_s,\sigma^2_s\right)$ & $(0.5,1.5)$  \\
    \hline \rowcolor[gray]{0.985}
    Number of backhaul subcarriers & $N_\text{BH}$ & 64 & Zipf parameter & $\zeta_1$ & $0.56$  \\
    \hline \rowcolor[gray]{0.990}
    Carrier frequency & - & 2 GHz & Relative cache size of BS & $M^\text{max}_b$ & MBS: 10\%, FBS: 3\%  \\
    \hline \rowcolor[gray]{0.995}
    Subcarrier bandwidth & $W_\text{S}$ & $312.5$ KHz & Maximum transmit powers & $P^\text{max}_\text{DC}$,$P^\text{max}_b$ & DC: $50$ W, MBS: $40$ W, FBS: $2$ W  \\
    \hline \rowcolor[gray]{0.999}
    Maximum delivery deadline & $T$ & $300$ Sec & Minimum rate of MU & $R^\text{min}_{u}$ & $3$ Mbps  \\
    \hline
\end{tabular}}
\end{center}
\end{table*}

The simulation structure is described in the following. In all shots (simulation loops), the position of MBS is fixed, whereas all FBSs and MUs are randomly distributed. Each result is averaged over various networks topologies and settings, i.e., CDI, location of FBSs and users, and size of contents in the CP (which is named as CP parameters). Then, we fix the CP parameters and change the IGRs and CSIs. After $50$ (number of time periods) shots, we again change the CP parameters. The number of the main simulation loops, i.e., changes of the CP parameters, is equal to $20$. Generally, each scenario is averaged over $1000$ samples to decrease the impact of randomness in our simulation study. In the DP, we apply the PF or the MMF delivery policy corresponding to the considered fairness scheme.

We compare the performance of our proposed CPSs with the following heuristic ones:
\begin{itemize}
  \item \textbf{Cache most popular (CMP)}: In this proactive strategy, each BS caches the most popular contents until its storage is full \cite{7558153,7805409}.
  \item \textbf{Uniform Random caching (URC)}: In this proactive strategy, each BS caches contents uniformly until its storage is full. This strategy does not depend on the PDI.
  \item \textbf{Popular Random caching (PRC)}: In this proactive strategy, each BS caches contents randomly until its storage is full. The caching probability in each BS is linear to the square root of requesting probability \cite{7562510}.
  \item \textbf{No caching (NC)}: In this case, the storage capacities are equal to zero. Hence, all the requested contents need to be prefetched at the DC over the backhaul links \cite{7558153,7322204}.
\end{itemize}

\subsection{Impact of the storage capacities}
\begin{figure}
\centering
\subfigure[Total backhaul latency vs. relative cache capacity of FBSs]{
   \includegraphics[scale=0.40]{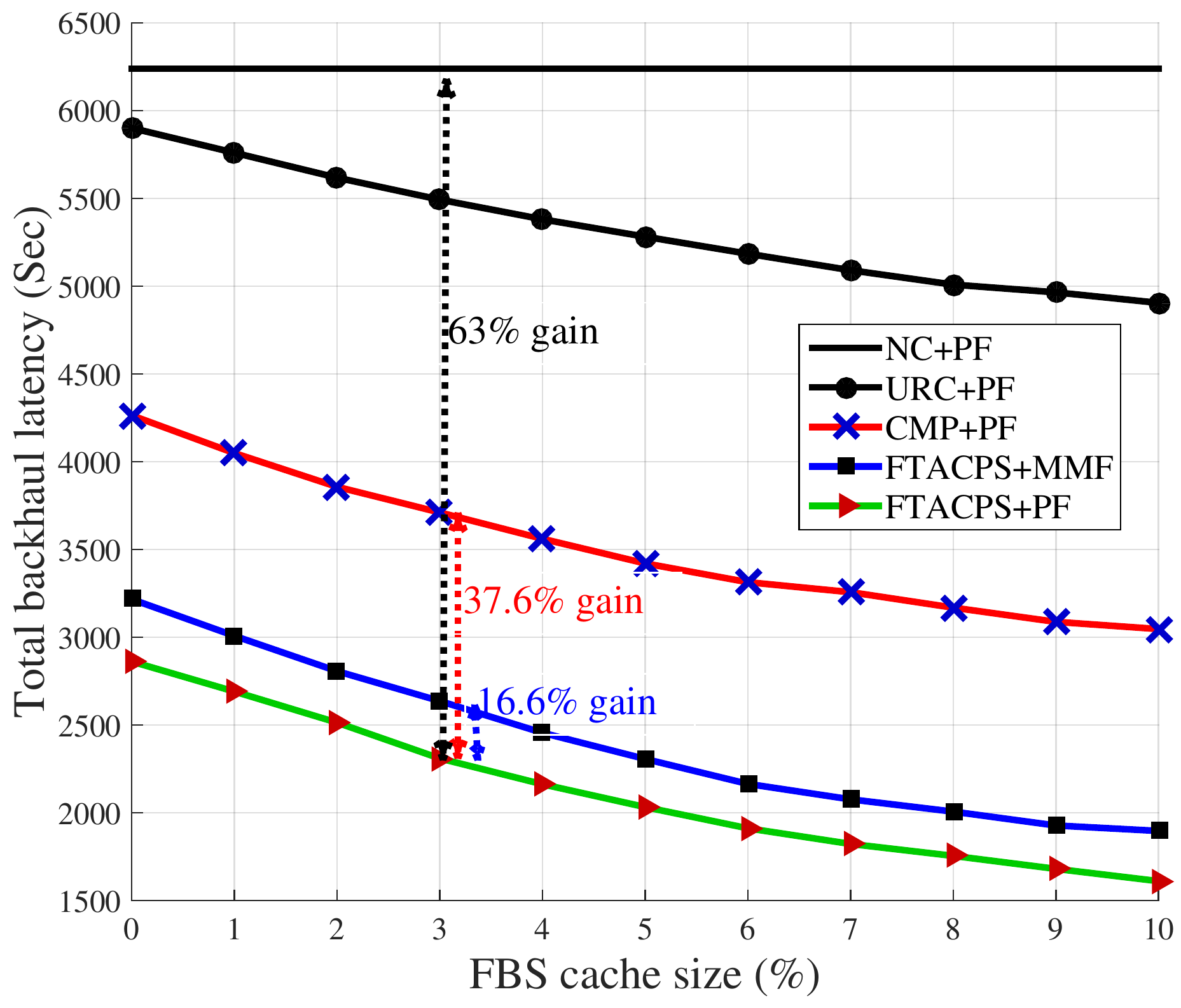}
   \label{Fig04CacheCapacityBHLatency}
}
\subfigure[Total latency of MUs vs. relative cache capacity of FBSs]{
   \includegraphics[scale=0.40]{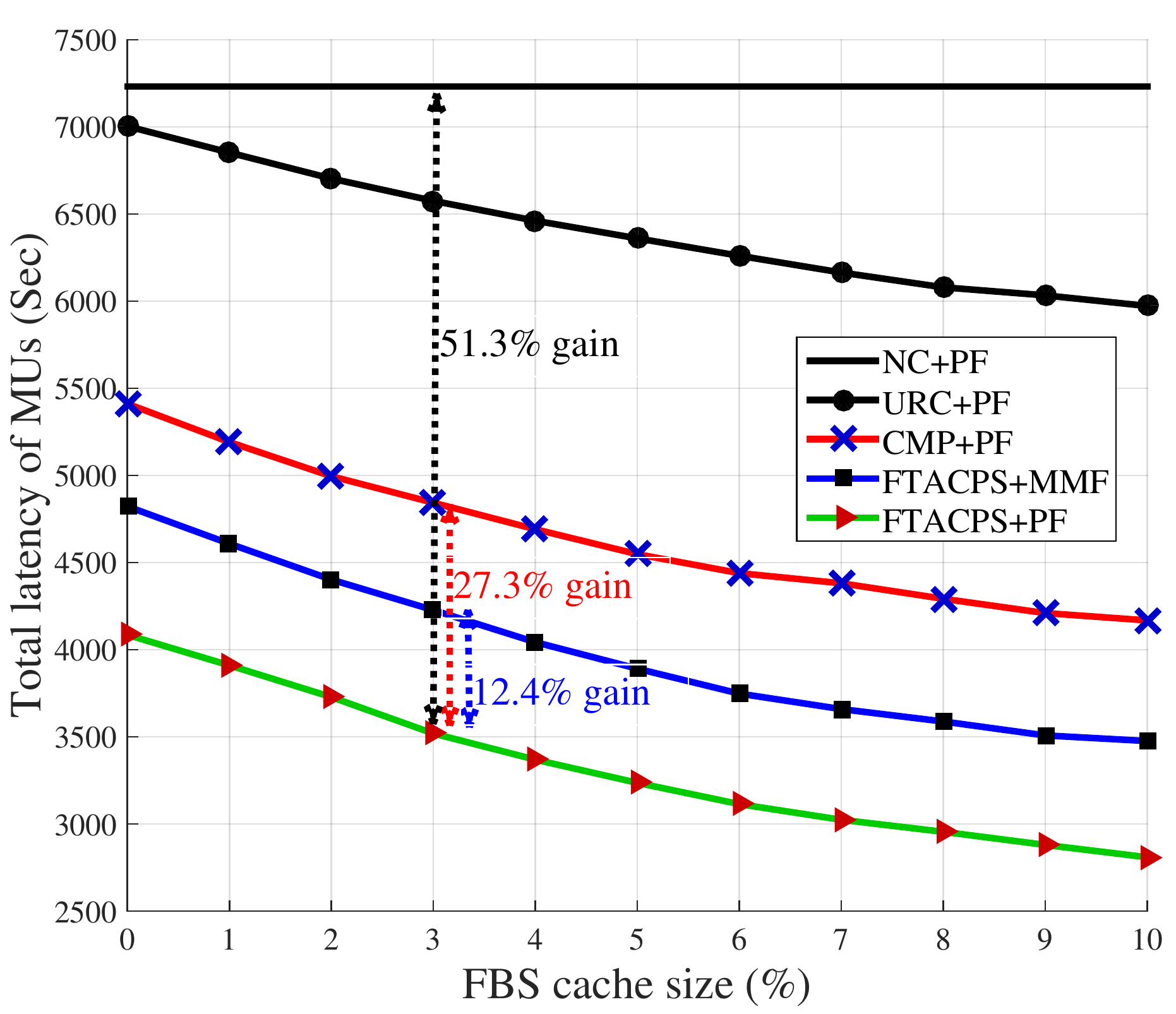}
   \label{Fig04CacheCapacityTotalLatency}
}
\caption{Impact of cache capacity of FBSs on the total latency of MUs in different fairness schemes.}
\label{Fig04}
\end{figure}

Fig. \ref{Fig04} illustrates the impact of the storage capacity of FBSs on the total latency of MUs for different CPSs in the PF and MMF fairness schemes.
As shown, when the relative cache capacities increase, the performance of all CPSs is improved in the system. This is because, more storage capacities lead to cache more contents in each FBS. In this regard, more requests can be served by RAN. Accordingly, the data traffic of backhaul links are significantly decreased in the network. Moreover, the per-content backhaul data rates increase in the system which cause a significant reduction on the total backhaul latency of MUs as shown in Fig. \ref{Fig04CacheCapacityBHLatency}. Subsequently, the total latency of MUs decreases in the network which is shown in Fig. \ref{Fig04CacheCapacityTotalLatency}.
From Fig. \ref{Fig04CacheCapacityBHLatency}, it can be seen that the caching technology with our proposed FTACPS strategy in the PF scheme has nearly $63\%$ performance gain in terms of total backhaul latency compared to the NC scheme which causes $51.3\%$ improvement on the total latency of MUs (see Fig. \ref{Fig04CacheCapacityTotalLatency}). Besides, as shown in Fig. \ref{Fig04CacheCapacityBHLatency}, our flexible FTACPS in the PF scheme improves the total latency of MUs close to almost $27.3\%$ compared to the CMP strategy. This result shows that it is more beneficial to exploit the flexible transmission opportunities in CPS in order to improve the delivery performance by integrally and jointly allocating physical resources as storage and radio. In other words, it is better that CPS be devised based on the network conditions with flexible access selection opportunities for all MUs compared to the heuristic strategies in which popular contents are cached everywhere or all contents are randomly cached. However, there is a significant performance gap between CMP and URC when $\zeta_1=0.56$.

\begin{figure*}
\centering
\includegraphics[scale=0.49]{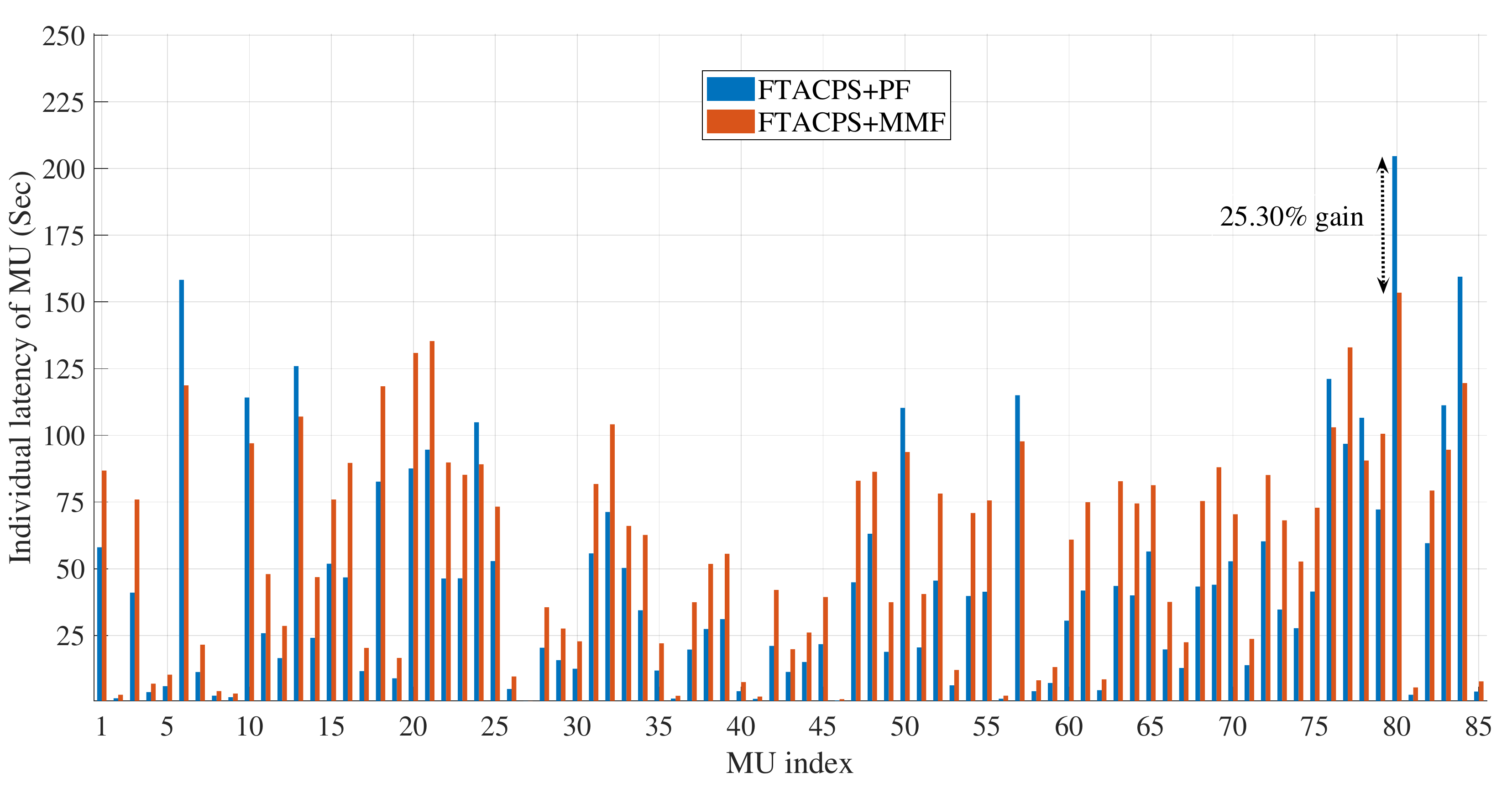}
\caption{Individual delay of MUs for different fairness schemes. The CP parameters are set based on Table \ref{Table parameters}.}
\label{Fig07Fairness}
\end{figure*}

Fig. \ref{Fig04CacheCapacityTotalLatency} shows near to $12.4\%$ reduction on the total latency of MUs in the PF scheme compared to MMF. This is because, MMF aims to only decrease the maximal latency between MUs which attracts all BSs to store popular contents with larger sizes in CP with a distributed manner as much as possible and allocate more radio resources for MUs with worse channel conditions in DP. We also present individual delay of MUs in the PF and MMF fairness schemes based on our proposed FTACPSs in Fig. \ref{Fig07Fairness}. In this figure, the network topology and user placement are set based on Fig. \ref{Fig03Locations} and Table \ref{Table parameters}. We fix all the CP parameters and generate various sets of IGRs and CSIs for different time periods. Actually, Fig. \ref{Fig07Fairness} is averaged over $50$ sets of IGRs and CSIs for fixed CP parameters. As seen in Fig. \ref{Fig07Fairness}, MMF has closed to almost $25.30\%$ performance gain in terms of maximal latency of MUs compared to PF. However, this scheme is weak to efficiently handle all MUs latencies in the system compared to PF.

\subsection{Effect of the Zipf Parameter}
\begin{figure*}
\centering
\subfigure[Total number of unique requests vs. Zipf parameter for different sets of contents.]{
   \includegraphics[scale=0.40]{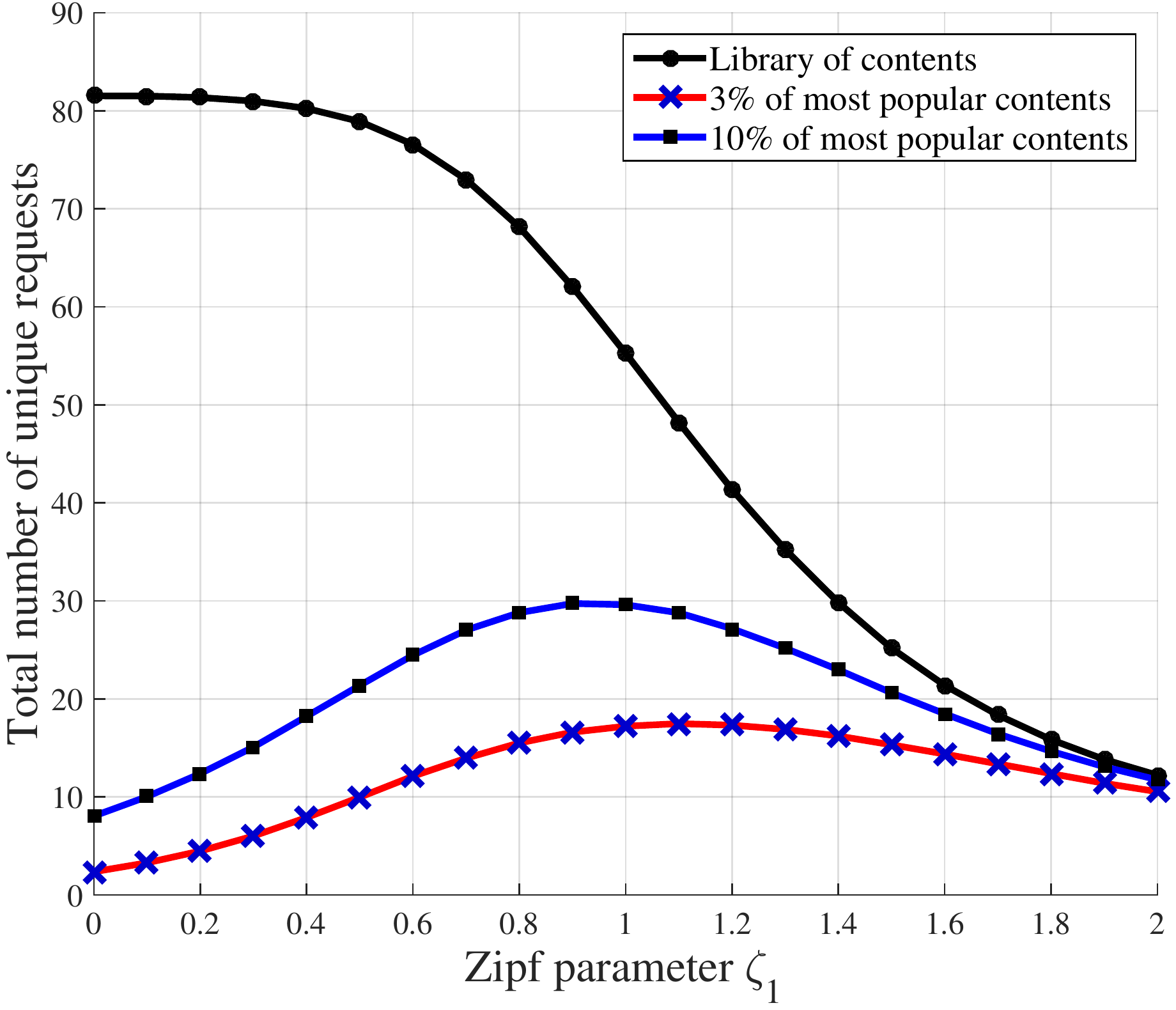}
   \label{Fig05TotalUniqueRequests}
}
%\subfigure[Request percentage of most popular contents vs. Zipf parameter for different sets of contents.]{
%   \includegraphics[scale=0.21]{Fig05RequestPercentage}
%   \label{Fig05RequestPercentage}
%}
\subfigure[Total backhaul latency vs. Zipf parameter.]{
   \includegraphics[scale=0.40]{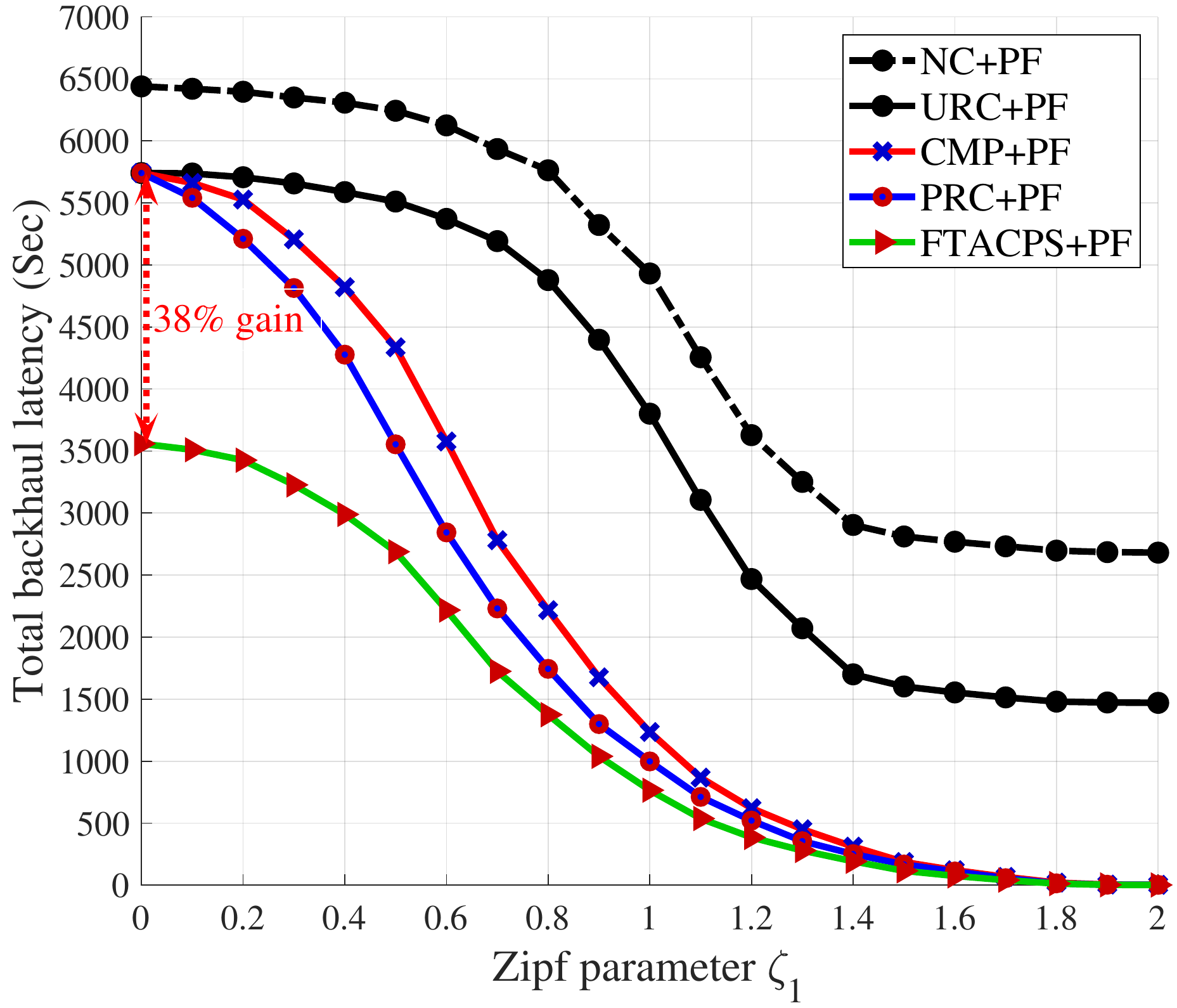}
   \label{Fig05ZipfBHLatency}
}
\subfigure[Total latency of MUs vs. Zipf parameter.]{
   \includegraphics[scale=0.40]{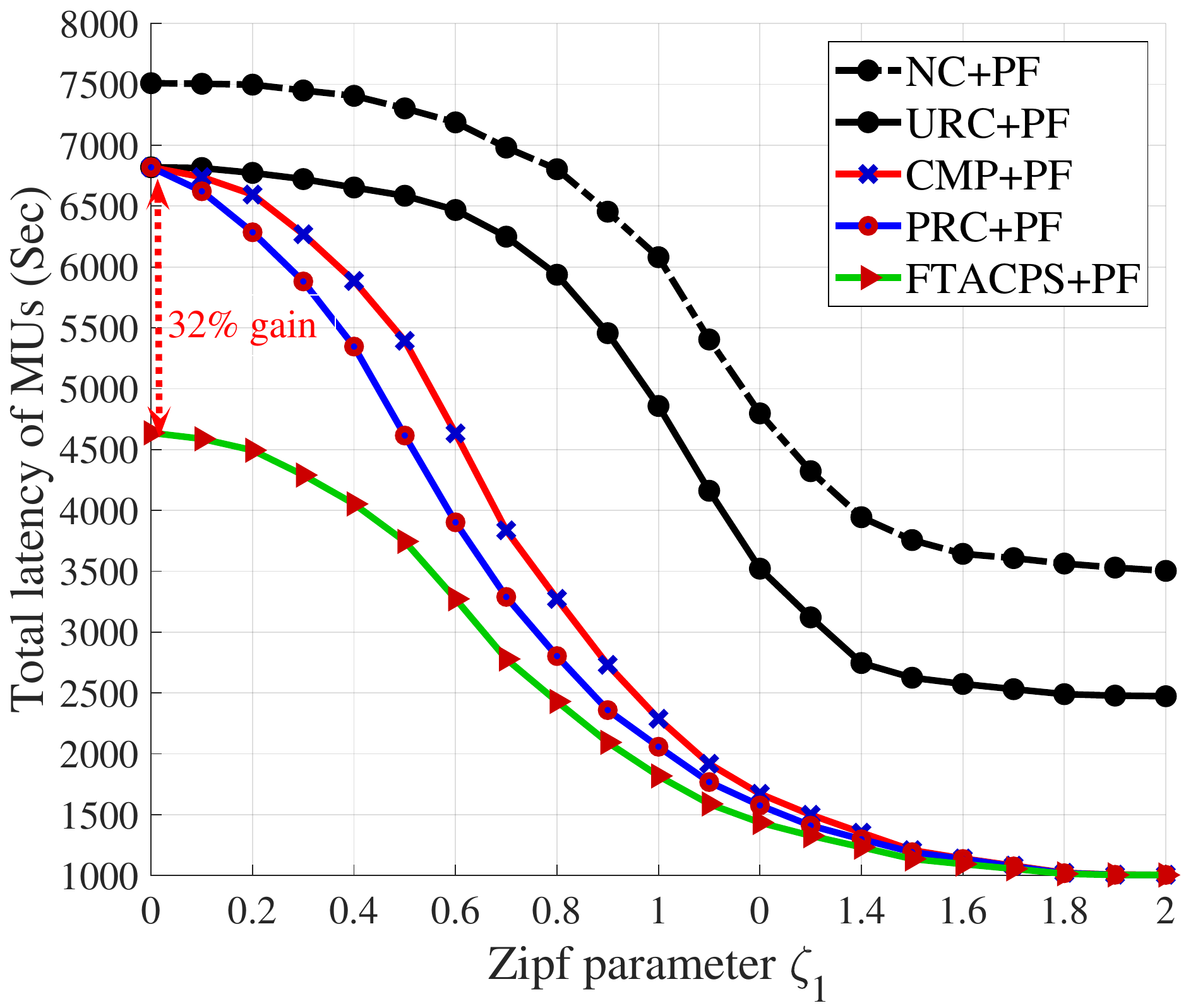}
   \label{Fig05ZipfTotalLatency}
}

\caption
{Effect of the Zipf parameter on performance of different CPSs in the PF scheme.}
\label{Fig05}
\end{figure*}

Here, we investigate the effect of the Zipf parameter on the performance of CPSs in the PF scheme shown in Fig. \ref{Fig05}.
Zipf parameter $\zeta_1$ tunes the skewness of PDI which directly affects the diversity of IGRs of MUs. When $\zeta_1$ tends to zero, the diversity of IGRs increases in the system. For instance, when $\zeta_1=0.4$, $U=85$ and $C=1000$, the average total number of unique requests are nearly $80.2$ where closed to $18.2$ of them are for $10\%$ of most popular contents which approximately can be stored in MBS (see Fig. \ref{Fig05TotalUniqueRequests}). In addition, nearly $7.9$ requests of the total unique requests are for $3\%$ of most popular contents which approximately can be stored in each FBS. In other words, only $22.71\%$ and $9.9\%$ of unique requests are for $10\%$ and $3\%$ of most popular contents, respectively. These results are averaged over $10000$ sets of IGRs for each value of $\zeta_1$. According to the above, when $\zeta_1$ tends to zero, the data traffic of backhaul links increases through the network. Subsequently, the per-content backhaul rates decreases which gradually increases the total backhaul latency in the network shown in Fig. \ref{Fig05ZipfBHLatency}. Hence, the total latency of MUs gradually increase (please see Fig. \ref{Fig05ZipfTotalLatency}). Besides, when $\zeta_1$ is large enough, the diversity of IGRs of MUs decreases which implies a decreases of the total number of unique requests. Moreover, the request percentage of most popular contents increases in the system which significantly improves the total backhaul and overall MUs latencies.

The CMP strategy is more affected by $\zeta_1$ than the other approaches, since this strategy is a baseline popular algorithm.
For small values of $\zeta_1$, the performance gaps between CMP, URC and PRC tend to zero which means these heuristic strategies are not beneficial for high-level request diversity situations. For instances, when $\zeta_1=0$ there is nearly $38\%$ and $32\%$ performance gaps between our proposed FTACPS strategy and CMP in terms of total backhaul and overall MUs latencies which are shown in Figs. \ref{Fig05ZipfBHLatency} and \ref{Fig05ZipfTotalLatency}, respectively. This performance gap also decreases to $27.3\%$ in terms of total latency of MUs when $\zeta_1$ achieves to $0.56$ (please see Fig. \ref{Fig04CacheCapacityTotalLatency}).
Actually, our proposed FTACPS is a good solution when $\zeta_1$ is not large. In this situation, a transmission-aware distributed CPS is integrally designed in the system to avoid duplicated prefetching contents in different co-located BSs to decrease the backhaul traffics as much as possible. It is noteworthy that the PRC strategy outperforms the systems performance compared to CMP, since this strategy uses both the distributed and most popular caching strategies in its algorithm structure.
On the other hand, when $\zeta_1$ is large enough, i.e., only a few popular contents are frequently requested by MUs, the performance gaps between CMP, PRC and our proposed FTACPS strategy tend to zero. As shown in Fig. \ref{Fig05ZipfTotalLatency}, CMP and PRC with their very low-complexity structures are good choices when PDI is too skewed.

\subsection{Impact of the Number of MUs}
\begin{figure}
\centering
\subfigure[Total access latency of MUs vs. number of MUs per FBS]{
   \includegraphics[scale=0.35]{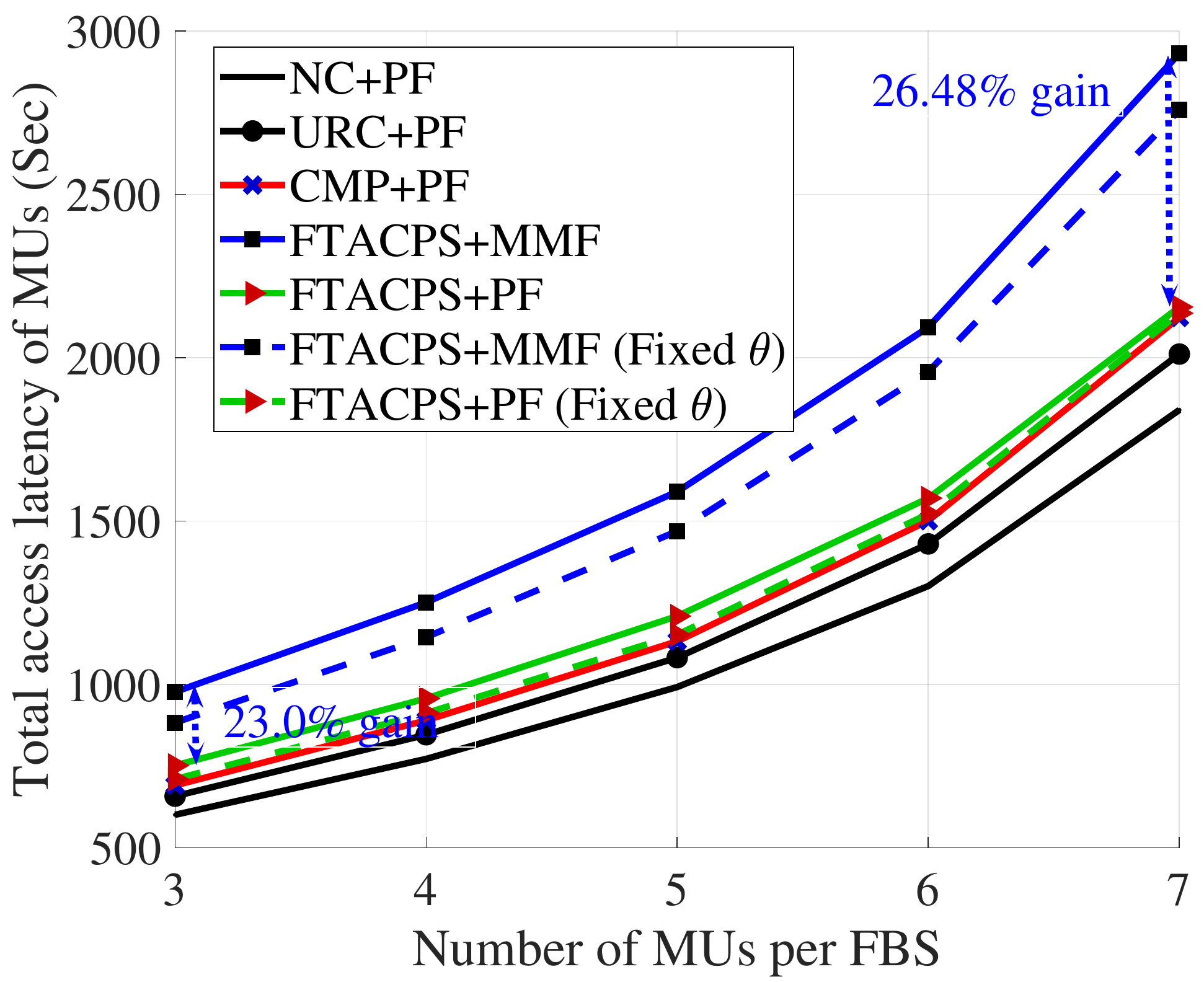}
   \label{Fig06NumberAccessLatency}
}
\subfigure[Total backhaul latency vs. number of MUs per FBS]{
   \includegraphics[scale=0.35]{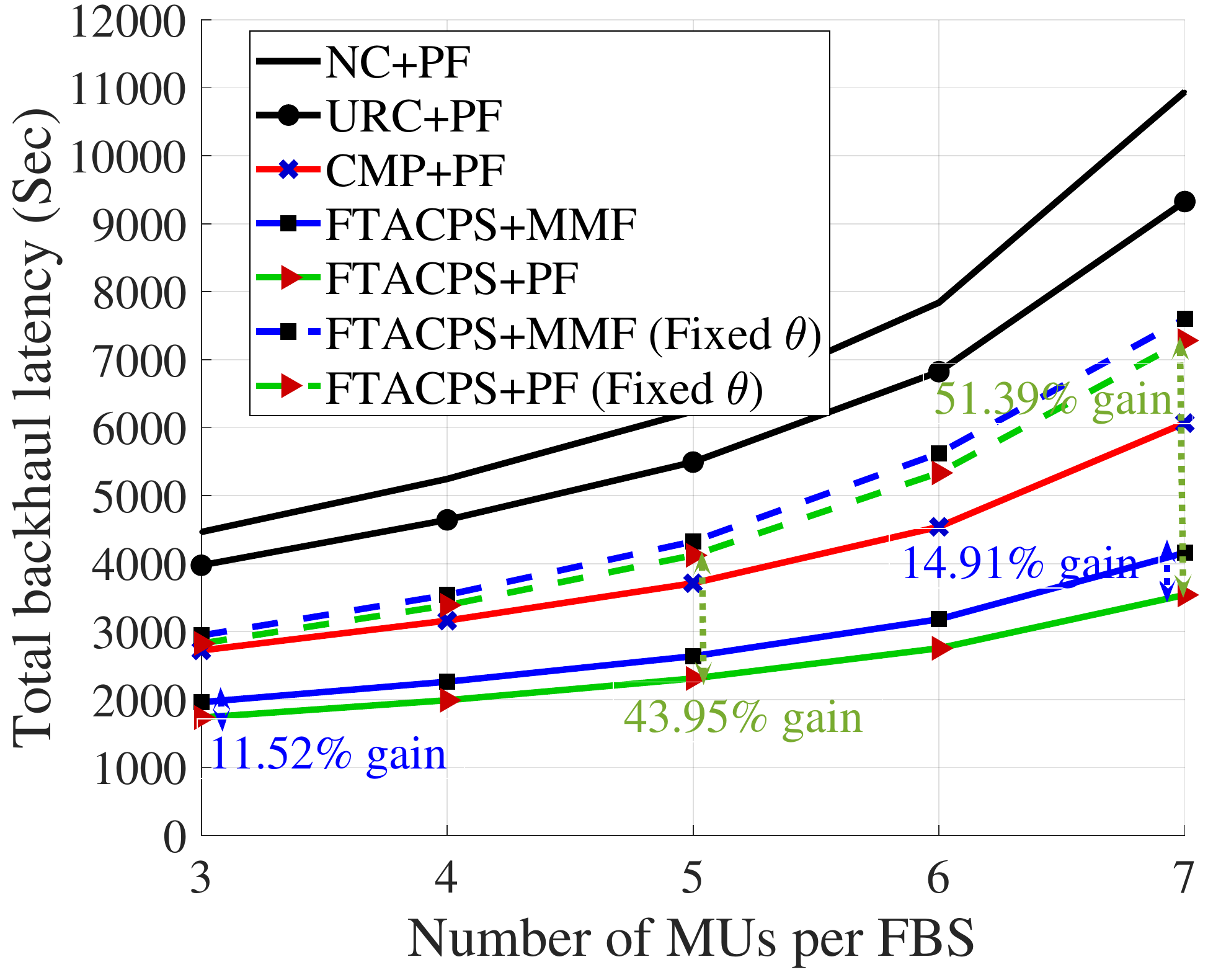}
   \label{Fig06NumberBHLatency}
}
\subfigure[Total latency of MUs vs. number of MUs per FBS]{
   \includegraphics[scale=0.35]{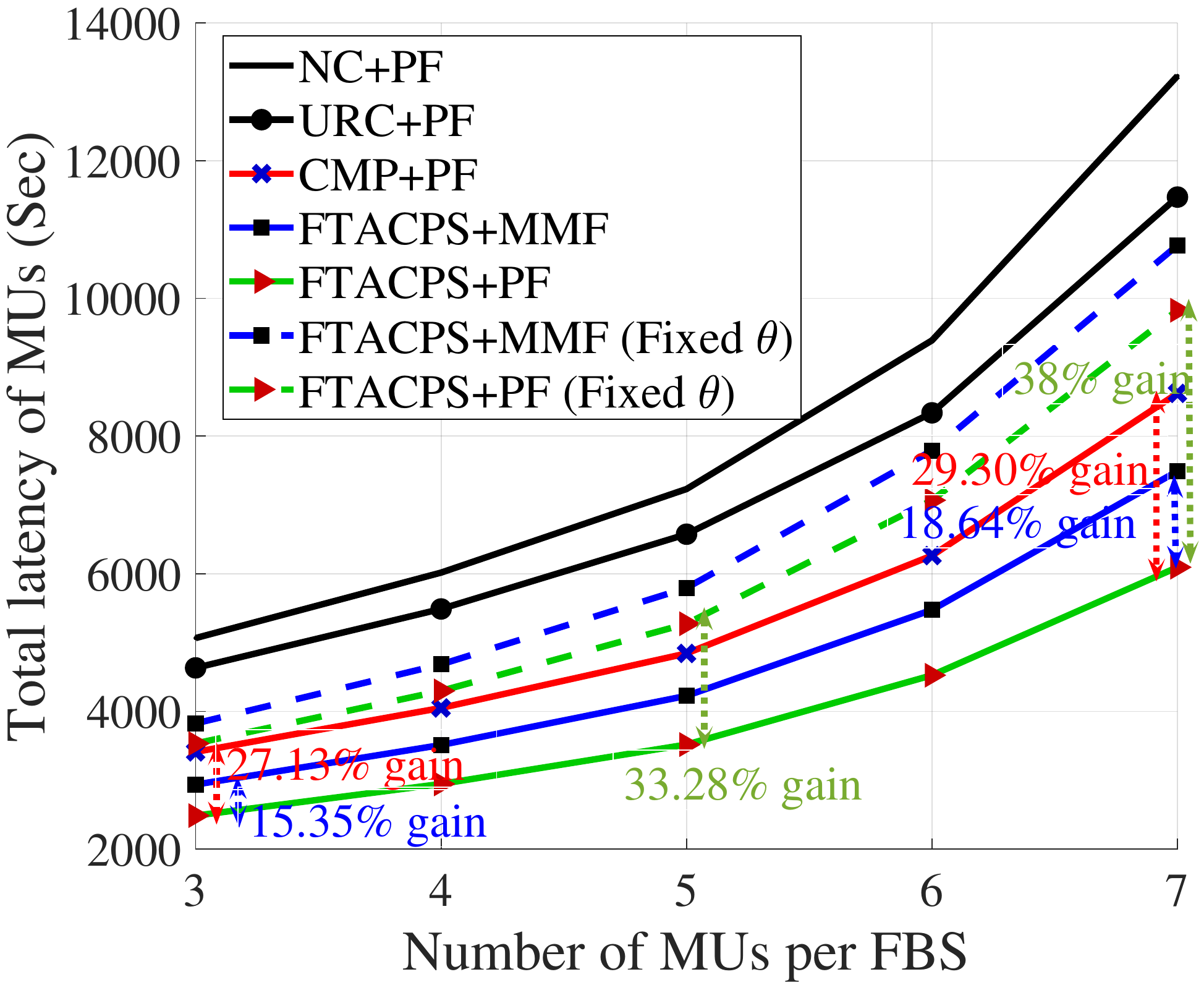}
   \label{Fig06NumberTotalLatency}
}
\subfigure[Computational time vs. number of MUs per FBS]{
   \includegraphics[scale=0.35]{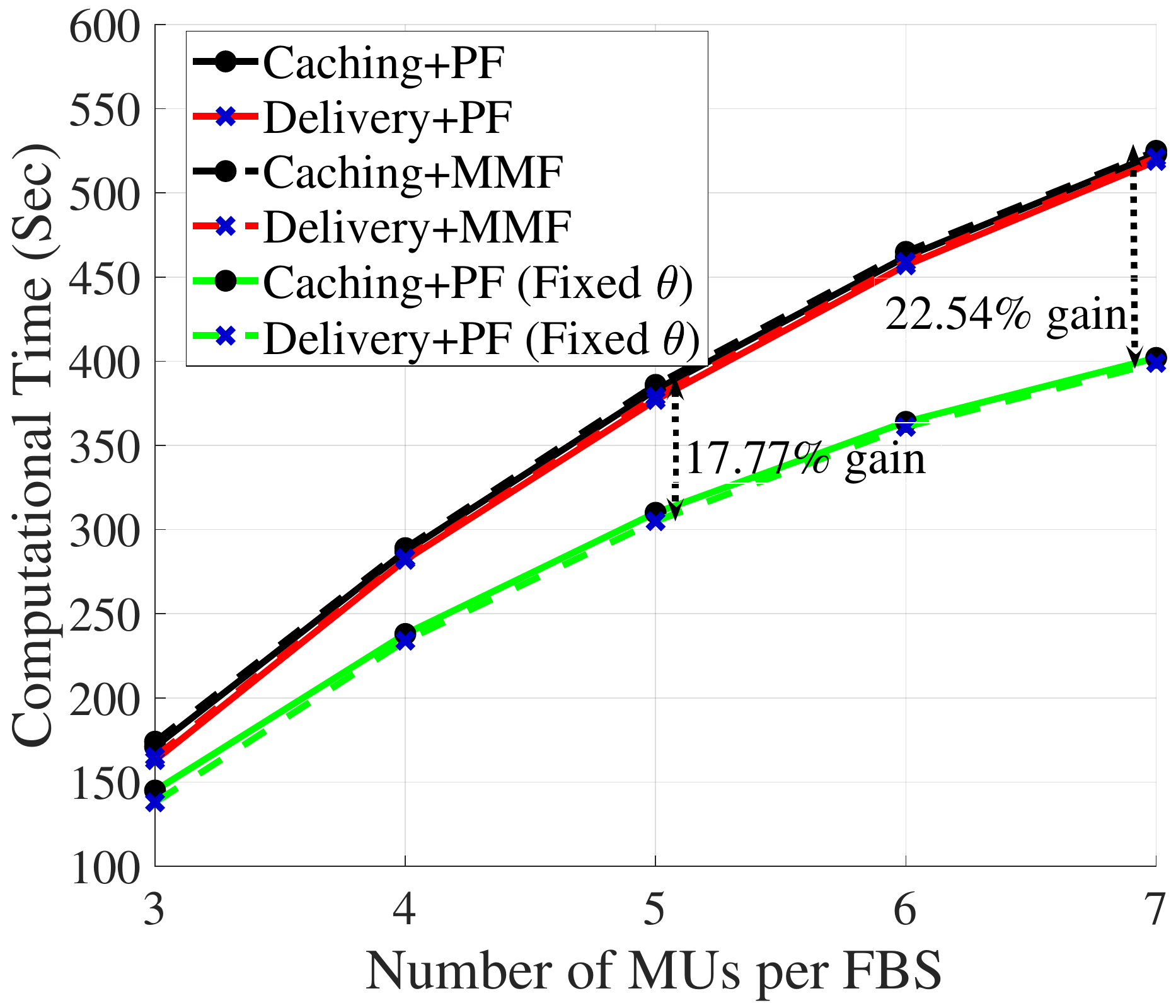}
   \label{Fig06RuntimeUser}
}
\caption{Impact of the number of femto-cell MUs on the total latency of MUs and computational time in different fairness schemes.}
\label{Fig06}
\end{figure}

Fig. \ref{Fig06} compares the performance of CPSs in different fairness schemes when the number of MUs per FBS gradually increases through the network.
As shown, with increasing number of MUs, the total access latencies exponentially increase in the system. This is because, the available limited radio access resources are distributed over more MUs in each BS in order to satisfy the QoS constraints which degrades the per-MU data rates in DP. Moreover, since each MU requests one content in the network, increasing $U$ gradually increases the data traffic in RAN. These per-MU data rate reductions and RAN traffic increments both together cause an intensification in the total access latency of MUs shown in Fig. \ref{Fig06NumberAccessLatency}. Besides, with increasing $U$ and subsequently increasing the number of received unique requests at BSs, the total backhaul data traffic increases in the network. Furthermore, the limited per-content backhaul rates are decreased in the system which cause an enormous increment on the total backhaul latency which is presented in Fig. \ref{Fig06NumberBHLatency}. From Figs. \ref{Fig06NumberAccessLatency} and \ref{Fig06NumberBHLatency}, it can be observed that when the number of MUs grows in the network, the total latency of MUs increases as shown in Fig. \ref{Fig06NumberTotalLatency}.

Interestingly, as shown in Fig. \ref{Fig06NumberAccessLatency}, the NC scheme has lower access latencies than the other approaches. This is because, in the NC scheme, all requested contents should be prefetched from DC in DP. Hence, MUs are generally associated to nearby BSs to increase only the access data rates. In other words, there is no information-centric access selection enforcement for MUs to alleviate the huge backhaul traffics, since all contents should be sent to BSs through backhaul links. In this line, the total backhaul latency of MUs in the NC scheme are significantly higher than other schemes (see Fig. \ref{Fig06NumberBHLatency}) which shows the effect of the caching technology in the system. Finally, the total latency of MUs in the NC scheme is more than that other approaches. From Fig. \ref{Fig06NumberAccessLatency}, it can be observed that there exist considerable performance gaps between the MMF and PF fairness schemes in terms of total access and backhaul latencies of MUs. For instances, when there exist at least $3$ MUs in the coverage of each FBS, our FTACPS in the PF scheme has nearly $23\%$ gain in terms of total access latency of MUs compared to the MMF scheme and this gap achieves up to $26.4\%$ when the mentioned number of MUs approaches $7$. The increment of performance gaps between different fairness schemes can be arrived by the weakness of the MMF scheme in order to handle the multi-level of more MUs latencies. The multi-level of MUs latencies comes from both the MUs channel conditions, specifically MUs distances to BSs in RAN, and non-homogeneity of contents size. From Fig. \ref{Fig06NumberTotalLatency}, it can be observed that there exists nearly $15.35\%$ performance gap between PF and MMF when each femto-cell has at least $3$ MUs in its coverage and this gap achieves to $18.64\%$ when the mentioned number of MUs achieves to $7$.
As seen in Fig. \ref{Fig06NumberAccessLatency}, the total access latency of MUs via our FTACPS in PF is bigger than CMP. This result comes from the fact that our distributed FTACPS provides more access selection opportunities in RAN than CMP to alleviate the backhaul traffic. In this line, MUs in nearby co-located BSs can be assigned to the BS that has the requested contents in order to offload the backhaul traffic instead of choosing the BS which provides the maximal data rates. From Fig. \ref{Fig06NumberTotalLatency}, it can be observed that the performance gap between our proposed FTACPS and CMP is $27.13\%$ when the number of each femto-cell MUs is $3$ and this gap achieves to $29.30\%$ when the number of each femto-cell MUs is $7$.

To evaluate the benefits of applying an efficient MU association policy in the system, we also consider other PF and MMF fairness schemes in Fig. \ref{Fig06} where our proposed FTACPSs and delivery policy are applied for a fixed $\boldsymbol{\theta}$. The MU association policy for both the CP and all time periods of the DP is predefined based on the following heuristic approach. Each MU selects the nearest FBS within $70$ m. If there was no FBS within $70$ m, the MU will be associated to MBS. From Fig. \ref{Fig06NumberBHLatency}, it can be seen that jointly allocating physical resources with an efficient MU association parameter in the PF scheme outperforms the total backhaul latency nearly $43.95\%$ when the number of MUs per FBS is $3$ and this improvement achieves to $51.39\%$ when the number of MUs per FBS is equal to $7$. In this way, the total latency of MUs is decreased closed to $33.28\%$ and $38\%$ when the number of MUs per FBS is $3$ and $7$, respectively (see Fig. \ref{Fig06NumberTotalLatency}).

Each optimization algorithm should strike a balance between the optimality and complexity, specifically in the dense small-cell networks. In this regard, we investigate the computational time of our proposed FTACPSs and delivery algorithm by using our numerical simulation environment in Fig. \ref{Fig06RuntimeUser}. The computational times provided in Fig. \ref{Fig06RuntimeUser} are averaged over $100$ samples.
As shown, all the proposed approaches have nearly the same computational time, since the complexity estimations presented in Table \ref{Table Complexity} are on the same level. It is noteworthy that the future 5G networks with a powerful dedicated SD controller can significantly accelerate the optimization process. Besides, fixing the user association parameter $\boldsymbol{\theta}$ decreases the computational time of our proposed approaches only $22.54\%$.

\section{Conclusion}\label{section Conclusion}
In this paper, we investigated the design of FTACPSs in the downlink of OFDMA-based HetNets. We proposed an optimization problem for each phase in the PF and MMF schemes. Specifically, in the PF scheme, we minimized the total weighted latency of MUs, and in the MMF scheme, we minimized the maximal latency of MUs in the system. In order to solve each optimization problem, we devised an AO algorithm and proved that the proposed algorithm converges to a local optimum solution. In simulation results, we showed that it is beneficial to allocate the physical resources as storage and radio jointly based on both the channel conditions and contents popularity. Moreover, we investigated the effect of different fairness schemes on the total latency of MUs.

\appendices
\section{PROOF OF PROPOSITION 1}\label{appendix proposition1}
After solving \eqref{problem rho total latency} and finding $(\boldsymbol{\rho}_{t_1},\boldsymbol{\beta}_{t_1},\boldsymbol{\theta}_{t_1})$ at iteration $t_1$ for a fixed $(\boldsymbol{p}_{t_1-1},\boldsymbol{\gamma}_{t_1-1})$ from the previous iteration $(t_1-1)$, we have
\begin{align}\label{ineq 001}
0 \leq D^{\text{Pr},\text{tot}}(\boldsymbol{\rho}_{t_1},\boldsymbol{\beta}_{t_1},\boldsymbol{\theta}_{t_1},\boldsymbol{p}_{t_1-1},\boldsymbol{\gamma}_{t_1-1})
\leq
D^{\text{Pr},\text{tot}}(\boldsymbol{\rho}_{t_1-1},\boldsymbol{\beta}_{t_1-1},\boldsymbol{\theta}_{t_1-1},\boldsymbol{p}_{t_1-1},\boldsymbol{\gamma}_{t_1-1}),
\end{align}
which can be derived due to the fact that solving the equivalent problem \eqref{problem rho total latency equivalent form 2} by using the standard optimization software CVX with MOSEK improves the objective function \eqref{obf problem rho total latency equivalent form 2} or the objective remains constant. Therefore, we have
\begin{align}\label{obj y improvement}
\sum\limits_{u \in \mathcal{U}} \omega_{u} \sum_{c=1}^{C} \Delta_c \sum\limits_{b \in \mathcal{B}}
\left( \theta^{(t_1)}_{b,u} \frac{s_c}{ R_{b,u}^{\text{Ac}} } +  y^{\text{BH},c,(t_1)}_{b,u} \right)
\leq
\sum\limits_{u \in \mathcal{U}} \omega_{u} \sum_{c=1}^{C} \Delta_c \sum\limits_{b \in \mathcal{B}}
\left( \theta^{(t_1-1)}_{b,u} \frac{s_c}{ R_{b,u}^{\text{Ac}} } +  y^{\text{BH},c,(t_1-1)}_{b,u} \right).
\end{align}
$y^{\text{BH},c}_{b,u}$ is the upper-bound value of $\theta_{b,u} (1-\rho_{b,c}) \frac{s_c} {  \beta_{b,c} R_{b}^{\text{BH}}  }$. Moreover, it can be easily shown that $y^{\text{BH},c}_{b,u}$ is lower bounded by zero. On the other hand, after solving \eqref{problem rho total latency equivalent form 2}, the gap between $ y^{\text{BH},c}_{b,u}$ and $\theta_{b,u} (1-\rho_{b,c}) \frac{s_c} {  \beta_{b,c} R_{b}^{\text{BH}}  }$ tends to zero. Hence, according to \eqref{latency each user probability}, we conclude that the gap between $\sum\limits_{u \in \mathcal{U}} \omega_{u} \sum_{c=1}^{C} \Delta_c \sum\limits_{b \in \mathcal{B}}
\left( \theta^{(t_1)}_{b,u} \frac{s_c}{ R_{b,u}^{\text{Ac}} } +  y^{\text{BH},c,(t_1)}_{b,u} \right)$ and $D^{\text{Pr},\text{tot}}(\boldsymbol{\rho}_{t_1},\boldsymbol{\beta}_{t_1},\boldsymbol{\theta}_{t_1},\boldsymbol{p}_{t_1-1},\boldsymbol{\gamma}_{t_1-1})$ also tends to zero. In doing so, based on \eqref{obj y improvement}, it is proved that the proposed algorithm for solving \eqref{problem rho total latency} improves the objective function \eqref{obf problem rho total latency} or the objective function remains fixed.
Besides, after finding $\boldsymbol{p}$ at iteration $t_1$, we can conclude that
\begin{align}\label{ineq 002}
0 \leq D^{\text{Pr},\text{tot}}(\boldsymbol{\rho}_{t_1},\boldsymbol{\beta}_{t_1},\boldsymbol{\theta}_{t_1},\boldsymbol{p}_{t_1},\boldsymbol{\gamma}_{t_1-1})
\leq
D^{\text{Pr},\text{tot}}(\boldsymbol{\rho}_{t_1},\boldsymbol{\beta}_{t_1},\boldsymbol{\theta}_{t_1},\boldsymbol{p}_{t_1-1},\boldsymbol{\gamma}_{t_1-1}),
\end{align}
which is proved in \emph{Proposition 2}.
Similarly, it can be shown that after finding $\boldsymbol{\gamma}$ at iteration $t_1$, we have
\begin{align}\label{ineq 003}
0 \leq D^{\text{Pr},\text{tot}}(\boldsymbol{\rho}_{t_1},\boldsymbol{\beta}_{t_1},\boldsymbol{\theta}_{t_1},\boldsymbol{p}_{t_1},\boldsymbol{\gamma}_{t_1})
\leq
D^{\text{Pr},\text{tot}}(\boldsymbol{\rho}_{t_1},\boldsymbol{\beta}_{t_1},\boldsymbol{\theta}_{t_1},\boldsymbol{p}_{t_1},\boldsymbol{\gamma}_{t_1-1}).
\end{align}
According to \eqref{ineq 001}, \eqref{ineq 002} and \eqref{ineq 003}, it is obvious that
\begin{multline}
0 \leq D^{\text{Pr},\text{tot}}(\boldsymbol{\rho}_\text{opt},\boldsymbol{\beta}_\text{opt},\boldsymbol{\theta}_\text{opt},\boldsymbol{p}_\text{opt},\boldsymbol{\gamma}_\text{opt})
\leq
\dots
\leq
D^{\text{Pr},\text{tot}}(\boldsymbol{\rho}_{t_1},\boldsymbol{\beta}_{t_1},\boldsymbol{\theta}_{t_1},\boldsymbol{p}_{t_1},\boldsymbol{\gamma}_{t_1})
\leq
\\
D^{\text{Pr},\text{tot}}(\boldsymbol{\rho}_{t_1},\boldsymbol{\beta}_{t_1},\boldsymbol{\theta}_{t_1},\boldsymbol{p}_{t_1},\boldsymbol{\gamma}_{t_1-1})
\leq
D^{\text{Pr},\text{tot}}(\boldsymbol{\rho}_{t_1},\boldsymbol{\beta}_{t_1},\boldsymbol{\theta}_{t_1},\boldsymbol{p}_{t_1-1},\boldsymbol{\gamma}_{t_1-1})
\leq
\\
D^{\text{Pr},\text{tot}}(\boldsymbol{\rho}_{t_1-1},\boldsymbol{\beta}_{t_1-1},\boldsymbol{\theta}_{t_1-1},\boldsymbol{p}_{t_1-1},\boldsymbol{\gamma}_{t_1-1})
\leq
\dots,
\end{multline}
which means after each iteration $t_1$, the objective function \eqref{obf total latency Problem caching} decreases or remains constant. Accordingly, when the number of main iterations increases, Alg. \ref{Alg iterative caching total} converges to a local optimum solution.

\section{EQUIVALENT TRANSFORMATION OF \eqref{problem power total latency}}\label{appendix EquivalentTransformation}
To tackle the fractional forms of \eqref{obf problem rho total latency} and \eqref{Constraint latency user}, we first define new variables $x_{b,u}^{\text{Ac},c}$ and $x^{\text{BH},c}_{b,u}$ such that
\begin{align}\label{xAc introducing constraint}
\theta_{b,u} \frac{s_c}{  R_{b,u}^{\text{Ac}}  }  \leq \frac{1}{x_{b,u}^{\text{Ac},c}},
\end{align}
for $x_{b,u}^{\text{Ac},c} >0$ which means that the access average latency of each MU $u$ for receiving content $c$ from BS $b$ should not exceed $\frac{1}{x_{b,u}^{\text{Ac},c}}$,
and
\begin{align}\label{xBH introducing constraint}
\theta_{b,u} (1-\rho_{b,c}) \frac{s_c} { \beta_{b,c} R_{b}^{\text{BH}}  } \leq \frac{1}{x^{\text{BH},c}_{b,u}},
\end{align}
for $x^{\text{BH},c}_{b} >0$ which means that the average latency of BS $b$ for receiving the un-cached content $c$ should not exceed $\frac{1}{x^{\text{BH},c}_{b,u}}$ for all MU $u$ associated to BS $b$. Based on \eqref{latency each user probability}, constraint \eqref{Constraint latency user} is transformed into the following constraints
\begin{align}\label{Constraint deadline with x Ac BH}
\sum\limits_{b \in \mathcal{B}} \sum\limits_{c \in \mathcal{C}} \Delta_c \left( \frac{1}{ x_{b,u}^{\text{Ac},c} } + \frac{1}{ x^{\text{BH},c}_{b,u} } \right) \leq T, \forall u \in \mathcal{U},
\end{align}
\eqref{xAc introducing constraint} and \eqref{xBH introducing constraint}. In addition, the objective function \eqref{obf problem power total latency} is transformed into    \\
$\sum\limits_{u \in \mathcal{U}} \omega_{u} \sum_{c=1}^{C} \Delta_c \sum\limits_{b \in \mathcal{B}} \left( \frac{1}{ x_{b,u}^{\text{Ac},c} } + \frac{1}{ x^{\text{BH},c}_{b,u} } \right)$.
In order to deal with the fractional terms $\frac{1}{ x_{b,u}^{\text{Ac},c} }$ and $\frac{1}{ x^{\text{BH},c}_{b,u} }$ in \eqref{Constraint deadline with x Ac BH} and the equivalent objective function, we first define new variables $\hat{x}_{b,u}^{\text{Ac},c}$ and $\hat{x}^{\text{BH},c}_{b,u}$ where
$\frac{1}{ x_{b,u}^{\text{Ac},c} } \leq \hat{x}_{b,u}^{\text{Ac},c}$ and $\frac{1}{ x^{\text{BH},c}_{b,u} } \leq \hat{x}^{\text{BH},c}_{b,u}$. On the one hand, \eqref{Constraint deadline with x Ac BH} and the equivalent objective function are transformed into following forms, respectively, as
$\sum\limits_{b \in \mathcal{B}} \sum\limits_{c \in \mathcal{C}} \Delta_c \left( \hat{x}_{b,u}^{\text{Ac},c} + \hat{x}^{\text{BH},c}_{b,u} \right) \leq T, \forall u \in \mathcal{U}$, and $\min \sum\limits_{b \in \mathcal{B}} \sum\limits_{u \in \mathcal{U}} \sum_{c=1}^{C} \omega_{u} \Delta_c \left( \hat{x}_{b,u}^{\text{Ac},c}  + \hat{x}^{\text{BH},c}_{b,u}  \right)$. On the other hand, constraints $\frac{1}{ x_{b,u}^{\text{Ac},c} } \leq \hat{x}_{b,u}^{\text{Ac},c}$ and $\frac{1}{ x^{\text{BH},c}_{b} } \leq \hat{x}^{\text{BH},c}_{b,u}$ are transformed into concave forms, respectively, as follows:
\begin{align}\label{Constraint deadline with xhat Ac}
\ln \left( \hat{x}_{b,u}^{\text{Ac},c} \right) + \ln \left( x_{b,u}^{\text{Ac},c} \right) \geq 0, \forall b \in \mathcal{B}, u \in \mathcal{U}, c \in \mathcal{C},
\end{align}
\begin{align}\label{objective with xhat BH}
\ln \left( x^{\text{BH},c}_{b,u} \right) + \ln \left( \hat{x}^{\text{BH},c}_{b,u} \right) \geq 0, \forall b \in \mathcal{B}, u \in \mathcal{U}, c \in \mathcal{C}.
\end{align}

\section{PROOF OF PROPOSITION 2}\label{appendix proposition3}
In order to prove that after each iteration $t_2$, the objective function \eqref{obf problem power total latency} is improved (lowered) or remains constant, we first prove that in the proposed SCA approach with the D.C. approximation method, the objective function \eqref{obf problem power total latency equivalent epi 2} is improved or remains constant after each iteration $t_2$. According to (\ref{g approximated}), $g_{b,u}^{n_\text{Ac}} (\boldsymbol{p}^\text{Ac}_{t_2})$ is approximated to its first order approximation series where $\nabla g_{b,u}^{n_\text{Ac}} (\boldsymbol{p}^\text{Ac}_{t_2-1})$ is the supergradient of $g_{b,u}^{n_\text{Ac}} (\boldsymbol{p}^\text{Ac}_{t_2-1})$ for the previous iteration $(t_2-1)$ \cite{7100916,6678362}. Accordingly, we can conclude that
\begin{align}\label{inequality gradients}
g_{b,u}^{n_\text{Ac}} (\boldsymbol{p}^\text{Ac}_{t_2}) \leq g_{b,u}^{n_\text{Ac}} (\boldsymbol{p}^\text{Ac}_{t_2-1}) + \nabla g_{b,u}^{n_\text{Ac}} (\boldsymbol{p}^\text{Ac}_{t_2-1}) (\boldsymbol{p}^\text{Ac}_{t_2} - \boldsymbol{p}^\text{Ac}_{t_2-1}).
\end{align}
On the other hand, $R_{b,u}^{n_\text{Ac}} (\boldsymbol{p}^\text{Ac}_{t_2})$ is approximated to a concave form $\hat{R}_{b,u}^{n_\text{Ac}} (\boldsymbol{p}^\text{Ac}_{t_2})$ by using \eqref{rate approximate}. It follows that
\begin{align}\label{inequality minimum rate downlink}
\sum\limits_{b \in \mathcal{B}} \sum_{n_\text{Ac}=1}^{N_\text{Ac}}
\hat{R}_{b,u}^{n_\text{Ac}} (\boldsymbol{p}^\text{Ac}_{t_2}) \geq R^\text{min}_{u} , \forall u \in \mathcal{U}.
\end{align}
In addition, we have
\begin{align}\label{inequality minimum rate downlink 2}
\hat{R}_{b,u}^{\text{Ac}} (\boldsymbol{p}^\text{Ac}_{t_2}) \geq  \theta_{b,u} s_c x_{b,u}^{\text{Ac},c,(t_2)}, \forall b \in \mathcal{B}, u \in \mathcal{U}, c \in \mathcal{C}.
\end{align}
Using \eqref{inequality gradients} and \eqref{inequality minimum rate downlink} it can be shown that
\begin{align}\label{conclision minimum rate inequality}
\sum\limits_{b \in \mathcal{B}} R_{b,u}^{\text{Ac}} (\boldsymbol{p}^\text{Ac}_{t_2}) =
\sum\limits_{b \in \mathcal{B}} \sum_{n_\text{Ac}=1}^{N_\text{Ac}}
\mathbb{E}_{\boldsymbol{h}} \bigg{ \{ }
\gamma_{b,u}^{n_\text{Ac}} \bigg(  f_{b,u}^{n_\text{Ac}} (\boldsymbol{p}^\text{Ac}_{t_2}) - g_{b,u}^{n_\text{Ac}} (\boldsymbol{p}^\text{Ac}_{t_2})
\bigg)
\bigg{ \} } \geq R^\text{min}_{u} , \forall u \in \mathcal{U}.
\end{align}
Moreover, using \eqref{inequality gradients} and \eqref{inequality minimum rate downlink 2}, we can conclude that
\begin{align}\label{conclision minimum rate inequality 2}
R_{b,u}^{\text{Ac}} (\boldsymbol{p}^\text{Ac}_{t_2}) \geq \theta_{b,u} s_c x_{b,u}^{\text{Ac},c,(t_2)}, \forall b \in \mathcal{B}, u \in \mathcal{U}, c \in \mathcal{C}.
\end{align}
In agreement with \eqref{conclision minimum rate inequality} and \eqref{conclision minimum rate inequality 2}, it is proved that after each iteration $t_2$, the convex approximated optimization problem \eqref{power Problem approximated} remains in the feasible region of \eqref{problem power total latency equivalent epi 2}. Moreover, according to \eqref{inequality gradients}, we have
\begin{multline}\label{ineq power obj01}
\sum\limits_{b \in \mathcal{B}} \sum\limits_{u \in \mathcal{U}} \sum_{c=1}^{C} \omega_{u} \bigg(
\sum_{n_\text{Ac}=1}^{N_\text{Ac}}
\mathbb{E}_{\boldsymbol{h}} \bigg{ \{ }
\gamma_{b,u}^{n_\text{Ac}} \bigg(  f_{b,u}^{n_\text{Ac}} ({\boldsymbol{p}^\text{Ac}_{t_2}}^*) - g_{b,u}^{n_\text{Ac}} ({\boldsymbol{p}^\text{Ac}_{t_2}}^*)
\bigg)
\bigg{ \} }
\bigg)
\leq    \\
\sum\limits_{b \in \mathcal{B}} \sum\limits_{u \in \mathcal{U}} \sum_{c=1}^{C} \omega_{u} \bigg(
\sum_{n_\text{Ac}=1}^{N_\text{Ac}}
\mathbb{E}_{\boldsymbol{h}} \bigg{ \{ }
\gamma_{b,u}^{n_\text{Ac}} \bigg(  f_{b,u}^{n_\text{Ac}} ({\boldsymbol{p}^\text{Ac}_{t_2}}^*) - g_{b,u}^{n_\text{Ac}} (\boldsymbol{p}^\text{Ac}_{t_2-1}) -
\nabla g_{b,u}^{n_\text{Ac}} (\boldsymbol{p}^\text{Ac}_{t_2-1})
({\boldsymbol{p}^\text{Ac}_{t_2}}^* - \boldsymbol{p}^\text{Ac}_{t_2-1})
\bigg)
\bigg{ \} }
\bigg).
\end{multline}
Furthermore, we can easily show that
\begin{multline}\label{ineq power obj02}
\sum\limits_{b \in \mathcal{B}} \sum\limits_{u \in \mathcal{U}} \sum_{c=1}^{C} \omega_{u} \bigg(
\sum_{n_\text{Ac}=1}^{N_\text{Ac}}
\mathbb{E}_{\boldsymbol{h}} \bigg{ \{ }
\gamma_{b,u}^{n_\text{Ac}} \bigg(  f_{b,u}^{n_\text{Ac}} ({\boldsymbol{p}^\text{Ac}_{t_2}}^*) - g_{b,u}^{n_\text{Ac}} (\boldsymbol{p}^\text{Ac}_{t_2-1}) -
\nabla g_{b,u}^{n_\text{Ac}} (\boldsymbol{p}^\text{Ac}_{t_2-1})
({\boldsymbol{p}^\text{Ac}_{t_2}}^* - \boldsymbol{p}^\text{Ac}_{t_2-1})
\bigg)
\bigg{ \} }
\bigg) =    \\
\min_{\boldsymbol{p}_{t_2}}
\sum\limits_{b \in \mathcal{B}} \sum\limits_{u \in \mathcal{U}} \sum_{c=1}^{C} \omega_{u} \bigg(
\sum_{n_\text{Ac}=1}^{N_\text{Ac}}
\mathbb{E}_{\boldsymbol{h}} \bigg{ \{ }
\gamma_{b,u}^{n_\text{Ac}} \bigg(  f_{b,u}^{n_\text{Ac}} (\boldsymbol{p}^\text{Ac}_{t_2}) - g_{b,u}^{n_\text{Ac}} (\boldsymbol{p}^\text{Ac}_{t_2-1}) -
\nabla g_{b,u}^{n_\text{Ac}} (\boldsymbol{p}^\text{Ac}_{t_2-1})
(\boldsymbol{p}^\text{Ac}_{t_2} - \boldsymbol{p}^\text{Ac}_{t_2-1})
\bigg)
\bigg{ \} }
\bigg)
\leq    \\
\sum\limits_{b \in \mathcal{B}} \sum\limits_{u \in \mathcal{U}} \sum_{c=1}^{C} \omega_{u} \bigg(
\sum_{n_\text{Ac}=1}^{N_\text{Ac}}
\mathbb{E}_{\boldsymbol{h}} \bigg{ \{ }
\gamma_{b,u}^{n_\text{Ac}} \bigg(  f_{b,u}^{n_\text{Ac}} (\boldsymbol{p}^\text{Ac}_{t_2-1}) - g_{b,u}^{n_\text{Ac}} (\boldsymbol{p}^\text{Ac}_{t_2-1}) -
\nabla g_{b,u}^{n_\text{Ac}} (\boldsymbol{p}^\text{Ac}_{t_2-1})
(\boldsymbol{p}^\text{Ac}_{t_2-1} - \boldsymbol{p}^\text{Ac}_{t_2-1})
\bigg)
\bigg{ \} }
\bigg)  =   \\
\sum\limits_{b \in \mathcal{B}} \sum\limits_{u \in \mathcal{U}} \sum_{c=1}^{C} \omega_{u} \bigg(
\sum_{n_\text{Ac}=1}^{N_\text{Ac}}
\mathbb{E}_{\boldsymbol{h}} \bigg{ \{ }
\gamma_{b,u}^{n_\text{Ac}} \bigg(  f_{b,u}^{n_\text{Ac}} (\boldsymbol{p}^\text{Ac}_{t_2-1}) - g_{b,u}^{n_\text{Ac}} (\boldsymbol{p}^\text{Ac}_{t_2-1})
\bigg)
\bigg{ \} }
\bigg).
\end{multline}
Based on \eqref{inequality minimum rate downlink 2} and \eqref{Constraint latency user epi2  power}, after solving \eqref{power Problem approximated} at each iteration $t_2$, the gap between $x_{b,u}^{\text{Ac},c,(t_2)}$ and $\frac{\theta_{b,u} \sum_{n_\text{Ac}=1}^{N_\text{Ac}}
\mathbb{E}_{\boldsymbol{h}} \bigg{ \{ }
\gamma_{b,u}^{n_\text{Ac}} \bigg(  f_{b,u}^{n_\text{Ac}} (\boldsymbol{p}^\text{Ac}_{t_2}) - g_{b,u}^{n_\text{Ac}} (\boldsymbol{p}^\text{Ac}_{t_2-1}) -
\nabla g_{b,u}^{n_\text{Ac}} (\boldsymbol{p}^\text{Ac}_{t_2-1})
(\boldsymbol{p}^\text{Ac}_{t_2} - \boldsymbol{p}^\text{Ac}_{t_2-1})
\bigg)
\bigg{ \} } }
{s_c}$ and the gap between $\frac{1}{x^{\text{BH},c,(t_2)}_{b,u}}$ and $\frac{ \theta_{b,u}  (1-\rho_{b,c}) s_c }
{\beta_{b,u} R_{b}^{\text{BH}} }$ tend to zero, respectively. Therefore, based on inequalities \eqref{ineq power obj01} and \eqref{ineq power obj02} and according to Appendix \ref{appendix EquivalentTransformation}, it can be concluded that after each iteration $t_2$, we have
\begin{align}\label{1barx ineq}
\sum\limits_{b \in \mathcal{B}} \sum\limits_{u \in \mathcal{U}} \sum_{c=1}^{C} \omega_{u} \Delta_c \left( \frac{1}{ x_{b,u}^{\text{Ac},c,(t_2)} } + \frac{1}{ x^{\text{BH},c,(t_2)}_{b,u} } \right) \leq \sum\limits_{b \in \mathcal{B}} \sum\limits_{u \in \mathcal{U}} \sum_{c=1}^{C} \omega_{u} \Delta_c  \left( \frac{1}{ x_{b,u}^{\text{Ac},c,(t_2-1)} } + \frac{1}{ x^{\text{BH},c,(t_2-1)}_{b,u} } \right).
\end{align}
By using the fact that after solving \eqref{power Problem approximated} at each iteration $t_2$, the gap between $\frac{1}{ x_{b,u}^{\text{Ac},c} }$ and $\hat{x}_{b,u}^{\text{Ac},c}$, and the gap between $\frac{1}{ x^{\text{BH},c}_{b,u} }$ and $\hat{x}^{\text{BH},c}_{b,u}$ tend to zero (based on inequalities $\frac{1}{ x_{b,u}^{\text{Ac},c} } \leq \hat{x}_{b,u}^{\text{Ac},c}$ and $\frac{1}{ x^{\text{BH},c}_{b,u} } \leq \hat{x}^{\text{BH},c}_{b,u}$, respectively), we can easily show that after each iteration $t_2$, we have
\begin{align}\label{xhat obj ineq}
\sum\limits_{b \in \mathcal{B}} \sum\limits_{u \in \mathcal{U}} \sum_{c=1}^{C} \omega_{u} \Delta_c  \left( \hat{x}_{b,u}^{\text{Ac},c,(t_2)}  + \hat{x}^{\text{BH},c,(t_2)}_{b,u}  \right) \leq \sum\limits_{b \in \mathcal{B}} \sum\limits_{u \in \mathcal{U}} \sum_{c=1}^{C} \omega_{u} \Delta_c  \left( \hat{x}_{b,u}^{\text{Ac},c,(t_2-1)}  + \hat{x}^{\text{BH},c,(t_2-1)}_{b,u}  \right).
\end{align}
Pursuant to \eqref{conclision minimum rate inequality} and \eqref{conclision minimum rate inequality 2}, it is proved that after each iteration $t_2$, the objective function \eqref{obf problem power total latency equivalent epi 2} is lowered or remains constant. Besides, based on \eqref{xAc introducing constraint} and \eqref{xBH introducing constraint}, and inequalities $\frac{1}{ x_{b,u}^{\text{Ac},c} } \leq \hat{x}_{b,u}^{\text{Ac},c}$ and $\frac{1}{ x^{\text{BH},c}_{b,u} } \leq \hat{x}^{\text{BH},c}_{b,u}$, it can be observed that $\hat{x}_{b,u}^{\text{Ac},c}$ and $\hat{x}^{\text{BH},c}_{b,u}$ are the upper-bound values of $\theta_{b,u} \frac{s_c}{  R_{b,u}^{\text{Ac}}  }$ and $\theta_{b,u} (1-\rho_{b,c}) \frac{s_c} { \beta_{b,u} R_{b}^{\text{BH}}  }$, respectively, and also are lower bounded by zero. By using \eqref{latency each user probability} and \eqref{xhat obj ineq}, it can be concluded that
$
0 \leq D^{\text{Pr},\text{tot}}(\boldsymbol{p}_{t_2})
\leq
D^{\text{Pr},\text{tot}}(\boldsymbol{p}_{t_2-1}).
$
On the other hand, pursuant to \eqref{conclision minimum rate inequality} and \eqref{conclision minimum rate inequality 2}, and using the fact that \eqref{problem power total latency equivalent epi 2} is an equivalent transformation form of \eqref{problem power total latency}, it is proved that after each iteration $t_2$, the convex approximated problem \eqref{power Problem approximated} remains in the feasible region of \eqref{problem power total latency}. Accordingly, the proposed algorithm for solving \eqref{problem power total latency} is improved or remains constant after each iteration $t_2$ and consequently converges to a local optimum solution.

%References:::
\hyphenation{op-tical net-works semi-conduc-tor}
\bibliographystyle{IEEEtran}
\bibliography{IEEEabrv,Bibliography}

\begin{thebibliography}{10}
\providecommand{\url}[1]{#1}
\csname url@samestyle\endcsname
\providecommand{\newblock}{\relax}
\providecommand{\bibinfo}[2]{#2}
\providecommand{\BIBentrySTDinterwordspacing}{\spaceskip=0pt\relax}
\providecommand{\BIBentryALTinterwordstretchfactor}{4}
\providecommand{\BIBentryALTinterwordspacing}{\spaceskip=\fontdimen2\font plus
\BIBentryALTinterwordstretchfactor\fontdimen3\font minus
  \fontdimen4\font\relax}
\providecommand{\BIBforeignlanguage}[2]{{%
\expandafter\ifx\csname l@#1\endcsname\relax
\typeout{** WARNING: IEEEtran.bst: No hyphenation pattern has been}%
\typeout{** loaded for the language `#1'. Using the pattern for}%
\typeout{** the default language instead.}%
\else
\language=\csname l@#1\endcsname
\fi
#2}}
\providecommand{\BIBdecl}{\relax}
\BIBdecl

\bibitem{wong_schober_ng_wang_2017}
V.~W.~S. Wong, R.~Schober, D.~W.~K. Ng, and L.-C. Wang, \emph{Key Technologies
  for {5G} Wireless Systems}.\hskip 1em plus 0.5em minus 0.4em\relax Cambridge
  U.K.: Cambridge University Press, 2017.

\bibitem{6871674}
E.~Bastug, M.~Bennis, and M.~Debbah, ``Living on the edge: The role of
  proactive caching in {5G} wireless networks,'' \emph{IEEE Communications
  Magazine}, vol.~52, no.~8, pp. 82--89, Aug. 2014.

\bibitem{7155502}
J.~Li, Y.~Chen, Z.~Lin, W.~Chen, B.~Vucetic, and L.~Hanzo, ``Distributed
  caching for data dissemination in the downlink of heterogeneous networks,''
  \emph{IEEE Transactions on Communications}, vol.~63, no.~10, pp. 3553--3568,
  Oct. 2015.

\bibitem{7530876}
W.~Jiang, G.~Feng, and S.~Qin, ``Optimal cooperative content caching and
  delivery policy for heterogeneous cellular networks,'' \emph{IEEE
  Transactions on Mobile Computing}, vol.~16, no.~5, pp. 1382--1393, May 2017.

\bibitem{6195469}
N.~Golrezaei, K.~Shanmugam, A.~G. Dimakis, A.~F. Molisch, and G.~Caire,
  ``Femtocaching: Wireless video content delivery through distributed caching
  helpers,'' in \emph{Proc. IEEE International Conference on Computer
  Communications (INFOCOM)}, Orlando, FL, USA, Mar. 2012, pp. 1107--1115.

\bibitem{7322204}
H.~Hsu and K.~Chen, ``A resource allocation perspective on caching to achieve
  low latency,'' \emph{IEEE Communications Letters}, vol.~20, no.~1, pp.
  145--148, Jan. 2016.

\bibitem{7828114}
F.~Cheng, Y.~Yu, Z.~Zhao, N.~Zhao, Y.~Chen, and H.~Lin, ``Power allocation for
  cache-aided small-cell networks with limited backhaul,'' \emph{IEEE Access},
  vol.~5, pp. 1272--1283, Jan. 2017.

\bibitem{CloudPrinciples}
T.~Q.~S. Quek, M.~Peng, O.~Simeone, and W.~Yu, \emph{Cloud radio access
  networks: Principles, technologies, and applications}.\hskip 1em plus 0.5em
  minus 0.4em\relax Cambridge U.K.: Cambridge University Press, 2017.

\bibitem{8008769}
J.~Liu, B.~Bai, J.~Zhang, and K.~B. Letaief, ``Cache placement in fog-{RAN}s:
  From centralized to distributed algorithms,'' \emph{IEEE Transactions on
  Wireless Communications}, vol.~16, no.~11, pp. 7039--7051, Nov. 2017.

\bibitem{8030120}
W.~C. Ao and K.~Psounis, ``Fast content delivery via distributed caching and
  small cell cooperation,'' \emph{IEEE Transactions on Mobile Computing},
  vol.~17, no.~5, pp. 1048--1061, May 2018.

\bibitem{7558153}
S.~Park, O.~Simeone, and S.~S. Shitz, ``Joint optimization of cloud and edge
  processing for fog radio access networks,'' \emph{IEEE Transactions on
  Wireless Communications}, vol.~15, no.~11, pp. 7621--7632, Nov. 2016.

\bibitem{7805409}
R.~G. Stephen and R.~Zhang, ``Green {OFDMA} resource allocation in
  cache-enabled {CRAN},'' in \emph{Proc. IEEE Online Conference on Green
  Communications (OnlineGreenComm)}, Piscataway, NJ, USA, Dec. 2016, pp.
  70--75.

\bibitem{5580131}
T.~Wang and L.~Vandendorpe, ``Iterative resource allocation for maximizing
  weighted sum min-rate in downlink cellular {OFDMA} systems,'' \emph{IEEE
  Transactions on Signal Processing}, vol.~59, no.~1, pp. 223--234, Jan. 2011.

\bibitem{1424595}
M.~Dianati, X.~Shen, and S.~Naik, ``A new fairness index for radio resource
  allocation in wireless networks,'' in \emph{IEEE Wireless Communications and
  Networking Conference (WCNC)}, vol.~2, New Orleans, LA, USA, Mar. 2005, pp.
  712--717.

\bibitem{Aquantitativemeasure}
R.~Jain, D.~M. Chiu, and W.~R. Hawe, \emph{A quantitative measure of fairness
  and discrimination for resource allocation in shared computer system}.\hskip
  1em plus 0.5em minus 0.4em\relax Hudson, MA: Eastern Research Laboratory,
  Digital Equipment Corporations, 1984.

\bibitem{7406764}
K.~Wang, H.~Li, F.~R. Yu, and W.~Wei, ``Virtual resource allocation in
  software-defined information-centric cellular networks with device-to-device
  communications and imperfect {CSI},'' \emph{IEEE Transactions on Vehicular
  Technology}, vol.~65, no.~12, pp. 10\,011--10\,021, Dec. 2016.

\bibitem{7514219}
J.~Zhang, X.~Zhang, and W.~Wang, ``Cache-enabled software defined heterogeneous
  networks for green and flexible {5G} networks,'' \emph{IEEE Access}, vol.~4,
  pp. 3591--3604, July 2016.

\bibitem{7744833}
R.~Huo, F.~R. Yu, T.~Huang, R.~Xie, J.~Liu, V.~C.~M. Leung, and Y.~Liu,
  ``Software defined networking, caching, and computing for green wireless
  networks,'' \emph{IEEE Communications Magazine}, vol.~54, no.~11, pp.
  185--193, Nov. 2016.

\bibitem{7925732}
Z.~Tan, X.~Li, F.~R. Yu, L.~Chen, H.~Ji, and V.~C.~M. Leung, ``Joint access
  selection and resource allocation in cache-enabled {HCN}s with {D2D}
  communications,'' in \emph{Proc. IEEE Wireless Communications and Networking
  Conference (WCNC)}, San Francisco, CA, USA, Mar. 2017, pp. 1--6.

\bibitem{7227022}
A.~Abboud, E.~Baştuğ, K.~Hamidouche, and M.~Debbah, ``Distributed caching in
  {5G} networks: An alternating direction method of multipliers approach,'' in
  \emph{Proc. IEEE International Workshop on Signal Processing Advances in
  Wireless Communications (SPAWC)}, June 2015, pp. 171--175.

\bibitem{7510807}
J.~Liu, B.~Bai, J.~Zhang, and K.~B. Letaief, ``Content caching at the wireless
  network edge: A distributed algorithm via belief propagation,'' in
  \emph{Proc. IEEE International Conference on Communications (ICC)}, Kuala
  Lumpur, Malaysia, May 2016, pp. 1--6.

\bibitem{6600983}
K.~Shanmugam, N.~Golrezaei, A.~G. Dimakis, A.~F. Molisch, and G.~Caire,
  ``Femtocaching: Wireless content delivery through distributed caching
  helpers,'' \emph{IEEE Transactions on Information Theory}, vol.~59, no.~12,
  pp. 8402--8413, Dec. 2013.

\bibitem{7150324}
M.~Ji, G.~Caire, and A.~F. Molisch, ``Wireless device-to-device caching
  networks: Basic principles and system performance,'' \emph{IEEE Journal on
  Selected Areas in Communications}, vol.~34, no.~1, pp. 176--189, Jan. 2016.

\bibitem{7536886}
J.~Tang, T.~Q.~S. Quek, and W.~P. Tay, ``Joint resource segmentation and
  transmission rate adaptation in cloud {RAN} with caching as a service,'' in
  \emph{Proc. IEEE International Workshop on Signal Processing Advances in
  Wireless Communications (SPAWC)}, Edinburgh, UK, July 2016, pp. 1--6.

\bibitem{VASILAKOS2016306}
X.~Vasilakos, V.~A. Siris, and G.~C. Polyzos, ``Addressing niche demand based
  on joint mobility prediction and content popularity caching,'' \emph{Computer
  Networks}, vol. 110, pp. 306--323, Dec. 2016.

\bibitem{Fairnessawarecooperative}
D.~Wei, K.~Zhu, and X.~Wang, ``Fairness-aware cooperative caching scheme for
  mobile social networks,'' in \emph{Proc. IEEE International Conference on
  Communications (ICC)}, June 2014, pp. 2484--2489.

\bibitem{5463218}
M.~Salem, A.~Adinoyi, M.~Rahman, H.~Yanikomeroglu, D.~Falconer, and Y.~Kim,
  ``Fairness-aware radio resource management in downlink {OFDMA} cellular relay
  networks,'' \emph{IEEE Transactions on Wireless Communications}, vol.~9,
  no.~5, pp. 1628--1639, May 2010.

\bibitem{7297864}
M.~Ji, G.~Caire, and A.~F. Molisch, ``The throughput-outage tradeoff of
  wireless one-hop caching networks,'' \emph{IEEE Transactions on Information
  Theory}, vol.~61, no.~12, pp. 6833--6859, Dec. 2015.

\bibitem{7417343}
X.~Peng, J.~C. Shen, J.~Zhang, and K.~B. Letaief, ``Backhaul-aware caching
  placement for wireless networks,'' in \emph{Proc. IEEE Global Communications
  Conference (GLOBECOM)}, Dec. 2015, pp. 1--6.

\bibitem{7158137}
F.~Zhang, C.~Xu, Y.~Zhang, K.~K. Ramakrishnan, S.~Mukherjee, R.~Yates, and
  T.~Nguyen, ``Edgebuffer: Caching and prefetching content at the edge in the
  mobilityfirst future internet architecture,'' in \emph{Proc. IEEE
  International Symposium on A World of Wireless, Mobile and Multimedia
  Networks (WoWMoM)}, June 2015, pp. 1--9.

\bibitem{Sobkowicz2013}
P.~Sobkowicz, M.~Thelwall, K.~Buckley, G.~Paltoglou, and A.~Sobkowicz,
  ``Lognormal distributions of user post lengths in internet discussions - a
  consequence of the {Weber-Fechner} law?'' \emph{EPJ Data Science}, vol.~2,
  no.~1, 2013.

\bibitem{6678362}
D.~T. Ngo, S.~Khakurel, and T.~Le-Ngoc, ``Joint subchannel assignment and power
  allocation for {OFDMA} femtocell networks,'' \emph{IEEE Transactions on
  Wireless Communications}, vol.~13, no.~1, pp. 342--355, Jan. 2014.

\bibitem{7456319}
M.~R. Abedi, N.~Mokari, M.~R. Javan, and H.~Yanikomeroglu, ``Secure
  communication in {OFDMA}-based cognitive radio networks: An incentivized
  secondary network coexistence approach,'' \emph{IEEE Transactions on
  Vehicular Technology}, vol.~66, no.~2, pp. 1171--1185, Feb. 2017.

\bibitem{Jorswieckfractional}
A.~Zappone and E.~Jorswieck, ``Energy efficiency in wireless networks via
  fractional programming theory,'' \emph{Foundations and Trends in
  Communications and Information Theory}, vol.~11, no. 3-4, pp. 185--396, June
  2015.

\bibitem{Boydconvex}
S.~Boyd and L.~Vandenberghe, \emph{Convex Optimization}.\hskip 1em plus 0.5em
  minus 0.4em\relax Cambridge U.K.: Cambridge University Press, 2009.

\bibitem{7100916}
N.~Mokari, F.~Alavi, S.~Parsaeefard, and T.~Le-Ngoc, ``Limited-feedback
  resource allocation in heterogeneous cellular networks,'' \emph{IEEE
  Transactions on Vehicular Technology}, vol.~65, no.~4, pp. 2509--2521, Apr.
  2016.

\bibitem{CVXmatlab}
C.~Research, ``{CVX}: {M}atlab software for disciplined convex programming,
  version 2.0,'' Aug. 2012 [Online] {A}vailable: http://cvxr.com/cvx.

\bibitem{MOSEKsolver}
Https://www.mosek.com/products/mosek.

\bibitem{6251827}
D.~W.~K. Ng, E.~S. Lo, and R.~Schober, ``Energy-efficient resource allocation
  in {OFDMA} systems with large numbers of base station antennas,'' \emph{IEEE
  Transactions on Wireless Communications}, vol.~11, no.~9, pp. 3292--3304,
  Sept. 2012.

\bibitem{1658226}
W.~Yu and R.~Lui, ``Dual methods for nonconvex spectrum optimization of
  multicarrier systems,'' \emph{IEEE Transactions on Communications}, vol.~54,
  no.~7, pp. 1310--1322, July 2006.

\bibitem{6883210}
K.~Poularakis, G.~Iosifidis, and L.~Tassiulas, ``Approximation algorithms for
  mobile data caching in small cell networks,'' \emph{IEEE Transactions on
  Communications}, vol.~62, no.~10, pp. 3665--3677, Oct. 2014.

\bibitem{7562510}
Y.~Chen, M.~Ding, J.~Li, Z.~Lin, G.~Mao, and L.~Hanzo, ``Probabilistic
  small-cell caching: Performance analysis and optimization,'' \emph{IEEE
  Transactions on Vehicular Technology}, vol.~66, no.~5, pp. 4341--4354, May
  2017.

\end{thebibliography}

\end{document}